\documentclass[10pt]{article}
\usepackage{bm}
\usepackage{mathrsfs}
\usepackage{multirow}
\usepackage{amssymb,amsfonts,amsmath}
\usepackage{mathrsfs}
\usepackage{mathtools}
\usepackage{graphicx}
\usepackage{subfig}
\usepackage{multirow}
\usepackage{color}
\usepackage{fullpage}
\usepackage{hyperref}
\usepackage{tabularx}
\usepackage{tikz}
\usepackage{pgfplots}
\usepackage{float}
\usepackage{color}
\usepackage{bbm}
\usepackage{cleveref}
\usepackage{setspace}
\usepackage{tikz-cd}
\usepackage{pst-node}
\usepackage[font={footnotesize}]{caption}
\usepackage{lipsum}
\usepackage{mwe}
\usepackage{comment}
\usepackage{epstopdf}
\usepackage{algorithmic}

\DeclarePairedDelimiter{\ceil}{\lceil}{\rceil}

\usetikzlibrary{calc,backgrounds}
\usetikzlibrary{shapes.geometric, arrows}

\makeatletter
\newcommand*{\rom}[1]{\expandafter\@slowromancap\romannumeral #1@}
\makeatother

\makeatletter
\newcommand{\mathleft}{\@fleqntrue\@mathmargin0pt}
\newcommand{\mathcenter}{\@fleqnfalse}
\makeatother

\def\Box{\leavevmode\vbox{\hrule
     \hbox{\vrule\kern4pt\vbox{\kern4pt}%
           \vrule}\hrule}}
\def\blackbox{\leavevmode\vrule height 5pt width 4pt depth 0pt\relax}
\def\endproof{\null\hfill {$\blackbox$}\bigskip}

\def\paragraph#1{{\bf #1\ }}

\newtheorem{lemma}{Lemma}[section]

\newtheorem{corollary}[lemma]{Corollary}

\newtheorem{definition}[lemma]{Definition}

\newtheorem{proposition}[lemma]{Proposition}

\newtheorem{remark}{Remark}[section]

\newtheorem{assumption}{Assumption}[section]

\title{A mathematical framework for modelling order book dynamics} 
\author{Rama Cont$^{(1)}$, Pierre Degond$^{(2)}$,   Lifan Xuan$^{(3)}$} 
\date{} 
\begin{document}

\maketitle

$\mbox{}$

\vspace{-1 cm}

\begin{center}

$^{(1)}$ Mathematical Institute, University of Oxford, \\
Oxford, UK\\
rama.cont@maths.ox.ac.uk

\vspace{0.2cm}

$^{(2)}$ Institut de Math\'ematiques de Toulouse ; UMR5219 \\
Universit\'e de Toulouse ; CNRS \\
UPS, F-31062 Toulouse Cedex 9, France \\
pierre.degond@math.univ-toulouse.fr 

\vspace{0.2cm}

$^{(3)}$  Department of Mathematics, Imperial College London, \\
London SW7 2AZ, UK \\
xlfliza@gmail.com

\end{center}

\begin{abstract}
We present a general framework for modelling the dynamics of limit order books, built on the combination of two modelling ingredients: the order flow, modelled as a general spatial point process, and market clearing, modelled via a deterministic ‘mass transport’ operator acting on distributions of buy and sell orders. At the mathematical level, this corresponds to a natural decomposition of the infinitesimal generator describing the evolution of the limit order book into two operators: the generator of the order flow and the clearing operator.
Our model provides a flexible framework for modelling and simulating order book dynamics and studying various scaling limits of discrete order book models. We show that our framework includes previous models as special cases and yields insights into the interplay between order flow and price dynamics.
\end{abstract}



\medskip
\noindent
{\bf Acknowledgements:} PD holds a visiting professor association with the Department of Mathematics, Imperial College London, UK.

\section{Introduction}
\label{sec.Introduction}

\begin{figure}[h]
\centering
\includegraphics[scale = 0.5]{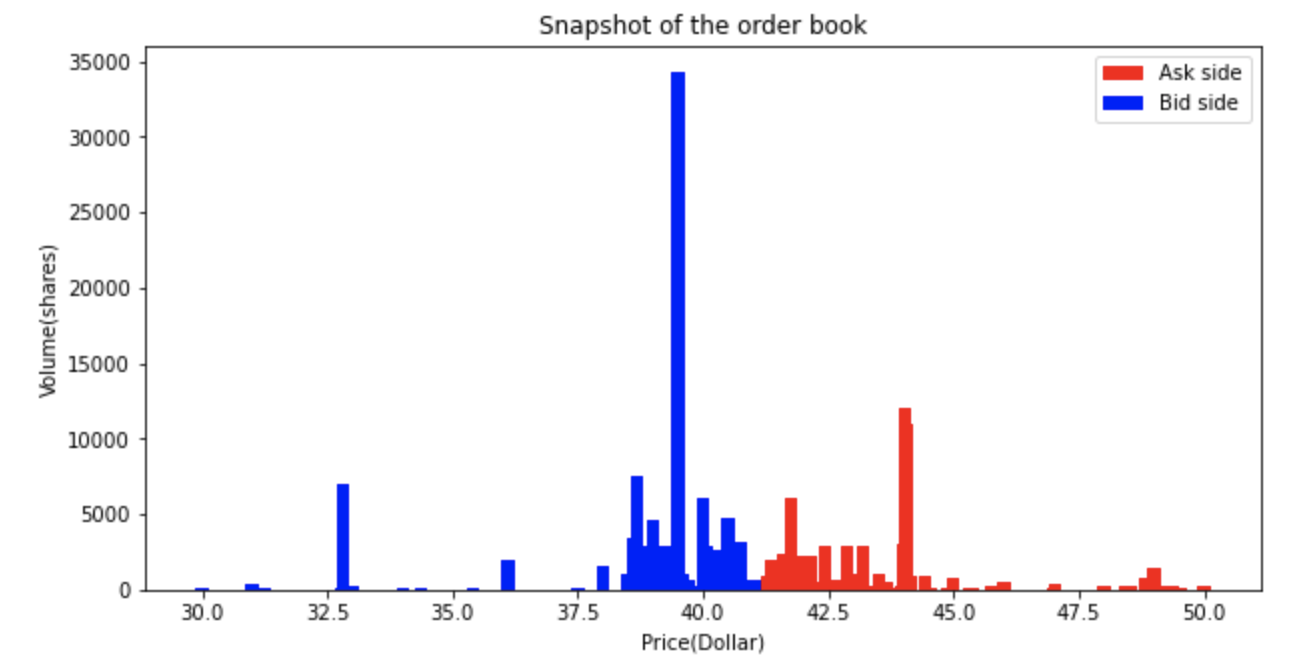}
\caption{MU(Micron) limit order book on 1st July 2019, with  limit buy orders in blue and  limit sell orders in red.}
\label{figure_limit_order_book}
\end{figure}

In the last few decades major financial markets have transitioned to electronic platforms where market participants may post buy and sell orders through a centralized {\it limit order book} (LOB)   \cite{cont2001}. These orders are then matched and executed according to   time and price priority rules.  The state of the limit order book, which represents outstanding buy and sell orders, constantly evolves through the arrival, execution and cancellation of buy and sell orders. Figure~\ref{figure_limit_order_book} shows a typical shape of the order book.

A quantitative understanding of the dynamics of limit order books is therefore important both from a theoretical standpoint for understanding the dynamics of supply and demand in order-driven markets, and from the standpoint of applications to intraday market modeling and optimal trade execution. 

The empirical study of limit order book dynamics has revealed various empirical regularities \cite{biais1995,bouchaud2002} across a range of different time scales  \cite{cont2001}. Faced with the challenge of accommodating these empirical features, many stochastic models for limit order book dynamics have been developed in the literature \cite{cont2012,cont2013, cont2010,daniels2003, huang2015,luckock2003,morariu2018state,vinkovskaya2014}. Queueing models \cite{cont2013, daniels2003, huang2015,luckock2003} represent the limit order book as a  system of interacting queues, driven by an order flow described as a spatial point process. Another class of models represents the state of the order book as a (pair of) densities, solution of a (stochastic) partial differential equation (SPDE) \cite{cont2019,hambly2020}. These two classes of models are connected by scaling limits such as fluid limits \cite{gao2018,horst2017weak,horst2017} and diffusion limits \cite{cont2012,cont2013,lakner2016} of discrete order book models  under suitable assumptions.

The sheer variety of stochastic models for limit order book dynamics makes it difficult to perform a comparative analysis, especially since these models  use different representations and assumptions as  starting points.
In the present work we propose an approach which  these models into a unifying overall framework, which can be useful for model comparison, model construction and also for investigating  asymptotic behaviour under various scaling assumptions. Specifically, we propose to decompose the dynamics of the limit order book process into two separate ingredients:
\begin{itemize}\item the incoming {\it order flow}, represented as a spatial point process, and 
\item the {\it market clearing} procedure, represented as an operator acting of a pair of distributions of buy and sell orders. 
\end{itemize}

\begin{figure}[h!]
\centering
\begin{tikzpicture}[scale=0.8, transform shape]
  \tikzstyle{state} = [draw, very thick, fill=white, rectangle, minimum height=3em, minimum width=7em, node distance=8em, font={\sffamily\bfseries}]
  \tikzstyle{stateEdgePortion} = [black,thick];
  \tikzstyle{stateEdge} = [stateEdgePortion,->];
  \tikzstyle{edgeLabel} = [pos=0.5, text centered, font={\sffamily\bfseries }];
   \node[state, name=CurrentState, text width=5em] {Current state:     $X_t\in\mathcal{L}$}; 
   \node[state, name=Intermediate, right of=CurrentState, xshift=6em,text width=6em]{Intermediate state: \break $X_t+\Delta X\in E$};
   \node[state, name=FinalState, right of=Intermediate, xshift=9em,text width=7em]{State at time $t+\Delta t$: \break $X_{t+\Delta t}\in\mathcal{L}$};
  
  \draw ($(CurrentState.east) + (0,.5em)$) 
      edge[stateEdge] node[edgeLabel, yshift=1em]{Order flow} node[edgeLabel, yshift=-1em]{Stochastic} 
      ($(Intermediate.west) + (0,.5em)$); 
  \draw ($(Intermediate.east)+(0,.5em)$) 
      edge[stateEdge] node[edgeLabel, yshift=1em, text width=8em]{Clearing operator} node[edgeLabel, yshift=-1em]{Deterministic} 
      ($(FinalState.west) + (0,.5em)$);  
      
 \coordinate (CloseA) at ($(CurrentState.south)+(0,-2em)$);
 \coordinate(CloseB) at ($(FinalState.south)+(0,-2em)$);
\draw (CurrentState.south) edge[stateEdgePortion] (CloseA);
 \draw (CloseA) edge[stateEdgePortion] node[edgeLabel, yshift=-1em]{Existing models} (CloseB);
 \draw (CloseB) edge[stateEdge] (FinalState.south);
\end{tikzpicture}
\caption{Decomposition of the order book dynamics.}
\label{fig.general_idea}
\end{figure}
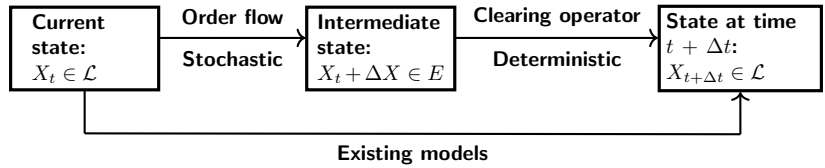

The dynamics of the order book is decomposed as shown in Figure \ref{fig.general_idea}: at time $t$, the current state of the order book is $X_t$ belonging to an appropriate state space $\mathcal{L}$. An order flow $\Delta X$ arrives in the order book after time $\Delta t$, changing the state of the order book from $X_t$ into an intermediate state $X_t+\Delta X$, which is an element of a larger space $E$. This transient state $X_t+\Delta X$ is then mapped to $X_{t+\Delta t}$ in the state space $\mathcal{L}$ by a clearing operator. We remark that the order flow in this framework can be any stochastic process. Since the order flow and the market clearing processes are independent, this framework can be adjusted to many different types of markets just by choosing the order flows and the clearing operators properly. Furthermore, this decomposition permits a modular approach in simulating LOBs where algorithms to simulate the order flow and the clearing operator can be developed independently. If the order flow is assumed to be a Markov process, then as the market clearing operator and the order flow are independent, the order book process is also Markovian. Therefore, the order book dynamics can be defined by its infinitesimal generator. Our framework allows us to write the infinitesimal generator as the composition of the generator of the order flow dynamics, which is simpler, and two operators, one related to the clearing operator, and one related to the embedding of the state space $\mathcal{L}$ of the LOB into the general space $E$ of the intermediate state. 

Outline: The paper is organised as follows: Section \ref{sec.Dynamics_of_the_limit_order_book} describes the building blocks involved in the dynamics of the limit order book: the order flow, the market clearing process and the order book process itself. A detailed mathematical description of the clearing operator is given in Section~\ref{sec.Market_clearing_via_order_matching}. Section~\ref{sec.the_markovian_framework} describes the application of this framework to a Markovian order book process, described through its infinitesimal generator and the associated backward Kolmogorov equation. Specifically, Section~\ref{sec.decomposition_of_the_generator_of_the_lob_process} uses this general framework to decompose the infinitesimal generator into  operators associated with order flow and clearing. Section~\ref{sec.dynamics_in_coordinates_centred_at_mid_price} adapts our framework to centred order books, i.e. order books with price coordinates centred at the  mid-price. In Section~\ref{sec.examples} we illustrate how this framework may be used to embed previous models from  the literature. 
In Section~\ref{sec.numerical_simulations}, we show how the decomposition shown in Figure \ref{fig.general_idea} translates into a modular approach to numerical simulations of limit order books. Section~\ref{sec.conclution} provides a conclusion and some perspectives.

\section{Dynamics of a limit order book}
\label{sec.Dynamics_of_the_limit_order_book}

The main idea of Figure \ref{fig.general_idea} is that a limit order book evolves through successive events involving an incoming order, immediately followed by   market clearing. We  now describe these steps in more details.

\subsection{Notations and definitions}
\label{sec.Notations_and_definitions}

Let $E:=\mathbb{N}^d\times\mathbb{N}^d$, where $d\in \mathbb{N}\backslash \{  0\}$ represents the maximum price of an order. Any element $ {X}=(X^+, X^-)\in E$ represents a configuration of limit orders: for $j\in\{ 1,\ldots,d\}$, $X^+_j$ (resp. $X^-_j$ ) represents the number of buy (resp. sell) orders submitted at price $j$.  

To any $Z\in \mathbb{N}^d$ we associate an integer-valued  measure on $\{ 1,\ldots,d\}$ defined as
\begin{equation}
\nu_Z=\sum_{k=1}^d Z_j\delta_{j}
\label{eq:vector_by_measure}
\end{equation}
and we define
\[
{\rm supp}(Z):={\rm supp}(\nu_Z)=\{{j\in\{1,2,\ldots,d \},Z_j>0} \}.
\]
Thus to any $X\in E$, we associate a pair $\nu_X=(\nu_X^+,\nu_X^-)$ of integer-valued measures on $\{ 1,\ldots,d\},$ where $\nu_X^\pm=\nu_{X^{\pm}}.$  We further define a truncation operator which will be used later to specify the clearing operator as follows: for $Z\in \mathbb{N}^d$ and $i\in \mathbb{N}$ we define $\tau^i(Z)\in \mathbb{N}^d$ and $\tau_i(Z)\in \mathbb{N}^d$ such that  $\nu_{\tau^i(Z)}= \nu_Z\ 1_{[i,\infty)}$ and $\nu_{\tau_i(Z)}= \nu_Z\ 1_{[0,i]}$. Then
 \[
\tau_{i}(z):=
\begin{cases}
(z_1,\ldots, z_i,0,\ldots,0) & \mbox{if}\qquad 1\leq i\leq d,\\
(0,0,\ldots,0) & \mbox{if}\qquad i\leq 0, \\
z & \mbox{if}\qquad i\geq d+1,\\
\end{cases}
\]
\[
\tau^{  i}(z):=
\begin{cases}
(0,\ldots,0, z_i,\ldots,z_d) & \mbox{if}\qquad1\leq i\leq d,\\
(0,0,\ldots,0) & \mbox{if}\qquad i\geq d+1, \\
z & \mbox{if}\qquad i\leq 0. \\
\end{cases}
\]

\begin{figure}[H]
\centering
\begin{tikzpicture}[scale=0.6]
   \begin{axis}[
        ybar,
        xmin=0,
        xmax=10,
        xtick={1,2,3,4,5,6,7,8,9,10},
        x tick style={draw=none},
        xticklabels={1,2,3,4,5,6,7,8,9},        
        ytick={1,2,3,4,5},
        ymin=0,
        ymax=5,
         every axis plot/.append style={
          bar width=10,
          bar shift=0pt,
          fill
        }
       ]
       \addplot [ybar,fill=blue] coordinates {
          (1,1)
          (2,3)
          (3,2)
          (4,1)  };
      \addplot  [ybar,fill=red] coordinates {      
          (6, 1)
          (8, 3)
          (9, 2) };
     \legend{buy orders, sell orders}
   \end{axis}
\end{tikzpicture}
\caption{An admissible configuration of a limit order book, where the blue bars denote buy orders and the red bars denote sell orders. In this case, $X^+=[1, 3,2,1,0,0,0,0,0]$, and $X^-=[0,0,0,0,0,1,0,3,2]$. Its ask price is 6 and bid price is 4.}
\label{fig.state_absolute_coordinate}
\end{figure}
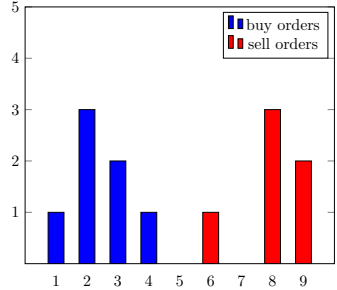

The state space of the LOB can be defined as follows. For $X\in E,$ we define
\begin{equation}
a(X):= \inf{\rm supp}(\nu_X^-),\qquad b(X)=\sup{\rm supp}(\nu_X^+), 
\label{eq:def_a_b}
\end{equation}
which will later represent the ask (resp. the bid) prices.
Here we use the convention
\begin{equation}
a(0)=\infty,\qquad b(0)=0.
\label{suuport_of_empty_set_absolute}
\end{equation}
The set of limit order book configurations is defined as:
\begin{equation}
\label{eq.state_space_absolute}
\begin{aligned}[b]
\mathcal{L}&:=\{ X=(X^+,X^-)\in E, \quad a(X)> b(X)\}.
\end{aligned}
\end{equation}
$X^+$ (resp. $X^-$) then represent the buy (resp. sell) orders in the limit order book and the condition (\ref{eq.state_space_absolute}) is interpreted as requiring that any outstanding buy order has a lower price than any outstanding sell order. The set ${\mathcal L}$ will be referred to as the set of ``admissible'' LOB configurations by opposition to the set $E$ which corresponds to fictional order book configurations in which the constraint $a(X)> b(X)$ is not enforced. Then $a(X)$ (resp. $b(X)$\ ) is the ask (resp. bid) price associated with the order book configuration $X$. Figure \ref{fig.state_absolute_coordinate} shows an admissible configuration of an order book. 

\subsection{Order flow}
\label{sec.order_flow} 
 
The state of the order book is modified through order arrivals, executions or cancellations. 
  
We distinguish  four types of 'elementary' events: limit buy and sell orders and cancellation of buy and sell orders. A limit buy/sell order is an order to buy/sell a specified amount of shares at a specified price and a cancellation of buy/sell order is an order to cancel a certain amount of standing buy/sell shares in an order book. Each  event is specified by a price $i\in \{1,\ldots,d  \}$ and a size $z\in\mathbb{N}\backslash \{0\}$. 
 Let \[
e_i:=(0,0,\ldots,1,0,\ldots,0)\in\mathbb{N}^d,
\]
where ``1'' is in the $i-$th position. 
Then an order book event of size $z$ at price $i$ maps a configuration $X\in E$ to a new configuration $X+\Delta X\in E$. This way $X\rightarrow X+\Delta X$ is described as below:
\begin{eqnarray}\label{eq.ordertypes}
  \begin{cases}
\mbox{limit buy order of size $z$:}&(X^+,X^-) \to ({X}^++ze_i,{X}^-)  ,\\
\mbox{limit sell order of size $z$:}& (X^+,X^-) \to ( {X}^+, {X}^-+ze_i) ,\\
\mbox{cancellation of a buy order of size $z$:}&(X^+,X^-) \to({X}^+-ze_i,{X}^-)\in E, \\
\mbox{cancellation of a sell order of size $z$:}&(X^+,X^-) \to( {X}^+, {X}^--ze_i) \in E.
\end{cases}  
\end{eqnarray}

Note that our description of limit orders includes both passive orders (not subject to immediate execution) and executable limit orders. Market buy (resp. sell) orders which are orders to buy (resp. sell) certain amount of shares at the best price may be seen as a special case of limit buy (resp. sell) orders with price $d$ (resp. $0$). If several orders arrive simultaneously, then their effects are additive. For example we may model a {\it price modification} for a limit order of size $z$ from $i$ to $i+1$ as the simultaneous cancellation of size $z$ at price $i$ and the posting of a new order of size $z$ at $i+1$.
We further remark that there is no restriction on the size $z\in\mathbb{N}\backslash \{0\}$ and the price level $i\in \{1,\ldots,d  \}$ of incoming limit orders; for cancellations, only existing orders may be cancelled i.e.  the size $z\leq |X^{\pm}_i|$. 

We model the arrival times and magnitudes of these order book events as a marked point process $(M_t,t\geq 0)$ on some probability space $(\Omega,{\cal F},\mathbb{P})$ taking values in $\mathbb{Z}^d\times \mathbb{Z}^d$. The jump times $t_1,t_2,..$ of $M$ represent the arrival times of orders and the jump amplitudes $\Delta M(t_k)\in \mathbb{Z}^d\times \mathbb{Z}^d $ represent the event sizes at $t_k$: the configuration $X(t_k-) $ of the order book is then perturbed by the incoming orders $\Delta M(t_k)$,  resulting in a new transient configuration  $$X(t_k-)\in {\cal L} \quad \to \quad X(t_k-)+\Delta M(t_k) \in E.$$

This process is what we call "the order flow". Throughout the paper, we will make the following assumption that the process remains in $E$ for all times, i.e. that
\begin{equation}
    \mathbb{P}\left(\forall t\geq 0,\quad  X(t-)+\Delta M(t) \in E\right)=1. \label{eq.statespacecondition}
\end{equation}
This condition is not restrictive: it simply means that sizes of cancellations cannot exceed existing queue sizes.

Note that in general the (transient) state may {\it not} belong to ${\cal L}$, i.e. $ X(t_k-)+\Delta M(t_k) \notin \mathcal{L} $ if, for instance, new orders can be executed with the outstanding orders. This is the case for example if we have an incoming limit sell order of size $z$ at a price $i<b(X)$, $\Delta M(t_k)= (0,z e_i)$.  

\subsection{Market clearing}
\label{sec.market_clearing}

Starting from an order book state $X(t-)\in\mathcal{L}$, the configuration $X(t-)+\Delta M(t)\in E$ immediately after an order flow event may not be, in general, an admissible LOB configuration, i.e. we may have $X(t-)+\Delta M(t)\notin {\cal L}$. The execution of orders, i.e. the matching between the sell and buy orders with compatible prices  then leads to an admissible  state  $X(t)\in\mathcal{L}$. This {\it market clearing} process may be mathematically  described in terms of a map ${\cal C}:E\mapsto \mathcal{L}$ which we call  a {\it clearing operator}:

\begin{definition}[Clearing operator]
A clearing operator is a map $\mathcal{C}: E\rightarrow \mathcal{L}$ whose set of invariant points is ${\cal L}$: 
\begin{equation}
    \forall X\in  {E}, \quad X\in \mathcal{L}\iff \mathcal{C}(X)=X.
\end{equation}
\label{def.clearing_absolute_general}
\end{definition}

Such a map then {\it projects} a general configuration of orders onto the set ${\cal L}$ of admissible LOB configurations.

The simplest example of such an operator is obtained from market clearing by {\it order matching}.
Matching of buy and sell orders with compatible prices does not change the configuration of orders if and only if there is no intersection between the buy and sell sides of the order book i.e. if $X\in\mathcal{L}$ and reciprocally. So it satisfies this definition.

The  dynamics of the limit order book may then be described  as a succession of order flow events, defined by the point process $M$, followed by market clearing. More specifically,   we will assume {\it continuous-time clearing}, which means the market is cleared after each new order book event. The state of the order book $X(t)=(X^+(t),X^-(t))$ is piecewise constant on $[t_{k-1}, t_k[$. At the k-th event,
\begin{itemize}
\item   the configuration $X(t_k-)=(X^+(t_k-),X^-(t_k-))\in\mathcal{L}  $ is perturbed by the incoming order flow $\Delta M(t_k)$,  resulting in a new transient configuration of orders   $X(t_k-)+\Delta M(t_k) \in E$. 
\item The market clearing operator acts on the perturbed/transient state  $X(t_k-)+\Delta M(t_k) \in E$, leading to a new order book configuration $ X(t_k)= {\cal C}(X(t_k-)+\Delta M(t_k) ) \in  \mathcal{L}$.
\end{itemize}
So the evolution of the order book at each order flow event may be pictured as follows:
$$ 
X(t_k-)\in\mathcal{L}  \quad  \xrightarrow{\textrm{Order flow}} \quad  X(t_k-)+\Delta M(t_k) \in E\quad\xrightarrow{\textrm{Clearing}} \quad X(t_k)= {\cal C}(X(t_k-)+\Delta M(t_k)) \in  \mathcal{L}.$$
\begin{remark}{\em Our modeling framework allows to examine clearing mechanisms other than continuous clearing. For example, one may also consider {\it frequent batch auctions} \cite{budish2014}, which correspond to batching the order flow over regular intervals $[T_k,T_{k+1})$ where $T_k=k\delta $ and clearing the batch at the end of each interval, which may be pictured as follows:
\begin{eqnarray*}
&&\hspace{-1cm}
 X(T_{k})\in\mathcal{L} \quad   \xrightarrow{\textrm{ Order flow}} \quad X(T_{k})+ M(T_{k+1})-M(T_{k}) \in E  \\
&&\hspace{6cm}
\xrightarrow{\textrm{Clearing}} \quad X(T_{k+1})= {\cal C}(X(T_{k})+ M(T_{k+1})-M(T_{k})) \in  \mathcal{L}.
\end{eqnarray*}
}
\end{remark}

\subsection{Market clearing via order matching}
\label{sec.Market_clearing_via_order_matching}

There are many ways to specify a market clearing operator. The simplest case corresponds to an {\it order-matching} method which pairs buy and sell orders with compatible price limits, in order to maximize the volume of transactions. This is how matching engines operate in many electronic exchanges.

Let $\mathcal{C}$ be the clearing operator corresponding to order matching. The quantity $Z(X) = X - {\mathcal C}(X)$ corresponds to executed order. For $Z\in \mathbb{N}^d$, denote by $|Z|=\sum_{i=1}^{d}Z_i$. Then, the order-matching clearing operator is defined as an operator having the following properties.
\begin{definition}\label{def.order-matching-clearing-operator-absolute}
Let $\mathcal{L}:E\rightarrow \mathcal{L}$ be a map with the following properties.
\begin{itemize} 
\item[(i)] The volume of executed orders is non-negative:
\begin{equation}\tag{A1}
    Z(X) \in E.
\end{equation}
\item[(ii)] The total volume of executed sell and buy orders are equal:
\begin{equation}\tag{A2}
    |Z^+(X)|=|Z^-(X)|.
\end{equation}
\item[(iii)] Any matched buy order has a price larger than any matched sell order:
\begin{equation}\tag{A3}
    \sup {\textrm{supp}} \, (Z^-(X))\leq \inf{\textrm{supp}} \, (Z^+(X)).
\end{equation}
\item[(iv)] Buy orders with higher prices have priority, and sell orders with lower prices have priority, i.e. any executed buy (resp. sell) order has price higher (resp. lower) than any non-matched buy (resp. sell) orders:
\begin{equation}\tag{A4}
    \sup{\textrm{supp}}\,(Z^-(X))\leq \inf{\textrm{supp}}\,(\mathcal{C}(X)^-) \quad \mbox{and} \quad \inf{\textrm{supp}}\,(Z^+(X))\geq \sup{\textrm{supp}}\,(\mathcal{C}(X)^+).
\end{equation}
\end{itemize}
Then $\mathcal{C}$ is called an order-matching clearing operator.
\end{definition}

\begin{figure}[h!]
\centering
\begin{tikzpicture}[scale=0.5]
   \begin{axis}[
        ybar,
        xmin=0,
        xmax=10,
        xtick={1,2,3,4,5,6,7,8,9,10},
        x tick style={draw=none},
        xticklabels={1,2,3,4,5,6,7,8,9},        
        ytick={1,2,3,4,5},
        ymin=0,
        ymax=5,
         every axis plot/.append style={
          bar width=10,
          bar shift=0pt,
          fill
        }
       ]
        \addplot  [ybar,fill=pink] coordinates {                
          (3, 3)
          (4, 2)
          (6, 1)};
       \addplot  [ybar,fill=cyan] coordinates {                
          (7, 2) };
       \addplot [ybar,fill=blue] coordinates {
          (1,1)
          (2,3)
          (3,2)
          (4,1)};
      \addplot  [ybar,fill=red] coordinates {                
          
          (8, 3)
          (9, 2)
          (5, 1) };
     \legend{new sell, new buy, buy orders, sell orders}
   \end{axis}
 \draw [->] (2.8,3.5) -- (2.8,2.9); 
 \draw (2.8,3.8) [black] node {clearing};
\end{tikzpicture}
\hspace{1cm}
\begin{tikzpicture}[scale=0.5]
   \begin{axis}[
        ybar,
        xmin=0,
        xmax=10,
        xtick={1,2,3,4,5,6,7,8,9,10},
        x tick style={draw=none},
        xticklabels={1,2,3,4,5,6,7,8,9},        
        ytick={1,2,3,4,5},
        ymin=0,
        ymax=5,
         every axis plot/.append style={
          bar width=10,
          bar shift=0pt,
          fill
        }
       ]
       \addplot [ybar,fill=blue] coordinates {
          (1,1)
          (2,3)
          (3,2)
          (4,1)  };
      \addplot  [ybar,fill=red] coordinates {      
          (6, 1)
          (8, 3)
          (9, 2)
          (5, 1) };
     \legend{buy orders, sell orders}
   \end{axis}
\end{tikzpicture}
\caption{Market clearing by order matching. Left: arrival of new orders. Right:  state $W=\mathcal{C}(X)$ of the order book after clearing. Market clearing is processed as follows: sell orders with prices lower than 4 are matched with buy orders with prices higher than 4 to achieve the maximum volume of transaction. Then the executed orders at each price level are determined by the price priority rule: one sell order at price 3, another sell order at price 4 and two buy orders at price 7.}
\label{fig.clearing_absolute}
\end{figure}
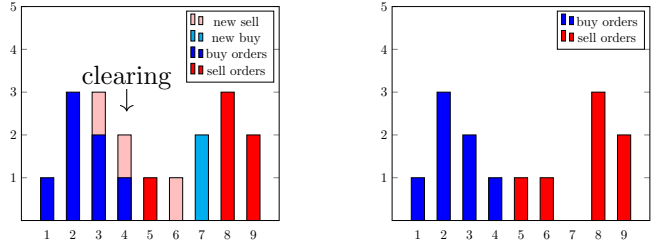

Figure~\ref{fig.clearing_absolute} shows an example of market clearing via order-matching, which we now define formally. For $X\in E$ and $k\in\{1,2,\ldots,d  \}$, define
\begin{equation}
B_{X}(k)=\nu^+_X([k,\infty)\ )=\sum_{i\geq k}X^+_i  \quad\mbox{  and   }\quad
S_{X}(k)=\nu^-_X([0,k]\ )=\sum_{i\leq k} X^-_i,
\label{eq.B_X_and_S_X}
\end{equation}
The function $k \to B_{X}(k) $ (resp. $k \to S_{X}(k) $) is non-increasing (resp. non-decreasing) so we can define the right (resp. left) inverse as  
\begin{equation}
B_{X}^{-1}(z)=\sup \{ i\in\{1,\ldots,d  \}:B_{X}(i)\geq z   \}\quad\mbox{  and   }\quad
S_{X}^{-1}(z)=\inf \{i\in\{1,\ldots,d  \}: S_{X}(i)\geq z   \}.
\label{eq.inverse_of_B_X_and_S_X}
\end{equation}

Define 
\begin{equation}
    g_{X}(k)=S_{X}(k)-B_{X}(k)
    \label{eq.g_x_absolute_coordinate}
\end{equation}
for $X\in E$ and $k\in \{ 1,\ldots,d \}$. The function $k \to g_X(k)$ is non-decreasing by definition. It is the difference between the total amount of sell orders at prices lower than $k$ and that of buy orders at prices higher than $k$. For any $X=(X^+,X^-)\in E$ we define a unique pair $(p_{B}(X),p_{A}(X))\in \mathbb{N}\times  \mathbb{N}$ such that
\begin{equation}
p_{A}(X) =\inf \{ k\in\{1,\ldots,d  \},  g_{X}(k)>0 \} \quad\mbox{  and   }\quad
p_{B}(X) =\sup \{ k\in\{1,\ldots,d  \},  g_{X}(k)<0 \}. 
\label{eq.p_a_p_b_absolute}
\end{equation}
Note that
\begin{equation}
X\in\mathcal{L}\Longleftrightarrow p_{B}(X)=b(X)\mbox{  and   }p_{A}(X)=a(X).
\label{eq.ask_price_and_bid_price_by_g_x}
\end{equation}
Thus, if $X\in\mathcal{L}$, then $p_A(X)$ and $p_B(X)$ are the ask and bid prices of $X$ respectively. Therefore, one can view $p_A(X)$ and $p_B(X)$ as extending the definition of $a(X)$ and $b(X)$ when $X \not \in {\mathcal L}$. Now we have the following proposition, whose proof can be found in Appendix~A. %

\begin{proposition}[Market clearing by order-matching]
The map $\mathcal{C}:E\rightarrow\mathcal{L}$ defined by $\mathcal{C}(X)=({\cal C}(X)^+,{\cal C}(X)^-)$ where
\begin{equation}
{\cal C}(X)^+=
\begin{cases}
\tau_{  p_{B}(X)}(X^+)& \mbox{  if }g_{X}(p_{B}(X))\leq-X^+_{p_{B}(X)}, \\
\tau_{  p_{B}(X)-1}(X^+)-\min\{0,g_{X}(p_{B}(X))\}e_{p_{B}(X)} &\mbox{  if } g_{X}(p_{B}(X))>-X^+_{p_{B}(X)}, \\
\end{cases}
\label{eq.discrete-clearingrule_buy}
\end{equation}
\begin{equation}
{\cal C}(X)^-=
\begin{cases}
\tau^{  p_{A}(X)}(X^-) & \mbox{  if }g_{X}(p_{A}(X))\geq X^-_{p_{A}(X)},\\
\tau^{  p_{A}(X)+1}(X^-)+\max\{0,g_{X}(p_{A}(X))\}e_{p_{A}(X)} & \mbox{  if } g_{X}(p_{A}(X))<X^-_{p_{A}(X)},
\end{cases}
\label{eq.discrete-clearingrule_sell}
\end{equation}
is the unique order-matching clearing operator in Def~\ref{def.order-matching-clearing-operator-absolute}.
\label{prop.ordermatching}
\end{proposition} 

\begin{remark}
If ${\cal C}(X)^+\neq 0$, it can be shown that $g_{X}(p_B(X))<0$. In this case, the min operator in  \eqref{eq.discrete-clearingrule_buy} can be removed i.e. $\min\{0,g_{X}(p_{B}(X))\} = g_{X}(p_{B}(X))$. A similar remark can be made when ${\cal C}(X)^-\neq 0$: then $g_{X}(p_A(X))>0$ and the maximum operator in  \eqref{eq.discrete-clearingrule_sell} can be removed. The cases where the min or max operators are needed correspond to cases where one side is wiped out after clearing.  
\end{remark}

The rationale for formulas \eqref{eq.discrete-clearingrule_buy} and \eqref{eq.discrete-clearingrule_sell} is as follows. Take \eqref{eq.discrete-clearingrule_buy} for ${\cal C}(X)^+$ as an example. By definition, we have $g_X(p_B(X)+1) \geq 0$. This means that, $i\geq p_B(X)+1$, the number of sell orders $S_X(i)$ exceeds the number of buy orders $B_X(i)$. Thus, all buy orders at such prices will be executed. On the other hand, we have $g_X(p_B(X)-1) < 0$. So, at prices smaller or equal to $p_B(X)-1$, the number of buy orders exceeds the number of sell orders, and hence all buy orders at such prices will remain unexecuted. The outstanding question is how many buy orders at price $p_B(X)$ do remain. For this, we examine the quantity 
$$Q =: g_X(p_B(X)) + X^+_{p_B(X)}= \sum_{i \leq p_B(X)} X_i^- - \sum_{i \geq p_B(X)+1} X_i^+. $$
This quantity represents the number of sell orders at prices less than or equal to $p_B(X)$ minus the number of buy orders at price larger than or equal to $p_B(X)+1$. So, we have the alternative: 
\begin{itemize}
    \item either $Q \leq 0$. Then, all sell orders at prices less than or equal to $p_B(X)$ can be executed against buy orders at price larger than or equal to $p_B(X)+1$. So, all buy orders at price $p_B(X)$ remain. Then, ${\cal C}(X)^+$ is just the truncation of $X^+$ by $0$ at prices larger than or equal to $p_B(X)+1$, i.e. ${\cal C}(X)^+ = \tau_{  p_{B}(X)}(X^+)$, which is the first alternative of \eqref{eq.discrete-clearingrule_buy}; 
    \item or $Q >0$. Then, we need to take some buy orders at price $p_B(X)$ to execute against the remaining sell orders at prices less than or equal to $p_B(X)$. This number of executed buy orders is exactly $Q$. In this case 
    $${\cal C}(X)^+ = \tau_{p_{B}(X)}(X^+) - Q e_{p_B(X)} = \tau_{p_{B}(X)-1}(X^+) + (X^+_{p_B(X)} - Q) e_{p_B(X)} = \tau_{p_{B}(X)-1}(X^+) -g_X(p_B(X)) e_{p_B(X)}, $$
    which is the second alternative of \eqref{eq.discrete-clearingrule_buy}. 
\end{itemize}


Now, we consider an important example which will be central to the study of Markovian order flow processes later. In the following proposition, we describe how market clearing by order matching operates in the case of elementary order flow events. 
The proof can be found in Appendix \ref{proof.2.5}.  

\begin{proposition}\label{prop.operation_clearing_absolute}
Let $X=(X^+,X^-)\in\mathcal{L}\cap E_+$, where $E_+:=\{ X\in E, \mbox{    such that    } X^+\neq 0\mbox{   and } X^-\neq 0 \}.$
An order flow event of size   $z<\min\{B_X(1), S_X(d)\}$ at price $k$ leads, after market clearing, to a new state $X'\in\mathcal{L}$ described as follows:
\begin{itemize}
    \item \textbf{Case 1:}  limit buy order with volume $z$   at price $k$:
    \begin{itemize}
    \item If $k\leq b(X)$, then $X'=(X^++ze_k,X^-)$ and $(b(X'),a(X'))=(b(X),a(X))$.
    \item if $b(X)<k<a(X)$, then $X'=(X^++ze_k,X^-)$ and $(b(X'),a(X'))=(k,a(X))$.
    \item if $a(X)\leq k<S_X^{-1}(z)$, then $X'=(X^++(z-S_X(k))e_k,\tau^{  k+1}(X^-))$ and  \\$(b(X'),a(X'))=(k,S_X^{-1}(S_X(k)+1))$.
    \item If $S_X^{-1}(z)\leq k$, then $X'=(X^+,\tau^{  S_X^{-1}(z)+1}(X^-)+(S_X(S_X^{-1}(z))-z)e_{S_X^{-1}(z)})$ and
    \\ $(b(X'),a(X'))=(b(X),S_X^{-1}(z+1))$.
    \end{itemize}
    \singlespacing
    \item \textbf{Case 2:}   limit sell order with volume $z$   at price $k$:
    \begin{itemize}
    \item If $a(X)\leq k$, then $X'=(X^{+},X^{-}+ze_k)$ and $(b(X'),a(X'))=(b(X),a(X))$.
    \item If $b(X)<k<a(X)$, then $X'=(X^{+},X^{-}+ze_k)$ and $(b(X'),a(X'))=(b(X),k)$.
    \item If $B_X^{-1}(z)<k\leq b(X)$, then $X'=(\tau_{  k-1}(X^+),X^-+(z-B_X(k))e_k)$ and \\$(b(X'),a(X'))=(B_X^{-1}(B_X(k)+1),k)$.
    \item If $k\leq B_X^{-1}(z)$, then $X'=(\tau_{  B_X^{-1}(z)-1}(X^+)+(B_X(B_X^{-1}(z))-z)e_{B_X^{-1}(z)},X^-)$ and \\ $(b(X'),a(X'))=(B_X^{-1}(z+1),a(X))$.
    \end{itemize}
    \singlespacing
    \item \textbf{Case 3:}   cancellation of $z$  buy orders at price $k$:
    \begin{itemize}
    \item If $k=b(X), z\leq X^+_k$, then $X'=(X^+-ze_k,X^-)$ and $(b(X'),a(X'))=(B_X^{-1}(z+1),a(X))$.
    \item If $k<b(X), z\leq X^+_k\mbox{ or }k=b(X), z<X^+_k$, then $X'=(X^+-ze_k,X^-)$ and $(b(X'),a(X'))=(b(X),a(X)$.
    \end{itemize}
    \singlespacing
    \item \textbf{Case 4:}   cancellation of $z$  sell orders at price $k$:
    \begin{itemize}
    \item If $k=a(X),z\leq X^-_k$, then $X'=(X^+-ze_k,X^-)$ and $(b(X'),a(X'))=(b(X),S_X^{-1}(z+1))$.
    \item If $k>a(X), z\leq X^-_k\mbox{ or }k=a(X), z<X^-_k$, then $X'=(X^+-ze_k,X^-)$ and $(b(X'),a(X'))=(b(X),a(X)$.
    \end{itemize}

\end{itemize}
\mathcenter
\end{proposition}

Again, let us give the rationale for these formulas. We take Case~1 as an example and leave the other cases to the reader. Case~1 corresponds to the arrival of a limit buy order at price $k$ and size $z$, i.e. the state before clearing is $(X^++ze_k,X^-)\in E$. There are four outcomes corresponding to the following subcases: 
\begin{itemize}
    \item Subcase 1: the price $k$ of the new order is less than or equal to the bid price $b(X)$. Then it cannot match any sell order and so, it cannot be executed. In other words, $(X^++ze_k,X^-)\in {\mathcal L}$ and this clearing operator acts as the identity on this configuration. Moreover, the new order does not change the bid price because its price is less than or equal to the  bid price $b(X)$. So, the ask and bid prices are unchanged as well.
    \item Subcase 2: the price $k$ is strictly larger than the bid price $b(X)$ and strictly smaller than the ask price $a(X)$. The situation is similar to the previous case because the order cannot be executed and the clearing operator acts as the identity on this configuration. However, the bid price changes to $k$ since the price of the new order is larger than $b(X)$. The ask price is kept unchanged.
    \item Subcase 3: the price $k$ is larger than or equal to $a(X)$ and strictly smaller than $S_X^{-1}(z)$. Indeed, $S_X^{-1}(z)$ is the smallest price such that the total number of sell orders of prices lower than or equal to this price is larger than or equal to $z$. If $k$ is strictly less than this price, this means that the new buy order can only be partially matched with sell orders of price lower than or equal to $k$. The number of such sell orders is $S_X(k)$ and thus, after clearing, it remains $z- S_X(k)$ buy orders at price $k$. All sell orders at price lower than or equal to $k$ are cleared. So the sell side of the order book after clearing is the truncation of the sell side before clearing with all sell orders at prices less or equal to $k$ set to $0$. The new bid price is $k$ while the new ask price corresponds to the lowest price of the sell orders of price strictly larger than $k$. Such price can be computed as $S_X^{-1}(S_X(k)+1))$.   
    \item Subcase 4: the price $k$ is larger than or equal to $S_X^{-1}(z)$. In this case, the buy order is totally executed because there are more sell orders at price lower than $k$ than the size $z$ of the order. So, after clearing, the buy side of the order book is back to its value before the arrival of the new order. On the sell side all orders at prices larger than or equal to $S_X^{-1}(z)+1$ remain. At price $S_X^{-1}(z)$, sell orders are matched with $z$ buy orders, so it remains $(S_X(S_X^{-1}(z))-z)$ orders at this price. The bid price is unchanged compared with its value before the arrival of the new order. The new ask price corresponds to the lowest price of the remaining sell orders of price larger than or equal to  $S_X^{-1}(z)$. Such price can be computed as $S_X^{-1}(z+1))$.  
\end{itemize}

We end with the following description of the set $\mathcal{C}^{-1}(\{X\})$ of order configurations which are mapped into $X\in \mathcal{L}$ through clearing (for a proof see Appendix \ref{proof.preimage}). The definition of $\mathcal{C}^{-1}(\{X\})$ will be used when we describe the order book dynamics when the order flow is Markovian.
\begin{proposition}[Pre-image of clearing operator]
\label{prop.pre_image_of_clearing_operator_absolute}
For $X\in \mathcal{L}$,  
\begin{eqnarray}
&&\hspace{1cm}
\mathcal{C}^{-1}(\{X\})=\{X+Z \, \, \big| \, \, Z\in E,  \,   |Z^+|=|Z^-|, \, \,  \label{eq.pre_image_set_expression}  \\
&&\hspace{2cm}
 \sup{\textrm{supp}} \, (Z^-) \leq \inf{\textrm{supp}} \, (Z^+) ,\quad 
\sup{\textrm{supp}} \, (Z^-) \leq a(X) , \,  \inf{\textrm{supp}} \, (Z^+) \geq b(X)   \}.
\nonumber
\end{eqnarray}
\end{proposition}

\newpage
\section{Markovian order flow}
\label{sec.the_markovian_framework}

We now consider the case where the order flow is described by a Markovian marked point process $M$, which corresponds to the case where the  intensity  of order arrivals and cancellations only depends on the current state of the order book. In this case the order book process $X$ is a Markov process with state space ${\cal L}$. The main result of this section is a decomposition of the infinitesimal generator of $X$ in terms of the infinitesimal generator of $M$ and of the clearing operator.

\subsection{Generator of the order flow}
\label{sec.generator_of_the_order_flow_process}
We consider a description of the order flow as a Markovian  point process, described by a transition kernel $p_o$: $E\times E\rightarrow \mathbb{R}_+$. This transition kernel can be written in terms of the  the intensities (or arrival rates) of buy orders, sell orders, cancellation of buy orders and cancellation of sell orders denoted by $\lambda_+$, $\lambda_-$, $C_+$ and $C_-$ respectively. These are functions $\{1,...,d\}\times E\times\mathbb{N}\rightarrow\mathbb{R}_+$ of the order price, current state and order size.

We impose three additional conditions on the functions $\lambda_{\pm}$ and $C_{\pm}$: 
\begin{itemize}
\item[(i)] If the size $z$ of the new order is too large, then one side of the order book might be wiped out. We want to discard this situation as we want to focus on a generic situation where the two sides of the order book are non-empty. 
Thus, we require $z<\min\{B_X(1),S_X(d)\}$.
\item[(ii)] 
We impose an upper bound $M/(1+z)^{\alpha}$ respectively on the four arrival rates and $\alpha>1$.
\item[(iii)] Cancellation of orders will not be placed when there are fewer existing orders at the corresponding price than the size of the cancellation order. Therefore, $C_{\pm}(i,X,z)=0$ when $z>X^{\pm}_i$. 
\end{itemize}
We summarise these three conditions into the following assumption:
\begin{assumption}\label{assumption.general_absolute}
$\lambda_+$, $\lambda_-$, $C_+$, $C_-$ satisfy the following two conditions:
\begin{enumerate}
\item[(i)] $\forall X\in E$, $\forall i\in \{1,\ldots,d\}$, $\lambda_{\pm}(i,X,z)=C_{\pm}(i,X,z)=0$ 
if $z\geq \min \{B_X(1),S_X(d) \}$.
\item[(ii)] $\forall X\in E$, $\forall i\in \{1,\ldots,d\}$ and $\forall z\in\mathbb{N}\backslash\{ 0 \}$, 
$\lambda_{\pm}(i,X,z)\leq M/(1+z)^{\alpha}$ and $C_{\pm}(i,X,z)\leq M/(1+z)^{\alpha}$, with $\alpha>1$.
\item[(iii.)] $\forall X\in E$, $C_{\pm}(i,X,z)=0$ when $z>X^{\pm}_i$. 
\end{enumerate}
\end{assumption}
Under Assumption~\ref{assumption.general_absolute}, we can now write the expression of the generator $\emph{L}_o$ of the order flow: $\emph{L}_o$ is an operator acting on the space $B(E)$ of bounded functions on the general state space $E$ as follows: for any $f\in B(E)$ and $X\in E$, $\emph{L}_o$ is defined by:
\begin{equation}\label{eq.generator_order_flow_absolute}
\emph{L}_of(X)=\lim_{h\rightarrow0}\frac{\mathbb{E}[f(X_h)|X_0=X]-f(X)}{h}.
\end{equation} 
From the standard theory of Markov process \cite{ethier}, $\emph{L}_o$ is given for any $f\in B(E)$ and $X\in E$ by
\begin{equation}
\emph{L}_of(X)=\sum_{Y \in E}[f(Y)-f(X)]p_o(X,Y).
\label{eq.generator_order_flow_expression}
\end{equation} 
If we insert the expression of $p_o$ in terms of the order arrival rates, we end up with the following expression: 
\begin{align}
\label{eq:express_Lo}
\emph{L}_of(X)&=\sum_{z=1}^{\infty}\sum_{i=1}^d \Big\{ \lambda_+(i,X,z)[f(X^+ + z e_i, X^-) - f(X)] + \lambda_-(i,X,z) [f(X^+ , X^- + z e_i) - f(X)]\\
\nonumber
&+C_+(i,X,z) [f(X^+ -z e_i, X^-) - f(X)] + C_-(i,X,z) [f(X^+, X^- - z e_i)-f(X)] \Big\}. 
\end{align}
We note that the sum over $z$ contains a finite number of terms only thanks to Assumption \ref{assumption.general_absolute}.

\subsection{Generator of the order book process}
\label{sec.generator_of_the_order_book_process}
We now define the order book process in the Markovian framework as follows:
\begin{definition}
A limit order book (LOB) $\{X_t\}_{t\geq0}$, $X_t: \mathcal{L}\rightarrow \mathcal{L}$ is a Markovian point process defined as follows: Suppose the sequence of event times $t_1,\ldots,t_{k-1}$ and the corresponding states $X_1,\ldots,X_{k-1}$ are defined where $X_j$ is the state of the LOB during $[t_{j-1},t_j)$ and $t_0=0$. Then $t_k$ and $X_k$ are defined as follows: $(t_k-t_{k-1})$ is distributed according to the exponential law of rate $\sum_{Y\in E}p_o(X_{k-1},Y)$ and $X_k \in {\mathcal L}$ is distributed according to the probability 
$$\frac{\sum_{Y\in\mathcal{C}^{-1}(\{X_k\})}p_o(X_{k-1},Y)}{\sum_{Y\in E}p_o(X_{k-1},Y)}, $$ 
where $\mathcal{C}$ is the clearing operator defined in Prop.~\ref{prop.ordermatching}.
\end{definition}
Then we can write the generator $\emph{L}$ of the order book process as follows:
\begin{proposition}
The generator of the order book process $\emph{L}$ (defined by a similar formula as Eq.~(\ref{eq.generator_order_flow_absolute})) is given for any $f\in B({\mathcal L})$ and $X\in\mathcal{L}$ by
\begin{equation}
\emph{L}f(X)=\sum_{Y\in\mathcal{L}}p(X,Y)[f(Y)-f(X)] \quad \textrm{ with } \quad p(X,Y)=\sum_{Z \in\mathcal{C}^{-1}(\{ Y \})}p_o(X,Z), \qquad  \forall  X, Y \in \mathcal{L}.
\label{eq.general_generator_absolute}
\end{equation}
\end{proposition}

This proposition states that the jump rate $p$ of the order book to a state $Y \in {\mathcal L}$ is given in terms of the jump rate $p_o$ of the order flow and of the pre-image $\mathcal{C}^{-1}(\{ Y \})$ of $\{Y\}$ by the clearing operator $\mathcal{C}$ .

\noindent
\textbf{Proof.} 
For any $X, Y\in\mathcal{L}$, the transition from $X$ to $Y$ through the LOB process occurs when the order flow generates a transition from $X$ to a transient state $\tilde{X} \in E$ which is mapped by the clearing operator to $Y$. Then the transition probability from $\tilde{X}$ to $Y$ is the sum of the transition probabilities of the transient order book process from $X$ to $\tilde{X}$ over all $\tilde{X}\in\mathcal{C}^{-1}(\{ Y \})$. Therefore,
\begin{equation}
\frac{p(X,Y)}{\sum_{Z}p_o(X,Z)}=\sum_{W\in\mathcal{C}^{-1}(\{ Y \})} \frac{p_o(X,W)}{\sum_{Z}p_o(X,Z)},
\label{eq.relation_transition_probability}
\end{equation}
which leads to \eqref{eq.general_generator_absolute}. \endproof

\subsection{Decomposition of the infinitesimal generator}
\label{sec.decomposition_of_the_generator_of_the_lob_process}

To simplify the calculation of the generator $\emph{L}$ of the LOB process, we decompose it into the generator $\emph{L}_o$ of the order flow, an operator $\tilde{\mathcal{C}}$ related with the clearing operator $\mathcal{C}$ and a restriction operator $\Xi$. The general idea is pictured in Fig.~\ref{fig.relation_order_flow_order_book_absolute}. We then introduce the expressions of the operators $\tilde{\mathcal{C}}$ and $\Xi$. 
\begin{figure}[h]
\centering
\[ \begin{tikzcd}[column sep=5em,row sep=3em,arrows=-latex]
B(E) \arrow{r}{\textit{L}_o}  & B(E) \arrow{d}{\Xi} \\%
B(\mathcal{L})\arrow[swap]{u}{\tilde{\mathcal{C}}} \arrow{r}{\textit{L}}& B(\mathcal{L})
\end{tikzcd}
\]
\caption{Relation between the generator of the order flow $\emph{L}_o$ and the generator of the order book process $\emph{L}$ using operators $\tilde{\mathcal{C}}$ and $\Xi$. The diagram is commutative. }
\label{fig.relation_order_flow_order_book_absolute}
\end{figure}

We define the {\it restriction} operator $\Xi$ which restricts a function on $E$ to a function on $\mathcal{L}$: \begin{definition}[Restriction operator]
Define $\Xi:B(E)\rightarrow B(\mathcal{L})$, $f\mapsto \Xi(f)$ as follows: for any $f\in B(E)$,
\begin{equation}
\label{eq.definition_of_Xi}
\Xi f = f|_{\mathcal{L}}.
\end{equation}
\end{definition}
\begin{definition}[Action of clearing operator on functions of the state]
Define $\tilde{\mathcal{C}}:B(\mathcal{L}) \rightarrow B(E)$, $f\mapsto \tilde{\mathcal{C}}(f)=f\circ\mathcal{C}$, i.e. for any $f\in B(\mathcal{L})$ and any $X\in E$:
\begin{equation}
\label{eq.definition_of_tilde_C}
\tilde{\mathcal{C}}f(X):=f(\mathcal{C}(X)).
\end{equation}
\end{definition}
By $\|\cdot\|_{\mathfrak{L}(\mathfrak{X},\mathfrak{Y})}$ we denote the operator norm of bounded operators from a Banach space $\mathfrak{X}$ with norm $\|\cdot\|_{\mathfrak{X}}$ to a Banach space $\mathfrak{Y}$ with norm $\|\cdot\|_{\mathfrak{Y}}$, i.e. for any $H\in\mathfrak{L}(\mathfrak{X},\mathfrak{Y})$,
\begin{equation}\label{eq.operator_norm}
\|H\|_{\mathfrak{L}(\mathfrak{X},\mathfrak{Y})}:=\sup_{v\in\mathfrak{X},v\neq0}\frac{\|H(v)\|_{\mathfrak{Y}}}{\|v\|_{\mathfrak{X}}}.
\end{equation}
For any $f\in B(E)$, let $\|f\|_{B(E)}=\sup_{X\in E}|f(X)|$ and for any $g\in B(\mathcal{L})$, let $\|g\|_{B(\mathcal{L})}=\sup_{X\in\mathcal{L}}|g(X)|$. Then $B(E)$ and $B(\mathcal{L})$ are Banach spaces. For any $X\in E$, define $\|X\|=\sum_{i=1}^d (|X^+_i|+|X^-_i|)$. Then $E$ and $\mathcal{L}$ are discrete subspaces of $\mathbb{R}^d\times\mathbb{R}^d$, which is normed by $\|X\|$. The following lemma shows that $\Xi$, $\emph{L}_o$ and $\tilde{\mathcal{C}}$ are bounded. 
\begin{lemma}
\label{lemma.boundness_identity__of_Xi_and_tilde_C}
$\Xi$, $\emph{L}_o$ and $\tilde{\mathcal{C}}$ are bounded linear operators. Moreover, $\Xi\tilde{\mathcal{C}}=I_{B(\mathcal{L})}$ where $I_{B(\mathcal{L})}$ is the identity of $B(\mathcal{L})$.
\end{lemma}

\noindent
\textbf{Proof.}
From the definition of $\Xi$, $\emph{L}_o$ and $\tilde{\mathcal{C}}$, it is easy to see that they are linear. Now, since $\mathcal{C}$ is surjective from $E$ to $\mathcal{L}$, we have
\mathcenter
\begin{equation}
\|\tilde{\mathcal{C}}f\|_{B(E)}=\sup_{X\in E}|f\circ \mathcal{C}(X)|=\sup_{X\in\mathcal{L}}|f(X)|= \| f \|_{B(\mathcal{L})} =1,
\label{eq.boundness_of_tilde_C}
\end{equation}
showing that $\tilde{\mathcal{C}}$ is a bounded operator from $B(\mathcal{L})$ to $B(E)$ with norm 1: $\|  \tilde{\mathcal{C}} \|_{\mathfrak{L}(B(\mathcal{L}),B(E))}=1$.
Then, with Eq. \eqref{eq:express_Lo} and
Assumption~\ref{assumption.general_absolute}, we have
\begin{eqnarray*}
&&
|\emph{L}_o f(X)|=\sum_{i=1}^d\sum_{z=1}^{\infty}\lambda_+(i,X,z)[|f(X^++ze_i,X^-)|+|f(X)|]+\lambda_-(i,X,z)[|f(X^+,X^-+ze_i)|+|f(X)|]\\
&& \hspace{1cm} +C_+(i,X,z)[|f(X^+-ze_i,X^-)|+|f(X)|]+C_-(i,X,z)[|f(X^+,X^--ze_i)|+|f(X)|] 
\leq 8dM \| f\|_{B(E)},
\end{eqnarray*}
showing that $\emph{L}_o$ is bounded with $\|\emph{L}_o\|_{\mathfrak{L}(B(E),B(E))}\leq 8mMd$, where $M, m$ are defined in Assumption~\ref{assumption.general_absolute}. For $\Xi$, we have
\[
\|\Xi f\|_{B(\mathcal{L})}=\sup_{X\in\mathcal{L}}|\Xi f(X)|=\sup_{X\in\mathcal{L}}|f(X)| \leq  \sup_{X\in E}|f(X)|=\|  f\|_{B(E)}  = 1.
\]
So $\Xi$ is bounded. Moreover, it is easy to show that $\| \Xi \|_{\mathfrak{L}(B(E),B(\mathcal{L}))}=1$. Furthermore, for any $f\in B(\mathcal{L})$ and $X\in\mathcal{L}$, we have
\[  
\tilde{\mathcal{C}}f(X)=f(\mathcal{C}X)=f(X).
\]
So $\Xi \tilde{\mathcal{C}} f(X)=f(X)$. Therefore $\Xi\tilde{\mathcal{C}}=I_{B(\mathcal{L})}$, which finishes the proof. \endproof

\begin{proposition}[Decomposition of the generator of the order book]
\label{prop.decomposition_generator_absolute}
We have
\begin{equation}
\emph{L}=\Xi\emph{L}_o\tilde{\mathcal{C}}.
\label{eq.decompositioin_between_generator_absolute}
\end{equation}
\label{prop:decomp_gene}
\end{proposition}

\noindent
\textbf{Proof.}
Let $f\in B(\mathcal{L})$, and $X\in\mathcal{L}$. Thanks to Eqs.~(\ref{eq.general_generator_absolute}) and of the fact that $\mathcal{C}(X) = X$, we have
\begin{eqnarray*}
&&\hspace{-0.5cm}
\Xi L_o \tilde {\mathcal C} f (X) = L_o \tilde {\mathcal C} f (X) = L_o (\tilde {\mathcal C} f) (X) = \sum_{Y\in E} p_o(X,Y) [\tilde {\mathcal C}f (Y) -\tilde {\mathcal C} f (X)] = 
\sum_{Y\in E}p_o(X,Y)[f(\mathcal{C}(Y))-f(\mathcal{C}(X))]\\
&&\hspace{-0.5cm}
=\sum_{Y\in E}p_o(X,Y)[f(\mathcal{C}(Y))-f(X)]=\sum_{Z\in\mathcal{L}}[f(Z)-f(X)]\sum_{Y\in\mathcal{C}^{-1}(\{ Z\})}p_o(X,Y) =\sum_{Z\in\mathcal{L}}[f(Z)-f(X)]p(X,Z)=\emph{L}f(X).
\end{eqnarray*}
\endproof

This proposition shows that the diagram in Figure~\ref{fig.relation_order_flow_order_book_absolute} is commutative. 
This decomposition permits the mathematical determination of the generator of order book dynamics as shown in the following corollary.
\begin{corollary}
For $f\in B(\mathcal{L})$ and $X\in\mathcal{L}$, we have
\begin{eqnarray}
\emph{L}f(X) &=&\Xi \emph{L}_o f(\mathcal{C} X) =\sum_{z=1}^\infty \Big\{ \sum_{i=1}^{a(X)-1} \lambda_+(i,X,z)[f(X^++ze_i,X^-)-f(X)] \label{eq.generator_order_book_exact_expression} \\
&&+\sum_{i=a(X)}^{S_X^{-1}(z)-1}\lambda_+(i,X,z) [f(X^++(z-S_X(i))e_i,\tau^{  i+1}(X^-))-f(X)] \nonumber \\
&&+\sum_{i=S_X^{-1}(z)}^{d} \lambda_+(i,X,z) [f(X^+,\tau^{  S_X^{-1}(z)+1}(X^-)+(S_X(S_X^{-1}(z))-z)e_{S_X^{-1}(z)})-f(X)] \nonumber \\
&&+\sum_{i=b(X)+1 }^d \lambda_-(i,X,z)[f(X^+,X^-+ze_i)-f(X)] \nonumber \\
&&+\sum_{i=B_X^{-1}(z)+1}^{b(X)} \lambda_-(i,X,z)[f(\tau_{  i-1}(X^+),X^-+(z-B_X(i))e_i)-f(X)] \nonumber \\
&&+\sum_{i=1}^{B_X^{-1}(z)} \lambda_-(i,X,z)[f(\tau_{  B_X^{-1}(z)-1}(X^+)+(B_X(B_X^{-1}(z))-z)e_{B_X^{-1}(z)},X^-)-f(X)] \nonumber \\
&&+ \sum_{i=1}^{b(X)}C_+(i,X,z)[f(X^+-ze_i,X^-)-f(X)]
+\sum_{i=a(X)}^d C_-(i,X,z)[f(X^+,X^--ze_i)-f(X)] \Big\}. \nonumber 
\end{eqnarray}\label{eq.order_book_generator_expression_absolute}
\label{col.generator_order_book_exact_expression}
\end{corollary}

\noindent
\textbf{Proof.} 
For any $f\in B(\mathcal{L})$ and $X\in\mathcal{L}$, by Proposition~\ref{prop.decomposition_generator_absolute} we have $\emph{L}f(X)=\Xi \emph{L}_o f(\mathcal{C} X)$. This, together with Eq.~\eqref{eq:express_Lo} and Assumption~\ref{assumption.general_absolute} implies that
\[
\begin{aligned}
\emph{L}f(X)=&\sum_{i=1}^d\sum_{z=1}^\infty 
\Big\{ 
\lambda_+(i,X,z) [f(\mathcal{C}(X^++ze_i,X^-))-f(X)]+\lambda_-(i,X,z) [f(\mathcal{C}(X^+,X^-+ze_i))-f(X)]\\
&+C_+(i,X,z) [f(\mathcal{C}(X^+-ze_i,X^-))-f(X)]+C_-(i,X,z) [f(\mathcal{C}(X^+,X^--ze_i))-f(X)] 
\Big\}. 
\end{aligned}
\]
Now we calculate the first term $\sum_{i=1}^d\sum_{z=1}^\infty \lambda_+(i,X,z) [f(\mathcal{C}(X^++ze_i,X^-))-f(X)]$ of the right-hand side of the above equation, leaving the other terms to the reader. We recall that, from Proposition~\ref{prop.operation_clearing_absolute}, we have 
\[
\mathcal{C}(X^++ze_i,X^-)=
\begin{cases}
(X^++ze_i,X^-) & \mbox{   if   } i<a(X),\\
(X^++(z-S_X(i))e_i,\tau^{  i+1}(X^-))& \mbox{   if   }   a(X)\leq i < S_X^{-1}(z),\\
(X^+,\tau^{  S_X^{-1}(z)+1}(X^-)+(S_X(S_X^{-1}(z))-z)e_{S_X^{-1}(z)})& \mbox{   if   } S_X^{-1}(z) \leq i \leq d.
\end{cases}
\]
Consequently, we have
\begin{eqnarray*}
&&\sum_{i=1}^d\sum_{z=1}^\infty \lambda_+(i,X,z) [f(\mathcal{C}(X^++ze_i,X^-))-f(X)]
=\sum_{z=1}^\infty \Big\{ \sum_{i=1}^{a(X)-1} \lambda_+(i,X,z)[f(X^++ze_i,X^-)-f(X)]\\
&&\hspace{3cm} +\sum_{i=a(X)}^{S_X^{-1}(z)-1}\lambda_+(i,X,z) [f(X^++(z-S_X(i))e_i,\tau^{  i+1}(X^-))-f(X)] \\
&&\hspace{3cm}+\sum_{i=S_X^{-1}(z)}^{d} \lambda_+(i,X,z) [f(X^+,\tau^{  S_X^{-1}(z)+1}(X^-)+(S_X(S_X^{-1}(z))-z)e_{S_X^{-1}(z)})-f(X)] \Big\},\\
\end{eqnarray*}
which corresponds to the first three lines of (\ref{eq.order_book_generator_expression_absolute}). Similar computation can be done and leads to the result. \endproof

As a consequence of Prop. \ref{prop:decomp_gene}, we have the following obvious proposition: 

\begin{proposition}\label{prop.generator_order_book_absolute_boundness_group}
$\emph{L}$ defined in Eq.~(\ref{eq.general_generator_absolute}) is a bounded linear operator on $B(\mathcal{L})$. Consequently, $\emph{L}$ is a generator of a continuous group $e^{t\emph{L}}$ on $B(\mathcal{L})$.
\end{proposition}

\subsection{Decomposition of the Kolmogorov backward equation of the LOB process} 
\label{sec.decomposition_of_kolmogorov_backward_equation}

The Kolmogorov backward equation for the order book is defined as follows: Given $f\in B(\mathcal{L})$, $X\in\mathcal{L}$ and $T\in\mathbb{R}_+$ a fixed constant, let $u(t,X):=\mathbb{E}[f(X_T)|X_t=X]$ for $t\leq T$. Then $u(t,X)$ is the solution of
\begin{equation}
\frac{\partial u(t,X)}{\partial t}=-\emph{L} u(t,X), \qquad u(T,X)=f(X).  
\label{eq.Kolmogorov_equation}
\end{equation}
The solution of this equation is: $u(t,\cdot)=e^{(T-t)\emph{L}}u(T,\cdot)$. Now, introducing a discrete time step $\Delta t$, then we have $u(t+\Delta t, X)=e^{-\Delta t\emph{L}}u(t,X)$. The next proposition shows that $u(t+\Delta t, X)$ can be approximated by $\Xi e^{-\Delta t\emph{L}_o}\tilde{\mathcal{C}}u(t,X)$ when $\Delta t$ is small enough. In other words, the solution of the Kolmogorov backward equation of the LOB process on a small time step $\Delta t$ can be approximated by the composition of the operator $\tilde{\mathcal{C}}$, the solution of the Kolmogorov backward equation for the order flow on the same time-step and  the operator $\Xi$ as pictured in Figure~\ref{fig.decomposition_kbe_absolute}. The iteration of these operations in time provide an approximation of the solution of the Kolmogorov-backward equation for the LOB as shown in Eq.~(\ref{eq.approximation_kbe_absolute}). 
\begin{figure}[h]
\centering
\[ \begin{tikzcd}[column sep=5em,row sep=3em,arrows=-latex]
B(E) \arrow{r}{e^{-\Delta t\textit{L}_o}}  & B(E) \arrow{d}{\Xi} \\%
B(\mathcal{L})\arrow[swap]{u}{\tilde{\mathcal{C}}} \arrow{r}{e^{-\Delta t\textit{L}}}& B(\mathcal{L})
\end{tikzcd}
\]
\caption{Relation between the solution of the Kolmogorov backward equation of the order flow $e^{-\Delta t\textit{L}_o}$ and that of the LOB process $e^{-\Delta t\textit{L}}$. This diagram is not exactly commutative, but the commutator is of order $O(\delta t)$   as shown in Eq.~(\ref{eq.approximation_kbe_absolute}).}
\label{fig.decomposition_kbe_absolute}
\end{figure}

\begin{proposition}
\label{prop.approximation_kbe_absolute}
We have
\mathcenter
\begin{equation}\label{eq.approximation_kbe_absolute}
\lim_{\Delta t\rightarrow 0}\| (e^{-\Delta t\emph{L}})^{\frac{T}{\Delta t}}- (\Xi e^{-\Delta t\emph{L}_o}\tilde{\mathcal{C}})^{\frac{T}{\Delta t}}\|_{B(\mathcal{L})}=0
\end{equation}
for any fixed constant $T\in\mathbb{R}_+$.
\end{proposition}

\noindent
\textbf{Proof.} Suppose $\frac{T}{\Delta t}=N$ is an integer. Using Taylor expansion, we have 
\[
e^{-\Delta t \emph{L}}=I_{B(\mathcal{L})}-\Delta t\emph{L}+\frac{\Delta t^2}{2}\emph{L}^2+o(\Delta t^2)\qquad\mbox{and}\qquad
e^{-\Delta t \emph{L}_o}=I_{B(E)}-\Delta t\emph{L}_o+\frac{\Delta t^2}{2}\emph{L}_o^2+o(\Delta t^2).
\]
Therefore, together with Lemma~\ref{lemma.boundness_identity__of_Xi_and_tilde_C} and Proposition~\ref{prop.decomposition_generator_absolute} we deduce
\[
\Xi e^{-\Delta t \emph{L}_o}\tilde{\mathcal{C}}=\Xi I_{B(E)}\tilde{\mathcal{C}}-\Delta t \Xi \emph{L}_o\tilde{\mathcal{C}}+\frac{\Delta t^2}{2}\Xi\emph{L}_o^2\tilde{\mathcal{C}}+o(\Delta t^2)=I_{B(\mathcal{L})}-\Delta t\emph{L}+\frac{\Delta t^2}{2}\Xi\emph{L}_o^2\tilde{\mathcal{C}}+o(\Delta t^2).
\]
Consequently, we define a linear map $G_{\Delta t}:B(\mathcal{L})\rightarrow B(\mathcal{L})$, $f\mapsto G_{\Delta t}f$ as follows:
\[
G_{\Delta t}=\frac{\Xi e^{-\Delta t \emph{L}_o}\tilde{\mathcal{C}}-e^{-\Delta t \emph{L}}}{\Delta t^2}.
\]
By its definition, it is easy to see that $G_{\Delta t}$ is bounded. Furthermore, 
\[
\| G_{\Delta t}  \|_{{\mathfrak L}(B(\mathcal{L}))}= \Big\| \frac{\Xi e^{-\Delta t \emph{L}_o}\tilde{\mathcal{C}}-e^{-\Delta t \emph{L}}}{\Delta t^2} \Big\|_{{\mathfrak L}(B(\mathcal{L}))}= \Big\| \frac{\Xi\emph{L}_o^2\tilde{\mathcal{C}}-\emph{L}^2}{2} \Big\|_{{\mathfrak L}(B(\mathcal{L}))}+o(1)<\infty,
\]
which implies that 
$$
\lim_{\Delta t\rightarrow 0}\| G_{\Delta t}  \|_{{\mathfrak L}(B(\mathcal{L}))}= \Big\| \frac{\Xi\emph{L}_o^2\tilde{\mathcal{C}}-\emph{L}^2}{2} \Big\|_{{\mathfrak L}(B(\mathcal{L}))}:=\beta.$$
So $\exists \delta>0$ such that $\forall \Delta t<\delta$, $\| G_{\Delta t}\|_{{\mathfrak L}(B(\mathcal{L}))}<2\beta.$
From
\begin{equation}
\Xi e^{-\Delta t \emph{L}_o}\tilde{\mathcal{C}}=e^{-\Delta t \emph{L}} +\Delta t^2 G_{\Delta t}.
\label{eq.relation_infinitesimal_evolution_expression}
\end{equation}
we expand $(\Xi e^{-\Delta t\emph{L}_o}\tilde{\mathcal{C}})^{N}$ as follows:
\begin{align*}
&(\Xi e^{-\Delta t\emph{L}_o}\tilde{\mathcal{C}})^{N}=(e^{-\Delta t \emph{L}} +\Delta t^2 G_{\Delta t})^{N}\\
&=(e^{-\Delta t\emph{L}})^N+\sum_{\substack{(p_1,\ldots,p_N)\in \{0,1\}^N,\\ (p_1,\ldots,p_N)\neq (1,\ldots,1)}} \prod_{i=1}^{N} e^{-p_i \Delta t \emph{L}} (\Delta t^2 G_{\Delta t})^{1-p_i}.
\end{align*}
Therefore, we have 
\begin{equation}
\begin{aligned}
&\| (e^{-\Delta t\emph{L}})^{N}- (\Xi e^{-\Delta t\emph{L}_o}\tilde{\mathcal{C}})^{N}\|_{{\mathfrak L}(B(\mathcal{L}))}= \Big\|   \sum_{\substack{(p_1,\ldots,p_N)\in \{0,1\}^N,\\ (p_1,\ldots,p_N)\neq (1,\ldots,1)}} \prod_{i=1}^{N} e^{-p_i \Delta t \emph{L}} (\Delta t^2 G_{\Delta t})^{1-p_i}  \Big\|_{{\mathfrak L}(B(\mathcal{L}))}\\
&\leq \sum_{\substack{(p_1,\ldots,p_N)\in \{0,1\}^N,\\ (p_1,\ldots,p_N)\neq (1,\ldots,1)}} \prod_{i=1}^{N} \| e^{-\Delta t \emph{L}}\|_{{\mathfrak L}(B(\mathcal{L}))}^{p_i} \, \|\Delta t^2 G_{\Delta t}\|_{{\mathfrak L}(B(\mathcal{L}))}^{1-p_i}\\
&\leq \sum_{i=1}^{N}\binom{N}{i} \|\Delta t^2 G_{\Delta t}\|_{{\mathfrak L}(B(\mathcal{L}))}^i \, \| e^{-\Delta t \emph{L}}\|_{{\mathfrak L}(B(\mathcal{L}))}^{N-i}.
\end{aligned}
\end{equation}
Since $\| e^{-\Delta t \emph{L}} \|_{{\mathfrak L}(B(\mathcal{L}))}\leq e^{\Delta t \| \emph{L}\|_{{\mathfrak L}(B(\mathcal{L}))}}$, and $e^{\Delta t \| \emph{L}\|_{{\mathfrak L}(B(\mathcal{L}))}}\geq 1$, for any $i\leq N$, we have
\[
\| e^{-\Delta t \emph{L}} \|_{{\mathfrak L}(B(\mathcal{L}))}^{N-i}\leq e^{\Delta t(N-i) \| \emph{L}\|_{{\mathfrak L}(B(\mathcal{L}))}}\leq e^{\Delta t N \| \emph{L}\|_{{\mathfrak L}(B(\mathcal{L}))}}=e^{T\| \emph{L}\|_{{\mathfrak L}(B(\mathcal{L}))}}=\alpha_T,
\]
where $\alpha_T$ does not depend on $\Delta t$ and where we have used that $N\Delta t=T$. This implies that for each $i\leq N$:
\begin{align*}
\binom{N}{i} \|\Delta t^2 G_{\Delta t}\|_{{\mathfrak L}(B(\mathcal{L}))}^i \, \| e^{-\Delta t \emph{L}}\|_{{\mathfrak L}(B(\mathcal{L}))}^{N-i}
\leq \alpha_T \frac{(N)^i}{i!}(\Delta t)^{2i} \| G_{\Delta t} \|_{{\mathfrak L}(B(\mathcal{L}))}^i =\alpha_T \frac{(\Delta t T\| G_{\Delta t} \|_{{\mathfrak L}(B(\mathcal{L}))})^i}{i!}.
\end{align*}
Consequently, 
\begin{equation}
\| (e^{-\Delta t\emph{L}})^{N}- (\Xi e^{-\Delta t\emph{L}_o}\tilde{\mathcal{C}})^{N}\|_{{\mathfrak L}(B(\mathcal{L}))}
\leq \alpha_T \sum_{i=1}^{\infty}\frac{(\Delta t T\| G_{\Delta t} \|_{{\mathfrak L}(B(\mathcal{L}))} )^i}{i!}=\alpha_T(e^{\Delta t T\| G_{\Delta t} \|_{{\mathfrak L}(B(\mathcal{L}))}}-1).
\end{equation}
Since $\| G_{\Delta t} \|_{{\mathfrak L}(B(\mathcal{L}))}<2\beta$ when $\Delta t<\delta$, letting $\Delta t\rightarrow 0$, we have
\begin{equation}
\lim_{\Delta t\rightarrow 0} \| (e^{-\Delta t\emph{L}})^{N}- (\Xi e^{-\Delta t\emph{L}_o}\tilde{\mathcal{C}})^{N}\|_{{\mathfrak L}(B(\mathcal{L}))}\leq \alpha_T \lim_{\Delta t\rightarrow 0} (e^{\Delta t T\| G_{\Delta t} \|_{{\mathfrak L}(B(\mathcal{L}))}}-1)=0,
\end{equation}
which finishes the proof. \endproof

\subsection{Decomposition of the adjoint of the generator \emph{L}}
\label{sec.decomposition_of_the_adjoint_of_generator}

We may also obtain a similar decomposition for the adjoint
  $\emph{L}^*$  of $\emph{L}$, which intervenes in the forward Kolmogorov equation for the state probability distribution. $\emph{L}^*$ can be written as 
the composition of the adjoint operators $\tilde{\mathcal{C}}^*$, $\emph{L}_o^*$ and $\Xi^*$. We first introduce the expressions of these three adjoint operators. Let $\mathcal{P}(E)$ and $\mathcal{P}(\mathcal{L})$ be the spaces of bounded measures on $E$ and $\mathcal{L}$, which are the duals of $B(E)$ and $B({\mathcal L})$ respectively. 

\begin{lemma}
The adjoint $\Xi^*:\mathcal{P}(\mathcal{L})\rightarrow\mathcal{P}(E)$ of $\Xi$ is such that for any $\nu\in\mathcal{P}(\mathcal{L})$ and any measurable subset $A$ of~$E$: 
\begin{equation}
\Xi^*\nu(A):=\nu(A\cap\mathcal{L}).
\label{eq:express_Xi*}
\end{equation}
The adjoint $\tilde{\mathcal{C}}^*:\mathcal{P}(E)\rightarrow\mathcal{P}(\mathcal{L})$ of $\tilde{\mathcal{C}}$ is such that for any $\mu\in \mathcal{P}(E)$, and any measurable subset $A$ of $\mathcal{L}$:
\begin{equation}
\tilde{\mathcal{C}}^*\mu(A):=\mu(\mathcal{C}^{-1}(A)) =\sum_{Z\in A}\mu(\mathcal{C}^{-1}(\{ Z \})).
\label{eq:express_C*}
\end{equation}
The adjoint $\emph{L}_o^*:\mathcal{P}(E)\rightarrow \mathcal{P}(E)$ of $\emph{L}_o$ is such that for any measure $\mu\in\mathcal{P}(E)$ and any measurable subset $A$ of $E$,
\begin{equation}
\label{eq.expression_adjoint_order_flow_absolute}
\emph{L}_o^*\mu(A)=\sum_{X\in A}\sum_{Y\in E}[p_o(Y,X)\mu(\{Y\})-p_o(X,Y)\mu(\{X\})].
\end{equation}
\label{lemma.adjoint_xi_clearing_generator_absolute}
\end{lemma}

\noindent
\textbf{Proof.} For any $f\in B(E)$ and $\nu\in\mathcal{P}(\mathcal{L})$, we have
\begin{equation}
\langle f,\Xi^*\nu\rangle = \langle \Xi f, \nu\rangle,
\label{eq.adjoint_Xi}
\end{equation}
where $\langle , \rangle$ is the duality between $B(E)$ and ${\mathcal P}(E)$ for the left-hand side and between $B({\mathcal L})$ and ${\mathcal P}({\mathcal L})$ for the right-hand side. Then we expand the right-hand side of the above equation as:
\begin{equation}
\begin{aligned}
&\langle \Xi f, \nu\rangle=\sum_{X\in\mathcal{L}} \Xi f(X)\nu(\{X \})=\sum_{X\in E}  f(X)\nu(\{ X\} \cap \mathcal{L}). 
\end{aligned}
\end{equation}
This together with Eq.~(\ref{eq.adjoint_Xi}) implies \eqref{eq:express_Xi*}. Similarly, 
using that $(\mathcal{C}^{-1}(\{ X \}))_{X\in\mathcal{L}}$ forms a partition of $E$, we have
\[
\langle \tilde{\mathcal{C}}f,\mu\rangle =\sum_{X\in E}\tilde{\mathcal{C}}f(X)\mu(\{ X \})=\sum_{X\in E}f\circ\mathcal{C}(X)\mu(\{ X \})
=\sum_{X\in\mathcal{L}}f(X)\sum_{X'\in\mathcal{C}^{-1}(\{X  \})}\mu(\{ X' \})=\sum_{X\in\mathcal{L}}f(X)\mu (\mathcal{C}^{-1}(\{X  \})),
\]
which implies \eqref{eq:express_C*}. Finally, Eq.~(\ref{eq.expression_adjoint_order_flow_absolute}) is classical and is left to the reader. This finishes the proof. \endproof

The relation between $\emph{L}^*$ and $\emph{L}_o^*$ is as follows:

\begin{corollary}
We have
\begin{equation}\label{eq.decompostion_adjoint_generator_absolute}
\emph{L}^*=\tilde{\mathcal{C}}^*\emph{L}_o^*\Xi^*.
\end{equation}
Moreover, $\emph{L}^*$ is bounded on $\mathcal{P}(\mathcal{L})$ and consequently is the generator of the group $e^{t\emph{L}^*}$ for $t\geq 0$. For any fixed $T\in\mathbb{R}_+$, we have:
\begin{equation}
\lim_{\Delta t\rightarrow 0}\| (e^{-\Delta t\emph{L}^*})^{\frac{T}{\Delta t}}- (\tilde{\mathcal{C}}^*e^{-\Delta t\emph{L}_o^*}\Xi^* )^{\frac{T}{\Delta t}}\|_{\mathcal{P}(\mathcal{L})}=0.
\end{equation}
\end{corollary}

We leave the proof to the reader as it relies on the same argument as in Prop.~\ref{prop.approximation_kbe_absolute}. We remark that Eq.~(\ref{eq.decompostion_adjoint_generator_absolute}) can be expanded into an explicit expression for $\emph{L}^*$. However, this expression is complex and not relatively enlightening so is not given here. Computing $\emph{L}^*$ directly from Eq.~(\ref{eq.decompostion_adjoint_generator_absolute}) is more convenient.

\section{Dynamics of the centred order book}
\label{sec.dynamics_in_coordinates_centred_at_mid_price}

Traders are mostly interested in features of the LOB in the vicinity of the best available prices, such as the spread (i.e. the difference between the ask and bid prices) or the number of standing orders. In order to better focus on this region of the LOB, it is more convenient to use a moving frame with its origin at the mid-price, i.e., the arithmetic average of the ask and bid prices. This section describes the LOB in this moving frame, also called a centred order book. There are two main differences with the fixed coordinates framework considered so far which make the description of the centred order book more complicated. One is the definition of the state space: since the mid-price is half-integer, we place the origin of the centred order book at the smallest larger integer to the mid-price (which is equal to the mid-price if the latter is an integer). A more detailed explanation can be found in the next section. The other difference is the need for an additional shift operator within the clearing operation, to make sure that the order book is always centred at the chosen reference price after clearing. In short, the clearing operator of the centred order book as a whole is a composition of a similar clearing operator to that used in the fixed coordinates framework,  and a shift operator.

\subsection{Centred limit order book}
\label{sec.state_space_centred_order_book}

Our centred order book framework focuses on orders queuing within a distance $\pm d'$ of the mid-price, where $d' \in {\mathbb N}\setminus \{0\}$ is given. Indeed, in some markets, it is not possible to submit orders at a distance larger than $d'$ from the mid-price, where $d'$ is a given number of ticks. In other situations, one may assume that orders will be canceled if they move out of a window of size $2d'$ centred at the mid-price. For this reason, we make the following assumption.

\begin{assumption}
\label{assumption.centred_order_book}
For an order book whose mid-price is $p_m$, orders with prices lower than $p_m-d'$ or higher than $p_m+d'$ are cancelled.
\end{assumption}

Define 
\[
E^m:=\{ X=(X^+,X^-)\in  \mathbb{N}^{2d'+1}\times\mathbb{N}^{2d'+1} \quad\mbox{such that } X^+\neq 0\mbox{  and  } X^-\neq 0\}.
\]
The state space $\mathcal{L}^m$ for the centred order book is
\begin{equation}
\label{eq.state_space_centred_order_book}
\begin{aligned}
\mathcal{L}^m:=\{ (X, p)\in E^m\times \mathbb{Z},  \qquad a(X)>b(X),\qquad a(X)+b(X)+\hat{p}=0 \},
\end{aligned}
\end{equation}
where $a(X)$ and $b(X)$ are defined by \eqref{eq:def_a_b} and where, by the definition of the supremum and the infimum of an empty set, we have $\inf{\rm supp}(0):=d'$ and $\sup{\rm supp}(0):=-d'$. Here, $\hat p$ is the remainder of the Euclidean division of $p$ by $2$, i.e. $\hat{p}=0$ if $p$ is even and $\hat{p}=1$ if $p$ is odd. 
For  $(X, p)\in\mathcal{L}^m$, $X^+$ is the bid side of the centred order book, $X^-$ is the ask side, and $p$ is the sum of the ask and bid prices, i.e. twice the value of the mid-price. 
The second condition in \eqref{eq.state_space_centred_order_book} is a centering condition.
Denote by $\ceil{\,}:\mathbb{R}\rightarrow\mathbb{Z}$ the round-off operator  
$$\ceil{q}:=\min \{i\in\mathbb{Z}, i\geq q  \}, \qquad \forall q \in {\mathbb Z}.$$ 
Then, $\ceil{\frac{p}{2}}$ is the origin of the centred  coordinate system and $X_i$ is the queue size at a distance  $i$ from the origin i.e. at price $i+\ceil{\frac{p}{2}}$.
We stress again that the origin of the order book $\ceil{\frac{p}{2}}$ is not exactly the mid-price $\frac{p}{2}$ but the two coincide  if $p$ is even and only differ by half a tick if $p$ is odd. This is to avoid centering the order book at a non-integer mid-price. 
Figure~\ref{fig.state_centred_order_book_even} (resp. \ref{fig.state_centred_order_book_odd}) shows the LOB in the absolute and centred coordinates when $p$ is even (resp. odd). We also stress that twice the mid price $p$ is now included in the state space, i.e. the state consists of (i) twice the mid price $p$ and (ii) the vector $X \in {\mathcal L}^m$ of queue sizes, and both will evolve in the course of the dynamics. As in the fixed coordinate framework, this evolution will be decomposed into order flow and clearing processes described below. 
 
\begin{figure}[H]
\centering
\begin{tikzpicture}[scale=0.65]
   \begin{axis}[
        ybar,
        xmin=,
        xmax=10,
        xtick={1,2,3,4,5,6,7,8,9,10},
        x tick style={draw=none},
        xticklabels={1,2,3,4,5,6,7,8,9},        
        ytick={1,2,3,4,5},
        ymin=0,
        ymax=5,
         every axis plot/.append style={
          bar width=10,
          bar shift=0pt,
          fill
        }
       ]
       \addplot [ybar,fill=blue] coordinates {
          (1,1)
          (2,3)
          (3,2)
          (4,1)  };
      \addplot  [ybar,fill=red] coordinates {      
          (6, 1)
          (8, 3)
          (9, 2) };
     \legend{buy orders, sell orders}
   \end{axis}
    \draw [->] (3.4,3.5) -- (3.4,0); 
    \draw (3.3,3.8) [black] node {Mid-price};
\end{tikzpicture}
\hspace{1cm}
\centering
\begin{tikzpicture}[scale=0.65]
   \begin{axis}[
        ybar,
        xmin=-4,
        xmax=4,
        xtick={-3,-2,-1,0,1,2,3},
        x tick style={draw=none},
        xticklabels={-3,-2,-1,0,1,2,3},        
        ytick={1,2,3,4,5},
        ymin=0,
        ymax=5,
         every axis plot/.append style={
          bar width=10,
          bar shift=0pt,
          fill
        }
       ]
       \addplot [ybar,fill=blue] coordinates {
          (-2,2)
          (-1,1) 
          (-3,3)};
      \addplot  [ybar,fill=red] coordinates {      
          (1, 1)
          (3, 3)
           };
     \legend{Buy orders, Sell orders}
   \end{axis}
    \draw [->] (3.4,3.5) -- (3.4,0); 
    \draw (3,4.4) [black] node {p=10};
    \draw (3,3.8) [black] node {(mid-price=5)};
\end{tikzpicture}
\caption{Left:   order book in the absolute coordinates. Right: same order book in centred coordinates with $d'=3$ levels on each side. The mid-price is 5,
$p=10$. $b(X)=\sup{\rm supp}(X^+)=-1$ and $a(X)=\inf{\rm supp}(X^-)=1$. $\hat{p}=0$ and $a(X)+b(X)+\hat{p}=0$, so Eq.~(\ref{eq.state_space_centred_order_book}) is satisfied. } 
\label{fig.state_centred_order_book_even}
\end{figure}
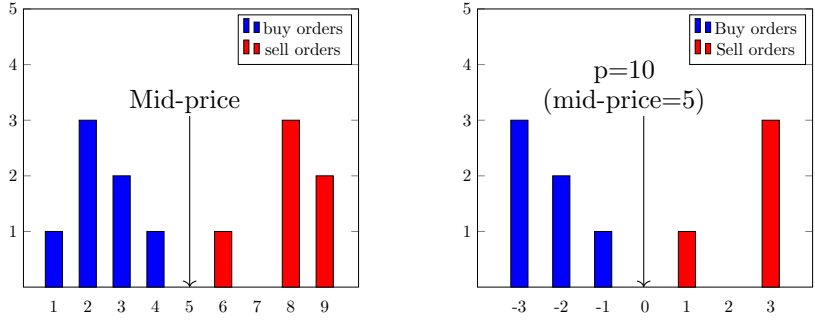
\begin{figure}[H]
\centering
\begin{tikzpicture}[scale=0.65]
   \begin{axis}[
        ybar,
        xmin=,
        xmax=10,
        xtick={1,2,3,4,5,6,7,8,9,10},
        x tick style={draw=none},
        xticklabels={1,2,3,4,5,6,7,8,9},        
        ytick={1,2,3,4,5},
        ymin=0,
        ymax=5,
         every axis plot/.append style={
          bar width=10,
          bar shift=0pt,
          fill
        }
       ]
       \addplot [ybar,fill=blue] coordinates {
          (1,1)
          (2,3)
          (3,2)
          (4,1)  };
      \addplot  [ybar,fill=red] coordinates {      
          (7, 1)
          (8, 3)
          (9, 2) };
     \legend{buy orders, sell orders}
   \end{axis}
    \draw [->] (3.7,3.5) -- (3.7,0); 
    \draw (3.7,3.8) [black] node {Mid-price};
\end{tikzpicture}
\hspace{1cm}
\centering
\begin{tikzpicture}[scale=0.65]
   \begin{axis}[
        ybar,
        xmin=-4,
        xmax=4,
        xtick={-3,-2,-1,0,1,2,3},
        x tick style={draw=none},
        xticklabels={-3,-2,-1,0,1,2,3},        
        ytick={1,2,3,4,5},
        ymin=0,
        ymax=5,
         every axis plot/.append style={
          bar width=10,
          bar shift=0pt,
          fill
        }
       ]
       \addplot [ybar,fill=blue] coordinates {
          (-3,2)
          (-2,1) 
          };
      \addplot  [ybar,fill=red] coordinates {      
          (1, 1)
          (2, 3)
          (3, 2)
           };
     \legend{buy orders, sell orders}
   \end{axis}
    \draw [->] (3.4,3.5) -- (3.4,0); 
    \draw (3,4.4) [black] node {p=11};
    \draw (3,3.8) [black] node {(Mid-price=5.5)};
\end{tikzpicture}
\caption{Left: order book in absolute price coordinates. Right:  Centred coordinates. $d'=3$, and the mid-price is~5.5,
$p=11$. $a(X)=-2$ and $b(X)=1$. $\hat{p}=1$. Therefore, $a(X)+b(X)+\hat{p}=0$, which shows that the centering condition in Eq.~(\ref{eq.state_space_centred_order_book}) is satisfied.}
\label{fig.state_centred_order_book_odd}
\end{figure}
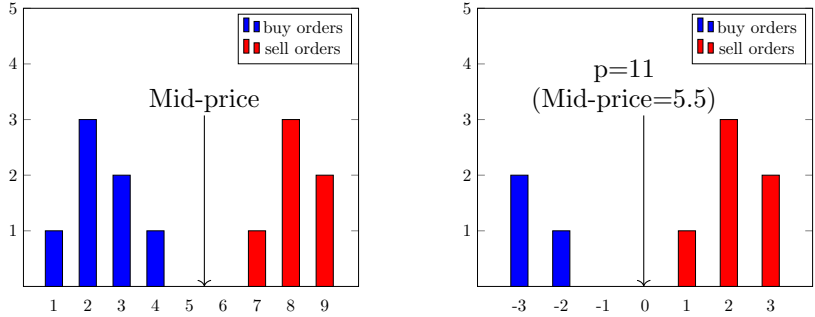

\subsection{Order flow process}\label{sec.order_flow_centred}
The order flow process in centred coordinates is similar to that in fixed coordinates. The extra variable $p$ in the centred framework is not changed by the order flow. 
So, the order flow for the queue size part of the state $X$ is described exactly like in Section \ref{sec.order_flow}, with $E$ replaced by $E^m$ and ${\mathcal L}$ by ${\mathcal L}^0$ defined by 
\begin{equation}
\mathcal{L}^0:=\{ X\in  E^m, \quad    a(X)>b(X)  \},
\label{eq.intermediate_space_centred}
\end{equation}
while the order flow leaves the value $p$ of the state unchanged. 

We assume that the order flow cannot wipe out either side of the order book,  which is expressed by 
\begin{assumption} We assume that $$\mathbb{P}\left(X(t_k-)+\Delta \hat{M}(t_k) \in E^m, \, \,  \forall k \in {\mathbb N} \right)=1,$$ 
where $\hat{M}$ is the marked Poisson point process introduced in Section \ref{sec.order_flow} expressed now in centred coordinates. 
\label{assumption.order_flow_centred_size}
\end{assumption}

\subsection{Clearing process in centred coordinates}\label{sec.clearing_operator_centred}
\begin{figure}[H]
\centering
\begin{tikzpicture}[scale=0.5]
   \begin{axis}[
        ybar,
        xmin=-4,
        xmax=4,
        xtick={-3,-2,-1,0,1,2,3},
        x tick style={draw=none},
        xticklabels={-3,-2,-1,0,1,2,3},        
        ytick={1,2,3,4,5},
        ymin=0,
        ymax=5,
         every axis plot/.append style={
          bar width=10,
          bar shift=0pt,
          fill
        }
       ]
      \addplot[ybar,fill=pink] coordinates {      
          (-2, 3)
          (2, 3)
          (0,1)};
       \addplot[ybar,fill=cyan] coordinates {      
          (2, 2)
          (1,2)};
       \addplot [ybar,fill=blue] coordinates {
          (-3,3)
          (-2,2)
          (-1,1) };
      \addplot  [ybar,fill=red] coordinates {      
          (1, 1)
          (3, 3) };
     \legend{new sell, new buy,buy orders, sell orders}
   \end{axis}
\end{tikzpicture}
\hspace{1cm}
\centering
\begin{tikzpicture}[scale=0.5]
   \begin{axis}[
        ybar,
        xmin=-4,
        xmax=4,
        xtick={-3,-2,-1,0,1,2,3},
        x tick style={draw=none},
        xticklabels={-3,-2,-1,0,1,2,3},        
        ytick={1,2,3,4,5},
        ymin=0,
        ymax=5,
         every axis plot/.append style={
          bar width=10,
          bar shift=0pt,
          fill
        }
       ]
       \addplot [ybar,fill=blue] coordinates {
          (-2,2)
          (-1,1) 
          (-3,3)};
      \addplot  [ybar,fill=red] coordinates {     
          (2, 1)
          (3, 3)};
     \legend{buy orders, sell orders}
   \end{axis}
\end{tikzpicture}
\hspace{1cm}
\centering
\begin{tikzpicture}[scale=0.5]
   \begin{axis}[
        ybar,
        xmin=-4,
        xmax=4,
        xtick={-3,-2,-1,0,1,2,3},
        x tick style={draw=none},
        xticklabels={-3,-2,-1,0,1,2,3},        
        ytick={1,2,3,4,5},
        ymin=0,
        ymax=5,
         every axis plot/.append style={
          bar width=10,
          bar shift=0pt,
          fill
        }
       ]
       \addplot [ybar,fill=blue] coordinates {
          (-3,2)
          (-2,1) };
      \addplot  [ybar,fill=red] coordinates {   
          (1, 1)
          (2, 3)};
     \legend{buy orders, sell orders}
   \end{axis}
\end{tikzpicture}
\caption{\textbf{Left:} Configuration of a centred order book after the arrival of new orders: the initial state of the centred order book before the arrival of new orders is 
in plain blue and red for buy and sell orders respectively. Three sell orders (light red), and two buy orders (light blue) arrive.  Initial mid-price is $\frac{p}{2}=5$. \textbf{Middle:} configuration $X'$ of the order book after execution of matched orders. We note that $a(X')+b(X')=\inf{\rm supp}(X'^-)+\sup{\rm supp}(X'^+)=1$. Thus, $p'=p+a(X')+b(X')=11$ and $\hat{p}'=1$. So $\hat{p}'+a(X')+b(X')=2\neq 0$ and the second condition of Eq.~(\ref{eq.state_space_centred_order_book}) is not satisfied.  \textbf{Right:} The center of the order book is shifted to the right by one unit to restore the second condition of Eq.~(\ref{eq.state_space_centred_order_book}).}
\label{fig.clearing_centred_order_book}
\end{figure}
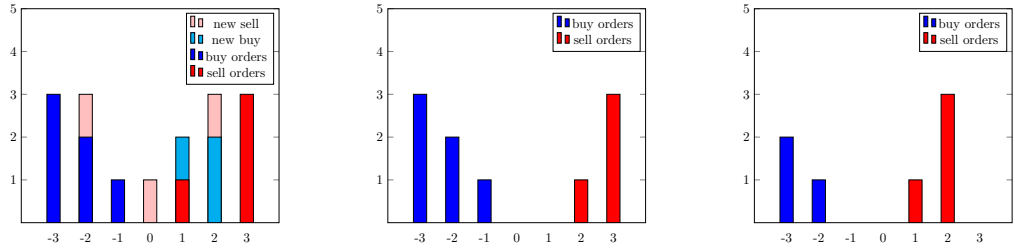

We now adapt the definitions of Section~\ref{sec.market_clearing} to centred coordinates.
\begin{definition}
\label{def.clearing_operator_centred_order_book} A clearing operator in centred coordinates is a map
$
\mathcal{C}^m: E^m\times\mathbb{Z} \rightarrow \mathcal{L}^m
$
such that $\forall (X,p)\in \mathcal{L}^m$, $\mathcal{C}^m (X,p)=(X,p)$.
\end{definition}
We now describe how clearing via order matching operates in centred coordinates. Figure~\ref{fig.clearing_centred_order_book} gives an example of clearing of a centred order book following an incoming order. Orders are first executed as they were in the fixed coordinate framework. Then, a translation of the origin of the order book to match the updated mid-price is needed. Thus, market clearing of the centred order book is the composition of an order-matching operator identical to the operator $\mathcal{C}$ introduced in Eqs.~(\ref{eq.discrete-clearingrule_buy}) and (\ref{eq.discrete-clearingrule_sell}), followed by a shift of the book to center it back to the smallest  integer larger to the mid-price. We now describe these two steps successively.

\smallskip
\textbf{Step 1: clearing process.} In this step, we define a clearing operator ${\mathcal C}'$: $E^m \to \mathcal{L}^0$. 
This operator is identical with the clearing operator ${\mathcal C}$ in the fixed coordinate framework after $E$ is changed into $E^m$ and ${\mathcal L}$ into ${\mathcal L}^0$. So, we refer the reader to Section  \ref{sec.Market_clearing_via_order_matching} for details.

\smallskip
\textbf{Step 2: centering process.} In the definition of the centering operator, we will need to pay attention to the fact that, because of Assumption~\ref{assumption.centred_order_book}, some orders at the boundary of the order book may drop outside its range after shifting and need to be cancelled. After the first step, the state $(X',p)$ of the order book belongs to ${\mathcal L}^0 \times {\mathbb Z}$. So the centering operator $J$ defined below maps ${\mathcal L}^0 \times {\mathbb Z}$ to ${\mathcal L}^m$. 
For $(p,p')\in\mathbb{Z}^2$ define  $[p,p']=\ceil{\frac{p'}{2}}-\ceil{\frac{p}{2}}$. Then the centering operator is defined as follows:

\begin{definition}[Centering operator]
Define $J:\mathcal{L}^0\times\mathbb{Z} \rightarrow \mathcal{L}^m$ by $J(Y,p)=(X,p')$ where
\begin{equation}
\begin{cases}
p'=p+\Delta p \quad\mbox{with} \quad\Delta p=a(Y)+b(Y)+\hat{p},\\
X=\sigma_{[p,p']}(Y).
\end{cases} 
\label{eq.expression_shifting_operator}
\end{equation}
and where
the map
$\sigma_i:\mathbb{N}^{2d'+1}\times\mathbb{N}^{2d'+1}\rightarrow \mathbb{N}^{2d'+1}\times\mathbb{N}^{2d'+1}$ is defined by $\sigma_i(Y)=(\sigma_i^+(Y),\sigma_i^-(Y))$ with
\begin{equation}
\begin{cases}
\sigma_i^{\pm}(Y)_j=0 &\mbox{  if  } j>\min\{d',d'-i\} \mbox{  or  } j<\max\{-d',-d'-i\},\\
\sigma_i^{\pm}(Y)_j=Y^{\pm}_{j+i} &\mbox{  otherwise.  }
\end{cases}
\label{eq.expression_sigma}
\end{equation}

\label{def.shifting_operator}
\end{definition}
The centering operator $J$ in the above definition consists of: (i) the computation of the new mid-price $p'$, and (ii) the corresponding shift of the coordinate system, $\sigma_i$ being the shift operator. The next lemma, whose proof is given in Appendix~\ref{proof.lemma.shifting_operator_image}, shows that the image of $J$ coincides with~$\mathcal{L}^m$:
\begin{lemma}
Im$(J)=\mathcal{L}^m$.
\label{lemma.shifting_operator_image}
\end{lemma}

\begin{proposition} The clearing operator in   centered coordinates is the composition of the order-matching operator $\mathcal{C}'$ and the centering operator $J$:
 $\mathcal{C}^m: E^m \times \mathbb{Z}\to {\cal L}^m$  defined by
\begin{equation}
\mathcal{C}^m(X,p)=J (\mathcal{C}'(X),p).
\label{eq.decomposition_clearing_operator_centred}
\end{equation}
is a clearing operator satisfying Definition~\ref{def.clearing_operator_centred_order_book}.
\end{proposition}

\noindent
\textbf{Proof.} From Lemma~\ref{lemma.shifting_operator_image} we obtain that the operator $\mathcal{C}^m$ is a map from $E^m$ to $\mathcal{L}^m$. Moreover, for any $(X,p)\in \mathcal{L}^m$, we claim that $\mathcal{C}^m(X,p)=(X,p)$. Indeed, define $(Z,p') =: \mathcal{C}^m(X,p)=J (\mathcal{C}'(X),p)$. Since $(X,p)\in \mathcal{L}^m$, we have $\mathcal{C}'(X) = X$ and $a(X) + b(X) + \hat p = \Delta p = 0$. 
Therefore from Eq.~(\ref{eq.expression_shifting_operator}), $p'=p+\Delta p=p$ and hence $Z=\sigma_0(X)=X$. 
This shows the claim and ends the proof. \endproof

To sum up, like in the fixed coordinate framework, the infinitesimal evolution of the centred order book can be pictured as follows:
\begin{eqnarray*}
&&
(X(t_k-),p(t_k-)) \in\mathcal{L}^m   \quad \xrightarrow{\textrm{Order flow}} \quad  (X(t_k-)+\Delta \hat{M}(t_k),p(t_k-))  \in E^m \times {\mathbb Z} \\
&& \hspace{5cm} \quad \xrightarrow{\textrm{Clearing}} \quad (X(t_k),p(t_k))=\mathcal{C}^m((X(t_k-)+\Delta \hat{M}(t_k),p(t_k-))) \in\mathcal{L}^m .
\end{eqnarray*}

\subsection{Markovian case: infinitesimal generator for the centred order book}
\label{sec.generator_of_the_centred_order_book}

We further assume that the order flow is a Markov point process. As a consequence, the LOB itself is a Markov process which can be described by its infinitesimal generator. Let $\lambda_+^m(i,(X,p),z)$, $\lambda_-^m(i,(X,p),z)$, $C_+^m(i,(X,p),z)$, $C_-^m(i,(X,p),z)$ be respectively the rate of arrival of a buy order, a sell order, a cancellation of buy order and a cancellation of sell order of size $z$ and price $i$ given that the initial state of the order book is $(X,p)$. We further assume that all intensities are bounded and that cancellations can only affect existing orders, which is expressed by:
\begin{assumption}
\label{assumption.centred_order_flow_markovian}
$\lambda_+^m$, $\lambda_-^m$, $C_+^m$, $C_-^m$ are such that
\begin{enumerate}
\item[(i)] $\forall (X,p)\in E^m\times \mathbb{Z}$, $\forall i\in \{-d',\ldots,d'\}$, $\lambda^m_{\pm}(i,(X,p),z)=C^m_{\pm}(i,(X,p),z)=0$ 
if $z\geq \min \{\sum_{i=0}^{d'}X^+_i,\sum_{i=-d'}^{-1}X^-_i \}$.
\item[(ii)] There exists $M>0$ such that $\forall (X,p)\in E^m \times {\mathbb Z}$, $\forall i\in \{-d',\ldots,d'\}$ and $\forall z\in\mathbb{N}$, $\lambda_{\pm}^m(i,(X,p),z)\leq M/(1+z)^{\alpha}$ and $C_{\pm}^m(i,(X,p),z)\leq M/(1+z)^{\alpha}$ with $\alpha>1$.
\item[(iii)] $\forall (X,p)\in E^m \times {\mathbb Z}$, $C_{\pm}^m(i,(X,p),z)=0$ when $z>X^{\pm}_i$.

\end{enumerate}
\end{assumption}
Under Assumptions~\ref{assumption.centred_order_book}, \ref{assumption.order_flow_centred_size} and  \ref{assumption.centred_order_flow_markovian}, the centred order book process is Markovian and defined as follows. Let $p_o^m:E^m\times\mathbb{Z}\rightarrow E^m\times\mathbb{Z}$ be the transition kernel of the order flow and define $\hat{\rho}(X,p)=\sum_{X'\in E^m}p_o^m((X,p),(X',p))$. We recall that $p_o^m((X,p),(X',p')) = 0$ if $p \not = p'$. 
\begin{definition}
A centred order book $\{(X_t,p_t)\}_{t\geq0}$, $(X_t,p_t): \mathcal{L}^m \rightarrow \mathcal{L}^m\}$ is a Markovian point process defined as follows: Suppose the sequence of event times $t_1,\ldots,t_{k-1}$ and the corresponding states $(X_1,p_1),\ldots,(X_{k-1},p_{k-1})$ are defined where $(X_j,p_j)$ is the state of the centred order book during $[t_{j-1},t_j)$ and $t_0=0$. Then $t_k$ and $(X_k,p_k)$ are defined as follows. $(t_k-t_{k-1})$ is distributed according to the law $\hat{\rho}(X_{k-1},p_{k-1})$ and $(X_k,p_k)$ is defined by $\mathcal{C}^m(Y,p)$ where $(Y,p)\in E^m\times\mathbb{Z}$ is distributed according to the probability $\frac{p_o^m((Y,p),(X_{k-1},p_{k-1}))}{\hat{\rho}(X_{k-1},p_{k-1})}.$
\end{definition}

Similarly to the case of absolute price coordinates, the generator $\emph{L}^m$ of the centred order book can be obtained as the composition of the generator of the order flow process $\emph{L}_o^m$ with an operator $\mathcal{C}^m$ related to the clearing operator and a restriction operator $\Xi^m$ as shown in Figure~\ref{fig.decomposition_centred_order_book}: 

\begin{figure}[H]
\centering
\[ \begin{tikzcd}[column sep=5em,row sep=3em,arrows=-latex]
B(E^m \times {\mathbb Z}) \arrow{r}{\textit{L}_o^m}  & B(E^m \times {\mathbb Z}) \arrow{d}{\Xi^m} \\%
B(\mathcal{L}^m)\arrow[swap]{u}{\tilde{\mathcal{C}}^m} \arrow{r}{\textit{L}^m}& B(\mathcal{L}^m)
\end{tikzcd}
\]
\caption{Relation between the generator of the order flow $\emph{L}_o^m$ and the generator of the order book process $\emph{L}^m$ in the centred coordinate framework. This figure is commutative.}
\label{fig.decomposition_centred_order_book}
\end{figure}

The generator $\emph{L}_o^m$ of the order flow in the centred coordinate framework can be written as follows: for any $f\in B(E^m \times {\mathbb Z})$ and any $(X,p)\in E^m \times {\mathbb Z}$,
\begin{equation}
\emph{L}_o^m f(X,p)=\sum_{Y\in E^m} p_o^m((X,p),(Y,p))[f(Y,p)-f(X,p)].
\label{eq.order_flow_generator_centred}
\end{equation}
The restriction operator $\Xi^m: B(E^m \times {\mathbb Z})\rightarrow B(\mathcal{L}^m)$ is defined for any $f\in B(E^m\times {\mathbb Z})$, by:
\begin{equation}
\Xi(f)=f|_{\mathcal{L}^m}, 
\end{equation} 
and $\tilde{\mathcal{C}}^m:B(\mathcal{L}^m)\rightarrow B(E^m\times {\mathbb Z})$ is the operator defined for any $f\in B(\mathcal{L}^m)$, by:
\begin{equation}
\tilde{\mathcal{C}}^m(f)=f\circ\mathcal{C}^m.
\end{equation}
\label{def.Xi_clearing_centred}
Then, we have the

\begin{proposition}[Generator of the centred order book]
The following decomposition holds: 
\begin{equation}
\label{eq.decomposition_generator_centred_order_book}
\emph{L}^m=\Xi^m\emph{L}_o^m\tilde{\mathcal{C}}^m.
\end{equation}
Consequently, $\emph{L}^m$ is a bounded operator and the generator of the group $e^{-t\emph{L}^m}$. Moreover, for any fixed constant $T\in\mathbb{R}_+$, we have 
\begin{equation}
\lim_{\Delta t\rightarrow 0}\| (e^{-\Delta t\emph{L}^m})^{\frac{T}{\Delta t}}- (\Xi^m e^{-\Delta t\emph{L}_o^m}\tilde{\mathcal{C}}^m)^{\frac{T}{\Delta t}} \|_{B(\mathcal{L}^m)}=0.
\end{equation}
\label{prop.decomposition_generator_centred}
\end{proposition}

The proof of this proposition is similar to that of Prop.~\ref{prop.decomposition_generator_absolute} and \ref{prop.approximation_kbe_absolute} and is ommitted. 

We can describe the formal adjoint $\emph{L}^{m*}$ of $\emph{L}^m$ in terms of the adjoint of $\emph{L}_o^m$, $\tilde{\mathcal{C}}^m$ and $\Xi^m$. Since we are not going to use the forward Kolmogorov equation in the paper, this development can be found in Appendix~\ref{sec.adjoint_of_the_generator_of_the_order_book_centred}.

\section{Examples}
\label{sec.examples}

In this section, we show that our general framework of LOB dynamics applies to a variety of models previously proposed in the literature. The first example (Model~1) deals with unit-size orders in the absolute coordinate framework. The second example (Model~2) extends it to order sizes belonging to a discrete set $\{1, \ldots, m\}$. The last example (Model~3) considers the centred coordinate framework with a state-dependent window width.

\subsection{Model 1: Poisson order flow with homogeneous order size}
\label{sec.example_1}

Cont \emph{et al.} \cite{cont2010} proposed a continuous-time stochastic model of an order book in which the order flow is a Poisson point process with six types of order book events, whose intensities are deterministic functions of the distance to the best (bid/ask) price. More precisely, letting $X\in\mathcal{L}$ be the initial state of the order book, then a limit buy (resp. sell) order of unit size arrives at the order book at price $p<a(X)$ with rate $\lambda(a(X)-p)$ (resp. at price $p>b(X)$ with rate $\lambda(p-b(X))$).  A market buy (resp. sell) order  of unit size arrives with rate $\mu$. A cancellation of a buy (resp. sell) order of unit size arrives at price $p<a(X)$ with rate $\theta(a(X)-p)|X_p|$ (resp. at price $p>b(X)$ with rate $\theta(p-b(X))|X_p|$). Here, $\lambda$ is such that
\begin{equation}
\lambda(i)=\frac{\beta}{i^{\alpha}}, \qquad \forall i\in\{1,\ldots,d  \},
\label{eq.model_1_intensity}
\end{equation}
while $\theta: \mathbb{N}\rightarrow \mathbb{R}_+$, $i\mapsto\theta(i)$ is a function of the price $i\in\{1, \ldots, d  \}$ of an order, $\beta$, $\alpha$ and $\mu$ are constants and $\beta$ and $\mu$ are positive.  
We now show that the model fits in our general framework. Denote the order book process in this model as $\{ X^1 \}_{t\geq0}$. Let new orders be only of unit size, i.e. the rate of arrival of new orders $\lambda_+(k, X, z)$, $\lambda_-(k, X, z)$, $C_+(k, X, z)$ and $C_-(k, X, z)$ are defined for any $X\in E$ as follows:
\begin{eqnarray*}
&&
\lambda_+(k, X, z)=
\begin{cases}
\frac{\beta}{(a(X)-k)^{\alpha}} & \mbox{   if   }k<a(X), z=1,\\
\mu &\mbox{   if   } k=a(X), z=1,\\
0&\mbox{   otherwise  },
\end{cases}
\hspace{1.3cm}
\lambda_-(k, X, z)=
\begin{cases}
\frac{\beta}{(k-b(X))^{\alpha}} & \mbox{   if   }k>b(X), z=1,\\
\mu & \mbox{   if   }k=b(X), z=1,\\
0&\mbox{   otherwise,  }
\end{cases} \\
&& \mbox{} \\
&&
C_+(k, X, z)=
\begin{cases}
\theta(b(X)-k)X_k^+ & \mbox{   if   }k<b(X), z=1,\\
0 & \mbox{   otherwise,   }
\end{cases}
\qquad 
C_-(k, X, z)=
\begin{cases}
\theta(k-a(X))X_k^- & \mbox{   if   }k>a(X), z=1,\\
0 & \mbox{   otherwise.   }
\end{cases}
\end{eqnarray*}
This model fits in our framework. Thanks to it, we can easily find the generator $\emph{L}^1$ of this LOB process. It is written for any $X\in\mathcal{L}$ and $f\in B(\mathcal{L})$ as follows:
\begin{eqnarray*}
\emph{L}^1f(X) &=&\sum_{i=1}^{a(X)-1}\frac{\beta}{(a(X)-i)^{\alpha}}[f(X^++e_i, X^-)-f(X)]
+\mu[f(X^+, X^--e_{a(X)})-f(X)]\\
&&+\sum_{i=b(X)+1}^{d}\frac{\beta}{(i-b(X))^{\alpha}}[f(X^+, X^-+e_i)-f(X)]
+\mu[f(X^+-e_{b(X)}, X^-)-f(X)]\\
&&+\sum_{i=1}^{b(X)-1}\theta(b(X)-i)X_k^+(t)[f(X^+-e_i, X^-)-f(X)]
\\
&&+\sum_{i=a(X)+1}^{d}\theta(i-a(X))X_k^-(t)[f(X^+, X^--e_i)-f(X)].
\end{eqnarray*}

\mathcenter

\subsection{Model 2: Poisson order flow with orders of different sizes}
\label{sec.poission_model_with_small_and_large_orders}

In real markets, orders do not always come with uniform size. In our framework, this situation can be described by choosing a proper order flow. In this subsection, we consider a simple example in which orders can be placed with sizes $z\in \{ 1,\ldots,m\}$, where $m\in\mathbb{N}$ is the maximum order size. 

Let $\lambda_{\pm}^{i}$ and $C_{\pm}^{i}$ be functions of $k\in\mathbb{N}$ and $X\in {\mathcal L}$ for $i=1, \ldots,m$. Then the rate of arrival of different orders in the order flow are defined for $k\in\{ 1, \ldots, d\}$, $z\in\mathbb{N}$ and $X\in {\mathbb L}$ as follows:\\
\begin{eqnarray}
\label{eq.model_2_buy}
&&
\lambda_+(k, X, z)=
\begin{cases}
\lambda_+^{z}(k,X) & \mbox{  if  }1\leq z\leq m,\\
0&\mbox{ otherwise,}
\end{cases}
\hspace{1.2cm}
\lambda_-(k, X, z)=
\begin{cases}
\lambda_-^{z}(k,X) &\mbox{  if  } 1\leq z\leq m,\\
0&\mbox{  otherwise,}
\end{cases}
\\
\nonumber &&\mbox{} \\
\label{eq.model_2_c_buy}
&&
C_+(k, X, z)=
\begin{cases}
C_+^{z}(k,X) & \mbox{  if  }1\leq z\leq m,\\
0&\mbox{  otherwise,}
\end{cases}
\qquad 
C_-(k, X, z)=
\begin{cases}
C_-^{z}(k,X) & \mbox{  if  }1\leq z\leq m,\\
0&\mbox{  otherwise.}
\end{cases}
\end{eqnarray}
The generator $\emph{L}_o$ of the order flow and that of the order book $\emph{L}$ can easily be deduced from \eqref{eq:express_Lo} and \eqref{col.generator_order_book_exact_expression} respectively, by restricting the sum over $z$ to $\{1, \ldots, m\}$. We will provide simulation results for this example in the next section.

\subsection{Model 3: Centred order book with state-dependent width}
Abergel et al \cite{abergel2013} modeled the dynamics of a centred order book up to a distance $K$ ticks from the opposite price on each side: for an order book $X\in E$, the   ask side in the model is $\textbf{a}=(a_1, ..., a_K):=(X^-_{b(X)+1},...,X^-_{b(X)+K})$ and the bid side is $\textbf{b}=(b_1, ..., b_K):=(X^+_{a(X)-K},...,X^+_{a(X)-1})$.
There are six types of order book events, namely limit buy (resp. sell) orders with intensity $\lambda^{L^+}_i$ (resp. $\lambda^{L^-}_i$) at relative price level $i$, market buy (resp. sell) orders with intensity $\lambda^{M^+}$ (resp. $\lambda^{M^-}$), cancellation of buy (resp. sell) orders with intensity $b_i\lambda^{C^+}_i$ (resp. $a_i\lambda^{C^-}_i$) at relative price level $i$. Once a price level moves out of the coordinate, its depth is set to be $a_{\infty}$ for the ask side and $b_{\infty}$ for the bid side. While this model is centred at two reference prices - ask and bid prices, it is in fact   centred to the mid-price with a state-dependent range of prices of the centred order book (which we refer to as the  width). We can embed this model in our framework as follows.

Let $d>K$ be the width in the centred order book, i.e. $(X+,X^-,p)\in \mathcal{L}^m $ where $X^{\pm}=(X^{\pm}_{-d},...,X^{\pm}_{d})$. Then because of the assumption that every price level more than $K$ ticks away from its opposite best available price is assigned with a default depth, we have $X^+_i=b_{\infty}$ for $i<a(X)-K$ and $X^-_i=a_{\infty}$ for $i>b(X)-K$. This sets the relation between the state in our centred order book  framework and the state defined in the model of \cite{abergel2013}. Therefore, we add an extra map $\varrho:\mathcal{L}^m\mapsto\mathcal{L}^m$ to make sure that after clearing, the state is as defined in \cite{abergel2013}. This map is defined by $\varrho(X,p)=(Y,p)$, where
\begin{equation*}
Y^+_i=
    \begin{cases}
    b_{\infty}, & \mbox{  if  }-d\leq i<a(X)-K,\\
    X^+_i, & \mbox{  if  } a(X)-K\leq i \leq d,
    \end{cases}
\qquad 
Y^-_i=
    \begin{cases}
     X^-_i, & \mbox{  if  } -d\leq i \leq a(X)-K,\\
     a_{\infty}, & \mbox{  if  } a(X)-K< i\leq d.
    \end{cases}
\end{equation*}
The clearing operator for this model is thus 
$\varrho \circ \mathcal{C}^m$. 
The intensity of the incoming order is as follows: for an order with price $z$ and size $k$, we have
\begin{eqnarray*}
&& 
\lambda_+(k,(X,p),z)=
\begin{cases}
\lambda^{L^+}_{a(X)-i} & \mbox{  if  } a(X)-K\leq i<a(X),\\
\lambda^{M^+}, & \mbox{  if  } i=d,\\
0, & \mbox{  otherwise  },
\end{cases}
\\
&& \mbox{} \\
&& 
\lambda_-(k,(X,p),z)=
\begin{cases}
\lambda^{L^-}_{b(X)+i} & \mbox{  if  }b(X)< i\leq b(X)+K, \\
\lambda^{M^-}, & \mbox{  if  } i=-d,\\
0, & \mbox{  otherwise  },
\end{cases}
\\
&& \mbox{} \\
&&
C_+(k,(X,p),z)=
\begin{cases}
\lambda^{C^+}_{a(X)-i}X^+_{a(X)-i} & \mbox{  if  }a(X)-K\leq  i<a(X), \\
0, & \mbox{  otherwise  },
\end{cases}
\\
&& \mbox{} \\
&& C_-(k,(X,p),z)=
\begin{cases}
\lambda^{-^+}_{b(X)+i}X^-_{b(X)+i} & \mbox{  if  }b(X)<  i\leq b(X)+K, \\
0, & \mbox{  otherwise  }.
\end{cases}
\end{eqnarray*}

The infinitesimal generator of the order book process for this model can easily be deduced from the theory of Section \ref{sec.state_space_centred_order_book}.

\section{Numerical simulations}
\label{sec.numerical_simulations}

The decomposition of LOB dynamics as in \ref{fig.general_idea} naturally 
lends itself to a {\it modular} approach where each of the components is implemented separately.
In this section we show  how this modularity and other features of   our approach lead to a flexible framework for  numerical simulations.The decomposition of LOB dynamics as in \ref{fig.general_idea} naturally 
lends itself to a {\it modular} approach where each of the components is implemented separately. In Sections~\ref{sec.monte_carlo_simulation} and \ref{sec.kolmogorov_backward_equation}, we apply this framework in association with a Monte Carlo simulation on the one hand and with the resolution of the Kolmogorov backward equation on the other hand, and compare the outcomes of these two simulations. We consider a test problem which consists of computing the probability of an ask price increase in two different situations, corresponding to short and long-time intervals.

\subsection{Monte Carlo Simulation}
\label{sec.monte_carlo_simulation}

In this subsection, we are interested in solving the following problem: conditioned on the initial state of the order book being $(X^+_0, X^-_0)$, find the conditional probability that the ask price increases at the first price movement. We use the fixed coordinate framework and choose an order flow according to Model~2 in Section~\ref{sec.poission_model_with_small_and_large_orders} with intensities $\lambda_{\pm}$ and $C_{\pm}$ defined in Eqs~(\ref{eq.model_2_buy}) and (\ref{eq.model_2_c_buy}). As for the clearing operator, we choose the order-matching operator defined in Eqs.~(\ref{eq.discrete-clearingrule_buy}) and (\ref{eq.discrete-clearingrule_sell}). 

\begin{figure}[H]
 \centering
\subfloat[Distribution of limit sell orders w.r.t. sizes.]{\includegraphics[width = 0.35\textwidth]{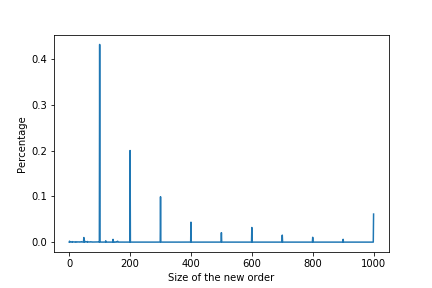}} 
\subfloat[Evolution of the ask and bid prices.]{\includegraphics[width = 0.35\textwidth]{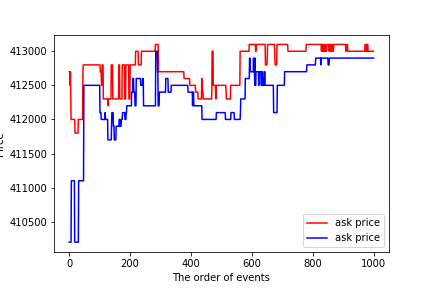}}

\caption{(a): distribution of limit sell orders w.r.t. sizes up to 100500 shares. The value at 1000 is the percentage of orders with sizes more than 1000 shares. The horizontal axis is the size of an order and the vertical axis, the percentage of orders of that size.  (b): evolution of the ask and bid prices before the 1000th order book event. The vertical axis is price in dollar times 10000 while the horizontal axis is the event index (which is roughly proportional to time but not exactly because the time interval between two events is random).}
\label{fig.data_analysis}
\end{figure}

To validate our numerical simulations, we compare its outcome with data from the Lobster Database (https://lobsterdata.com). We use these data to calibrate the coefficients of the model (such as the distribution of news orders as a function of their prices and sizes) as well as to  estimate the targeted output, namely the  conditional probability of an ask price increase at the first price movement. The data concern the shares of the company Micron Technology (MU) on the day 2019/07/01. The order size distribution and the evolution of the ask and bid prices with respect to order book events are respectively shown in Figure~\ref{fig.data_analysis}~(a) and (b).

\begin{figure}[H]
\centering
\subfloat[Sell order intensities]{\includegraphics[width = 0.35\textwidth]{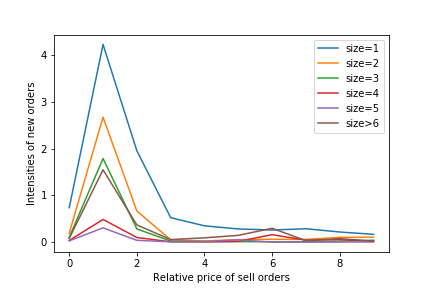}} 
\subfloat[Cancellation of sell orders]{\includegraphics[width = 0.35\textwidth]{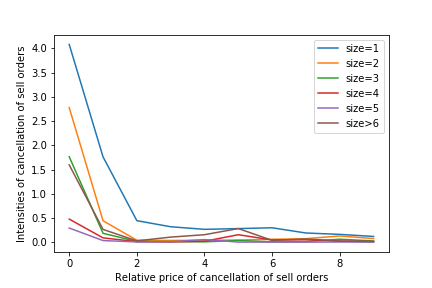}}
\caption{(a): intensities of sell orders at different prices relative to the bid price. (b): intensities of cancellation of sell orders at different prices relative to the ask price. In the two figures, the different curves corresponds to different order sizes, from 1 to 6 units. Units along the vertical axes are  (second)$^{-1}$. }
\label{fig.order_flow_size_relative_price}
\end{figure}

In Figure~\ref{fig.data_analysis}~(a), we see that most of the orders are of size 100 shares but with still a significant number of orders of sizes 200 shares or more. Consequently we will count order sizes in units of 100 shares, i.e., 4 units means 400 shares. In Figure~\ref{fig.data_analysis}~(b) we notice that the tick size is $100$ (i.e. $0.01$ dollar) and the difference between the ask and bid prices is one tick size most of the time (88\% of the time). Therefore for the sake of simplicity, we initialise the state with an initial spread equal to 1 tick size. Moreover, as observed in the literature \cite{biais1995, bouchaud2002}, price influences the intensity of an order. Therefore we also investigate the relation between the relative price of an order to its best available price and the intensity of this order. Figure~\ref{fig.order_flow_size_relative_price}~(a) (resp. (b)) shows the intensities of sell orders (resp. cancellation of sell orders) at different relative prices  to the bid price (resp. ask price). In both figures, the different curves corresponds to different order sizes, from 1 to 6 units.  From Figure~\ref{fig.order_flow_size_relative_price}~(a), we see that most sell orders arrive at a price one tick higher than the bid price and that the cancellation of sell orders arrive at prices equal to the ask price whatever order sizes. 

We let the dimension $d$ of $X^+$ and $X^-$ be~6, because the average size of the orders is around 3 units and $B^{-1}_{X_t}(3)$ is less than or equal to 3 ticks, with an exception of 0.0006\% and $S^{-1}_{X_t}(3)$ is less than or equal to 3 ticks, with an exception of 0.08\%. We take the initial states of the LOB to be $X_0^+=[b_2, b_1, z_1, 0, 0, 0]$ and $X_0^-=[0, 0, 0, z_2, a_1, a_2]$, and let $z_i$ range in $\{ 1 ,\ldots, 6 \}$, and $a_i$, $b_i$ range in the set extracted from the data of queue sizes (or depths) at prices 1 and 2 ticks from the ask and bid prices for $i=1, 2.$ We are interested in the dependence of the results on $z_1$ and $z_2$, which are respectively the volume of orders at the ask and bid prices and called the bid and ask depths.  We consider two different order flow processes, respectively given by Model~A and Model~B as follows.

In Model~A, the intensity of orders, i.e. the rate of order arrival is null except for order sizes equal to 1 and for queue sizes below or above a certain threshold $n$ according to the type of order. The intensity of each order type is calibrated from the data by equating it to the total number of this type of orders divided by the maximal price $d=6$, and by the total time $T$. Given the set of data at hand, this calibration leads to 
\\
\noindent\begin{minipage}{.4\linewidth}
\[
\lambda_+(k, X, z)=
\begin{cases}
11.6 & \mbox{  if  }X^+_k<n \textrm{ and } z=1,\\
0&\mbox{ otherwise,}
\end{cases}
\]
\end{minipage}
\begin{minipage}{.4\linewidth}
\[
\lambda_-(k, X, z)=
\begin{cases}
10.7 &\mbox{  if  } X^-_k<n \textrm{ and } z=1,\\
0&\mbox{  otherwise.}
\end{cases}
\]
\end{minipage}
\\
\noindent\begin{minipage}{.4\linewidth}
\[
C_+(k, X, z)=
\begin{cases}
10.8 & \mbox{  if  }X^+_k\geq n \textrm{ and } z=1,\\
0&\mbox{ otherwise,}
\end{cases}
\]
\end{minipage}
\begin{minipage}{.4\linewidth}
\[
C_-(k, X, z)=
\begin{cases}
9.7 &\mbox{  if  } X^-_k\geq n \textrm{ and } z=1,\\
0&\mbox{  otherwise,}
\end{cases}
\]
\end{minipage}
\\

\noindent 
where $n=300$. 

\begin{figure}[H]
\centering
\subfloat[Data]{\includegraphics[width = 0.25\textwidth]{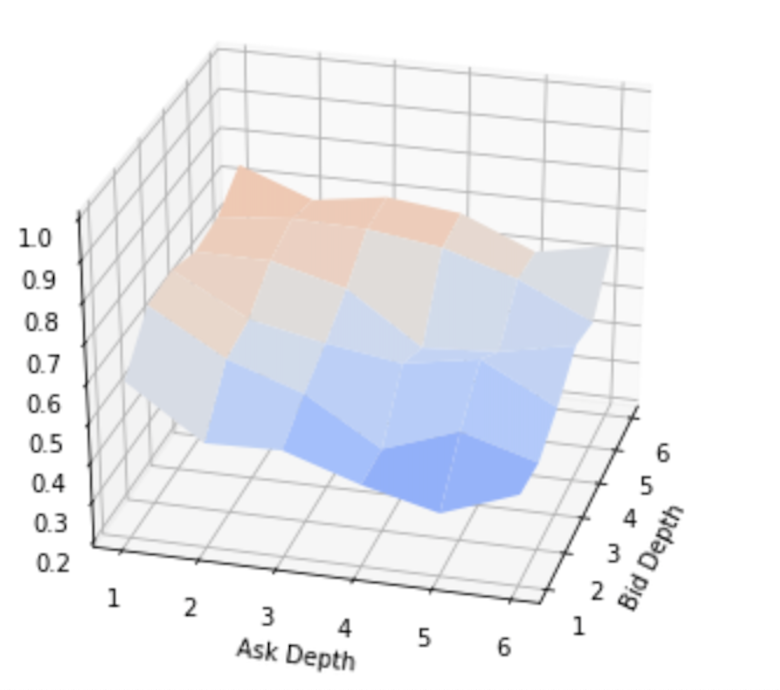}}
\subfloat[Model A]{\includegraphics[width = 0.25\textwidth]{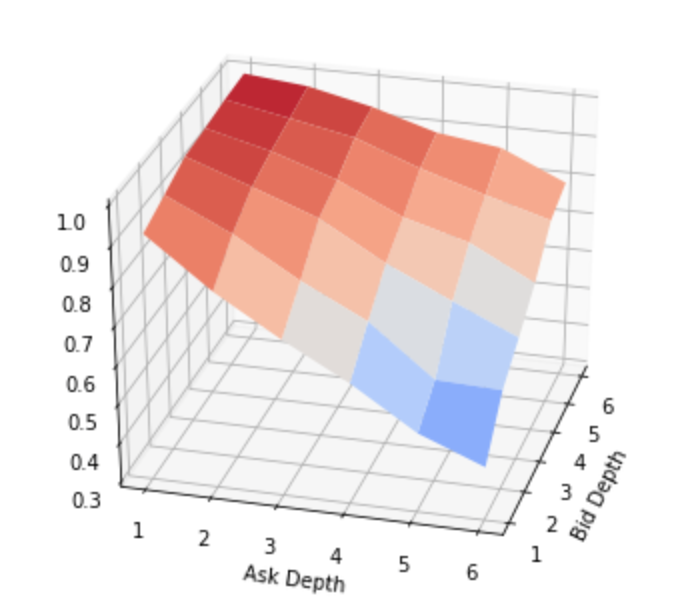}}
\subfloat[Model B]{\includegraphics[width = 0.25\textwidth]
{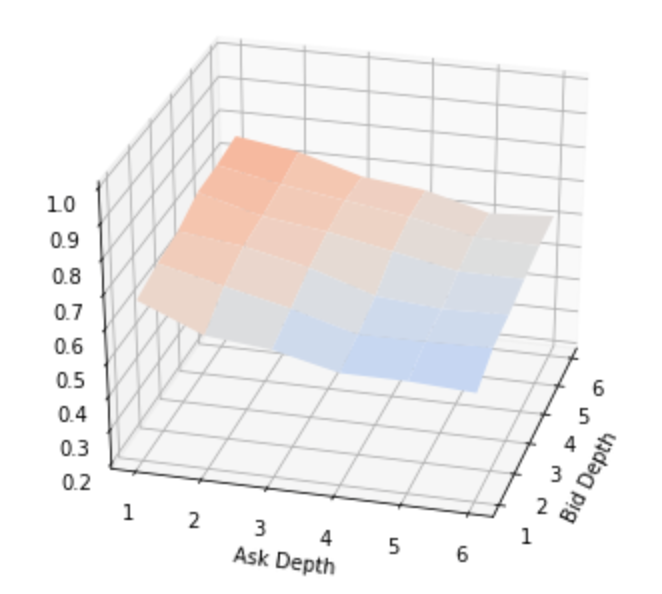}}
\subfloat{\includegraphics[width=0.09\textwidth]{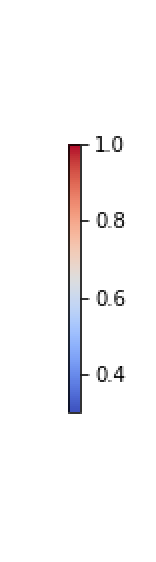}}
\caption{(a) conditional probability of an ask price increase as a function of the ask and bid depths: (a) estimated from the data; (b) simulated using Model A (uniform order sizes); (c) simulated using Model B (non-uniform order sizes). }
\label{fig.order_flow_size_relative_price_outcome}
\end{figure}

In Model~B, a more complex order flow where the intensities of orders depend on order sizes ranging in $\{ 1,\ldots,6 \}$ and on the relative price to the ask or bid price according to the type of order is considered. The intensity of each type of orders is approximated by the total number of this type of orders at each relative price divided by the total time $T$. Again, calibration against data gives rise to the following formulas:
\begin{eqnarray*}
\lambda_+(k,X,z ) &=&
\begin{cases}
\beta_{z,a(X)-k} &\mbox{  if  } X^+_k<n-z \textrm{ and }
0\leq a(X)-k\leq 5,\\
0&\mbox{  otherwise,}
\end{cases} \\
&& \mbox{} \\
\lambda_-(k,X,z ) &=&
\begin{cases}
\alpha_{z,k-b(X)} &\mbox{  if  } X^-_k<n-z \textrm{ and } 0\leq k-b(X)\leq 5,\\
0&\mbox{  otherwise,}
\end{cases} \\
&& \mbox{} \\
C_+(k,X,z )&=&
\begin{cases}
\gamma_{z,b(X)-k} &\mbox{  if  } X^+_k>0 \textrm{ and } 0\leq b(X)-k\leq 5,\\
0&\mbox{  otherwise,}
\end{cases} \\
&& \mbox{} \\
C_-(k,X,z )&=&
\begin{cases}
\mu_{z,k-a(X)} &\mbox{  if  } X^-_k>0 \textrm{ and } 0\leq k-a(X)\leq 5,\\
0&\mbox{  otherwise,}
\end{cases} 
\end{eqnarray*}
with matrices $\alpha$, $\mu$, $\beta$, $\gamma$ as follows:
\begin{equation*}
\alpha = \begin{pmatrix}
0.74 & 4.23 & 1.96 & 0.53& 0.35&0.28\\
0.19& 2.68& 0.67& 0.05& 0.03& 0.05\\
0.10& 1.79& 0.28& 0.02& 0.01& 0.01\\
0.04& 0.49& 0.10& 0.01& 0.01& 0.02\\
 0.03& 0.31& 0.04& 0.01& 0.01& 0.05\\
0.08& 1.55& 0.37& 0.06& 0.09& 0.15\\
\end{pmatrix},
\quad \mu = \begin{pmatrix}
4.08&  1.76&  0.44& 0.32& 0.27& 0.28\\
2.78&  0.44& 0.04&  0.04&  0.04& 0.04\\
1.76&  0.18&  0.02&  0.01& 0.01& 0.04\\
0.48& 0.09&  0.01& 0.01&  0.02& 0.15\\
0.29&  0.03&  0.01&  0.01&  0.05& 0.003\\
1.60& 0.26&  0.03& 0.11&  0.15& 0.28\\
\end{pmatrix},
\end{equation*}
\begin{equation*}
\beta = \begin{pmatrix}
0.46& 3.95& 1.75& 0.35& 0.30& 0.39\\
0.12& 2.65& 0.65& 0.07& 0.04& 0.05\\
0.08 &1.48 & 0.24 & 0.02 & 0.01 & 0.01\\
0.02& 0.49& 0.09&0.01& 0.01&0.01\\
0.018&0.30& 0.04&0.01&0.02& 0.04\\
0.05&1.26&0.37&0.09& 0.11&0.15\\
\end{pmatrix},
\quad \gamma = \begin{pmatrix}
3.72& 1.49& 0.41& 0.28& 0.34&0.42\\
2.73& 0.44& 0.06&0.05&0.05& 0.05\\
1.41&0.17&0.02& 0.01& 0.01&0.02\\
0.47& 0.08&0.01&0.01&0.02&0.16\\
0.27&0.03& 0.01& 0.01&0.04&0.002\\
1.29& 0.28&0.05& 0.12&0.15&0.28\\
\end{pmatrix},
\end{equation*}
 and $n=300.$ To avoid confusion, here $\alpha_{1,1}=0.74$, $\alpha_{1,2}=4.23$ and $\alpha_{2,1}=0.19$.
Details about the calibration of the coefficients (using a kernel estimation method) and the implementation of the Monte-Carlo method can be found in \cite{Lifan_phdthesis}. The conditional probability of the ask price increase at the next price movement as a function of the ask and bid depths estimated from the data or computed by means of Model~A and Model~B using the Monte Carlo method are respectively shown in Figure~\ref{fig.order_flow_size_relative_price_outcome} (a), (b) and (c).

We see that the Monte Carlo simulations of both models agree qualitatively well with the experimental data. In particular, two main properties of the data are recovered in the simulation: when the ask depth increases, the conditional probability decreases; when the bid depth increases, the conditional probability increases. Secondly, we notice that Model~B fits the data better than Model~A. The results of Model~B also agree quantitatively well with the experimental data. This indicates that both the order size and the relative price of the orders influence the price dynamics, which agrees with other observations from the literature \cite{biais1995,bouchaud2002}. In particular, comparisons between simulation results of a model where the order intensities are only dependent on the relative prices (which can be found in \cite{Lifan_phdthesis}) and the results of Model~A and Model~B show that large orders play a very important role in the evolution of the order book. 

\begin{figure}[H]
\centering
{\includegraphics[width = 0.7\textwidth]{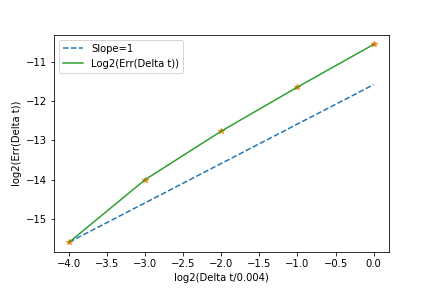}}
\caption{Convergence of the forward Euler scheme for the backward Kolmogorov equation \eqref{eq.Kolmogorov_equation} using Method 1. We observe a superlinear convergence rate.}
\label{fig.final_comparison_and_convergece}
\end{figure}

 The discrepancy between the simulations and the data can be attributed to several facts. First, the spread (i.e. the difference between the ask and bid prices)  may have significant influence on the conditional probability of the ask price increase at the next price movement. Indeed, if the spread is larger than 1, the probability of a limit sell order falling in the spread increases, which means that the probability of an ask price decrease gets larger. Since in the considered data, the spread is larger than 1 about 10\% of the time, this may explain why the conditional probability of an ask price increase is smaller in the data than with the model.  Another reason could be that the sample size is still not large enough. Finally, more realistic simulations should involve more complex state-dependent intensities of new orders, taking into consideration the influence of the ask and bid depths of the order book on the trader's choice of the order size. 

\subsection{Kolmogorov backward equation}
\label{sec.kolmogorov_backward_equation}

We now give an example of how  the Kolmogorov backward equation (\ref{eq.Kolmogorov_equation}) may be used to compute transition probabilities for price and order book dynamics. 
I
Given the initial state of the order book being $Y=(Y^+, Y^-)$, we seek to compute the conditional probability that the ask price at time $T$ is larger than the initial ask price:
${\mathcal P} =: \mathbb{P}[a(X_T)>a(Y)| X_0=Y] = \mathbb{E}[f_Y(X_T)| X_0=Y]$, where $X_t$ denotes a trajectory of the LOB and $f_Y(X) = 1_{\{ X \in {\mathcal L} \, | \, a(X)>a(Y)\}}$ with $1_S$ denoting the indicator function of the set $S$. We have ${\mathcal P} = w_Y(0,Y)$ where $w_Y(t,X) = \mathbb{E}[f_Y(X_T)| X_t=X]$ is the solution of the Kolmogorov backward equation \eqref{eq.Kolmogorov_equation} with terminal condition $w_Y(T,X) = f_Y(X)$.

We   discretize the time interval $[0, T]$ into $N_T=\frac{T}{\Delta t}$ steps of length $\Delta t$. Letting $t_n=n\Delta t$ where $n=0, \ldots, N_T$ with $t_0=0$ and $t_{N_T}=T$, we let the sequence $\{w(t_{n-1}, X)\}_{n=1}^{N_T}$ be such that $w(t_{n-1}, X)=e^{\Delta t \emph{L}}w(t_n, X)$, where $w(T, X)=f_{Y}(X)$. We present two methods for approximating $w(t_{n-1}, X)$. 

\smallskip
\noindent
{\em Method~1:} we use a forward Euler discretization of $e^{\Delta t \emph{L}}$ and write $$\tilde{w}(t_{n-1}, X)=(I_{B(\mathcal{L})}+\Delta t \emph{L})\tilde{w}(t_{n}, X),$$ where $\tilde{w}(t_{n}, X)$ is the time discretized approximation of $w(t, X)$ at time $t_n=n\Delta t$.

\smallskip
\noindent
{\em Method~2:} we use the decomposition from Proposition~\ref{prop.approximation_kbe_absolute} first and write $\hat{w}(t_{n-1}, X)=\Xi e^{\Delta t \emph{L}_o}\tilde{\mathcal{C}} \hat{w}(t_n, X)$. Then we apply a forward Euler discretization to $e^{\Delta t \emph{L}_o}$ and get: 
\begin{equation}
\hat{w}(t_{n-1}, X)=\Xi (I_{B(E)}+\Delta t \emph{L}_o)\tilde{\mathcal{C}} \hat{w}(t_n, X).
\label{eq.kolmogorov_discretise}
\end{equation}

\smallskip
\noindent
We note that, in view of the formula $\Xi \tilde{\mathcal{C}}=I_{B(\mathcal{L})}$, the two methods actually coincide. But this is a special feature resulting from the use of the forward Euler discretization. If, for instance, we used a second order approximation of the exponential, Method~1 would lead to 
$$\tilde{w}(t_{n-1}, X)=(I_{B(\mathcal{L})}+\Delta t \emph{L}+\frac{\Delta t^2\emph{L}^2}{2})\tilde{w}(t_{n}, X),$$ 
whereas Method~2 would give 
$$\hat{w}(t_{n-1}, X)=\Xi (I_{B(E)}+\Delta t \emph{L}_o+\frac{\Delta t^2\emph{L}_o^2}{2})\tilde{\mathcal{C}} \hat{w}(t_n, X).$$ The two methods are different because $\Xi \emph{L}_o^2\tilde{\mathcal{C}}\neq \emph{L}^2 $. Indeed $\emph{L}^2=\Xi \emph{L}_o\tilde{\mathcal{C}}\Xi \emph{L}_o\tilde{\mathcal{C}} $ and $\tilde{\mathcal{C}} \Xi\neq I_{B(E)}$ ($\tilde{\mathcal{C}} \Xi$ is a  projection of $B(E)$ on $B(\mathcal{L})$ when $B(\mathcal{L})$ is considered as a subspace of $B(E)$). In this work, we will restrict to the use of the forward Euler discretization, and will implement Eq.~(\ref{eq.kolmogorov_discretise}). We refer to \cite{Lifan_phdthesis} for the implementation details. 
\begin{remark}
In some cases where the operator $e^{\Delta t \emph{L}_o}$ can be explicitly computed, then the use of Method~2 would only involve the splitting error as quantified in Proposition~\ref{prop.approximation_kbe_absolute}.
\end{remark}

 The intensities of the order flow are the same as in Section~\ref{sec.monte_carlo_simulation} except that $n =10$. For simplicity, the initial state $X_0$ is assumed to be $X_0^+=[2, 4, z_1, 0 ,0 ,0]$ and $X_0^- = [0, 0, 0, z_2, 4, 2]$. We first check that the numerical approximation of the conditional probability of an ask price increase in~$0.2$ s by means of the Kolmogorov equation of Model~A solved by the forward Euler method (\ref{eq.kolmogorov_discretise}) is of order~1, as it should. 
 In Figure~\ref{fig.final_comparison_and_convergece}, we plot the error in log scale between a numerical solution computed with a time step $\Delta t$ and a reference solution obtained by approximating the Kolmogorov equation with a very small time step $\Delta t_{min}=1.25\times 10^{-4}$. The error is thus defined by
 \begin{equation}
Err(\Delta t) = |\hat{w}(0,X_0,\Delta t)-\hat{w}(0,X_0, \Delta t_{min})|,
\end{equation}
where $\hat{w}(0, X_0, \Delta t)$ is the numerical value of the conditional probability with time step $\Delta t$. We can see by comparing the curve joining the simulation points with the line of slope~1 that the approximation is at least of order~1. 

Now, the simulations show that the depths at prices which are not the ask and bid prices have extremely small influence on the conditional probability of an ask price increase in $0.2$ s.

Then, for simplicity, we fix these depths and  compute this conditional probability for varying ask and bid depths, by means of the Kolmogorov equation (with time step $\Delta t \, \sim \, 0.0005$ s.) on the one hand and by Monte Carlo simulations on the other hand, in the conditions of Model~B above. The results obtained by these two methods are shown side-by-side in Table~\ref{table.comparison_2}.
We notice the excellent agreement between the Kolmogorov and Monte Carlo simulations, as it should, because they are two approximations of the same process. To generate the whole table, the Monte Carlo method for 500 replications takes around~8 hours, while the Kolmogorov equation requires around~17~hours on our computer server. However, we notice that the time complexity of Monte Carlo simulations grows faster with time than that of the Kolmogorov equation. Indeed, suppose we have estimated the conditional probability of an ask price increase at time $T$. For any time $T'>T$, we only need to simulate the Kolmogorov equation from time $T$ to $T'$, which means that the time complexity of the Kolmogorov equation grows linearly. With the Monte Carlo method, we need to perform the simulation for the whole time interval $[0,T']$ again. Therefore the time complexity of the Monte Carlo method grows quadratically with time. In conclusion, the Monte-Carlo method is faster if we need to evaluate the probability of an ask price incresase after a single time interval, while the Kolmogorov equation is preferable if we need to evaluate it sequencially for multiple times. 

\begin{table}
\fontsize{7}{9}\selectfont
\center
\begin{tabular}{ | m{1.8em} | m{2.2em} | m{6.4em}| m{6.4em} | m{6.4em} |m{6.4em} |m{6.4em} |m{6.4em} |}
\hline
Ask & & Bid =1 & Bid =2 & Bid =3 & Bid =4 & Bid =5 & Bid =6 \\
 \hline
\multirow{2}{*}{1}& MC & $0.233\pm 0.009$ & $0.238 \pm 0.009$ & $0.244 \pm 0.009$ & $0.247\pm 0.009$ & $0.250\pm 0.009$ & $0.249 \pm 0.009$\\
\cline{2-8}
& KBE & 0.226 & 0.235 &  0.242 &  0.247 & 0.250 & 0.249 \\
 \hline         
\multirow{2}{*}{2}& MC & $ 0.186\pm  0.009$ & $0.192 \pm 0.009 $ & $0.196 \pm 0.009$ & $0.200 \pm 0.009 $ & $0.201 \pm 0.009$ & $ 0.201 \pm  0.009$\\
\cline{2-8}
 & KBE &0.183 & 0.190 &0.197 &  0.201 & 0.204 &0.203 \\
 \hline      
\multirow{2}{*}{3}& MC & $0.152\pm 0.008$ & $0.157 \pm  0.008$ & $ 0.161 \pm 0.008$ & $0.163  \pm 0.008 $ & $0.164 \pm 0.009$ & $ 0.164  \pm  0.008$\\
\cline{2-8}
& KBE &0.151 & 0.157 & 0.163  & 0.167 & 0.169 & 0.168\\
 \hline       
\multirow{2}{*}{4}& MC & $  0.115\pm  0.007$ & $0.118\pm 0.007 $ & $0.122 \pm 0.008 $ & $ 0.124 \pm 0.007$ & $0.125 \pm 0.007$ & $0.124 \pm 0.008$\\
\cline{2-8}
 & KBE &0.120 & 0.125 & 0.130 & ,0.134 &  0.136  & 0.135 \\
 \hline    
\multirow{2}{*}{5}& MC & $0.193 \pm  0.007$ & $0.097\pm  0.006$ & $0.099 \pm 0.007$ & $0.100\pm 0.007$ & $ 0.102\pm  0.007 $ & $0.102 \pm 0.007$\\
\cline{2-8}
& KBE &0.100  & 0.105  &0.109  & 0.112 & 0.114 & 0.113 \\
 \hline       
\multirow{2}{*}{6}& MC & $0.099 \pm 0.007$ & $0.101 \pm 0.007$ & $0.104 \pm 0.007$ & $0.105 \pm 0.007$ & $ 0.107 \pm 0.007$ & $0.106 \pm   0.007$\\
\cline{2-8}
& KBE &0.107 & 0.112 & 0.116 & 0.119 & 0.121 &0.120  \\
 \hline
\end{tabular}
\caption{Conditional probability of an ask price increase after~0.2s as a function of the bid and ask depths for Model~B. Each box includes the results of the Monte Carlo simulation on the top line and that of the Kolmogorov equation on the bottom line. The results of the Monte Carlo simulation show the mean and the standard deviation over $N^*=500$ experiments. }
\label{table.comparison_2}
\end{table}

\begin{figure}[h]
 \centering
\subfloat[Data]{\includegraphics[width = 0.31\textwidth]{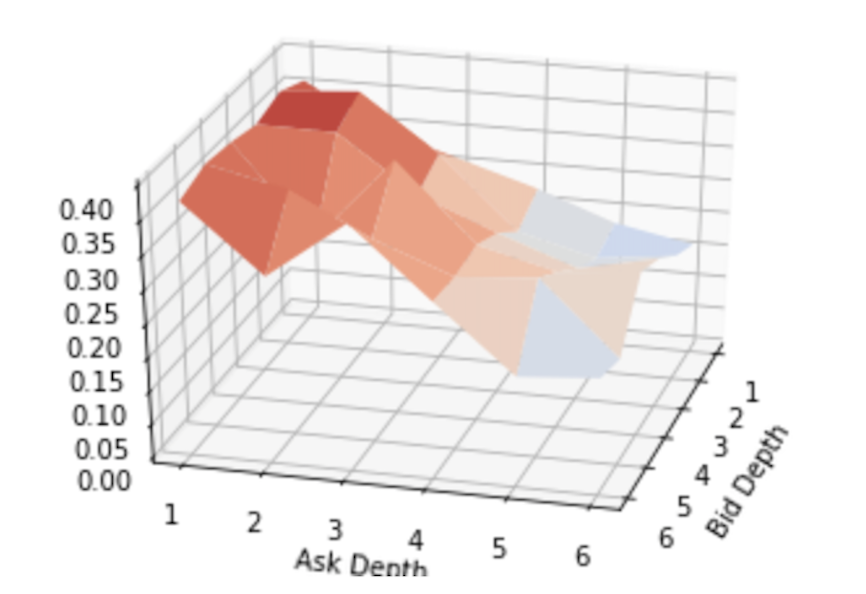}} 
\subfloat[Kolmogorov]{\includegraphics[width = 0.3\textwidth]{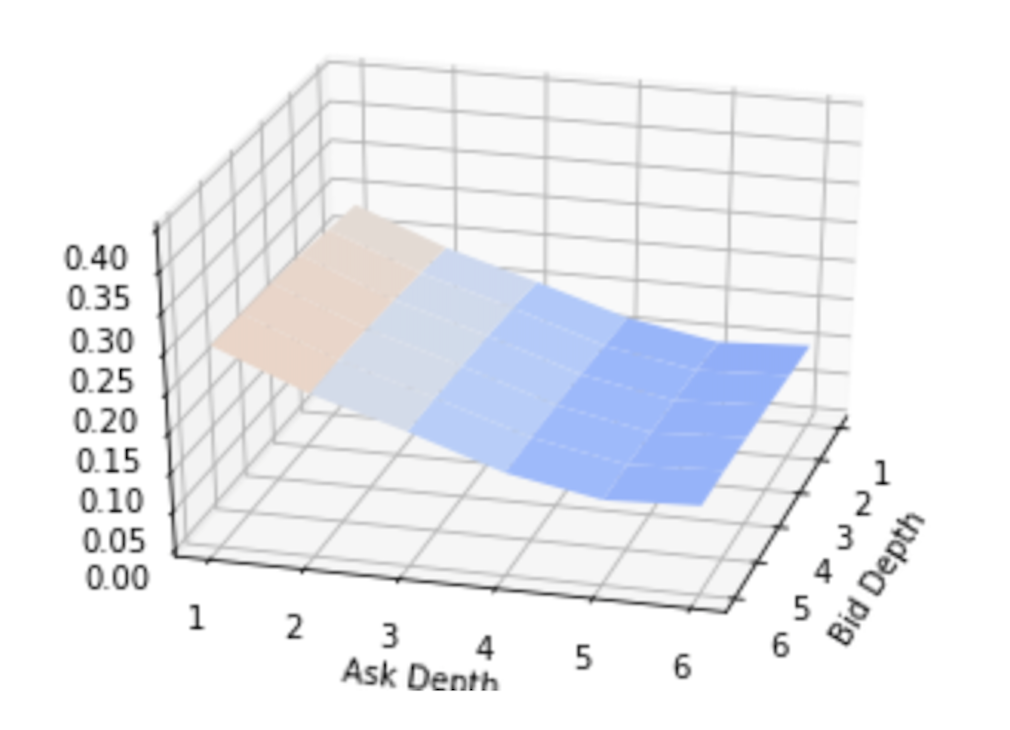}}
\subfloat[Monte Carlo]{\includegraphics[width = 0.3\textwidth]{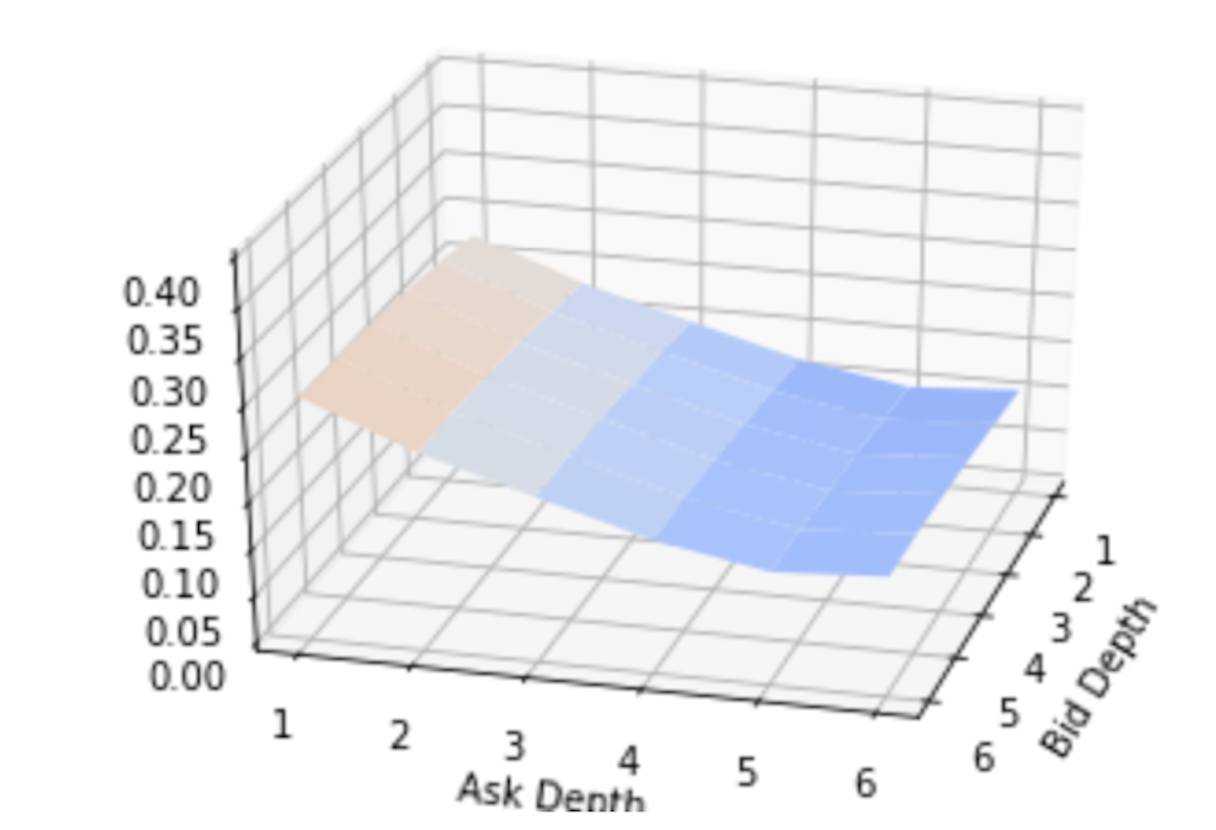}}
\subfloat[]{\includegraphics[width = 0.05\textwidth]{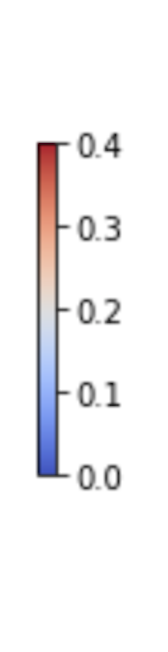}}

\caption{Conditional probability of an ask price increase after 0.2s as a function of the ask and bid depths. (a):~estimation of the conditional probability from the data. (b) and (c): respectively simulation results of the Kolmogorov equation and the Monte Carlo method. The x-axis and y-axis are respectively the ask and bid depths ranged in $\{ 1,\ldots,6 \}$ ticks.}
\label{fig.Kolmogorov_data}
\end{figure}

The conditional probability of an ask price increase in~$0.2$ s observed from the data is shown in Figure~\ref{fig.Kolmogorov_data} (a). Once again, it was estimated using a kernel estimation method the details of which can be found in \cite{Lifan_phdthesis}. From this figure, we see that there is no significant dependence of the conditional probability on the bid depth. This feature can also be noticed on the results of Monte-Carlo simulations (using Model~B for the data) (Figure~\ref{fig.Kolmogorov_data}, (b)) and on results of simulations of the Kolmogorov equation (Figure~\ref{fig.Kolmogorov_data}, (c)) (also using Model B). However, there is a discrepancy between the models and the data, namely the 
conditional probability is smaller in the model than in the data. We believe the reasons for this discrepancy can be explained in the same way as in the previous section.

\section{Conclusion and perspectives}\label{sec.conclution}

We have introduced a general framework for limit order book processes using a decomposition method: the evolution of the order book is modelled as the composition of an order flow and a market clearing process. The order flow can be any stochastic process, making it possible to apply our framework to a large number of different order book models. The market clearing and the order flow processes are independent. If both are Markovian, the order book process is also Markovian. Then the infinitesimal generator and Kolmogorov equations can be used to describe the dynamics of the order book. The determination of the generator $\emph{L}$ of the order book can be done by writing $\emph{L}$ as the composition of three simpler operators. This framework can also be applied when the order book is centred at the mid-price. Several examples of applications of this framework to order book models from the literature are then discussed to demonstrate the large applicability of out framework. Finally numerical simulations using a Monte Carlo method and a direct resolution of the Kolmogorov equation are provided. Future works involve, for instance, applying this framework to study the asymptotic behaviour of the order book. The analysis of the long-time behaviour of the order book can be achieved by time scaling. Another interesting research direction is to view the clearing operator as an optimal transport operator, so as to apply optimal transport theory to analyse order book dynamics.


\newpage
\appendix
\bigskip
\begin{center}
\textbf{\LARGE Appendices: technical proofs}
\end{center}

\section{Proof of Proposition \ref{prop.ordermatching} (Clearing operator for the order-matching method)}

\noindent
\textbf{Existence.} We first show that the map $\mathcal{C}$ defined in Prop.\ref{prop.ordermatching} satisfies properties (i)--(iv). Notice that $$\min\{0,g_{X}(p_B(X))\}\leq 0 \mbox{   and   }\max\{0,g_{X}(p_A(X))\}\geq 0.$$ Therefore, all entries of ${\cal C}(X)^+$ and ${\cal C}(X)^-$ are non-negative so that $X_{cl}\in E$. Moreover, $b({\cal C}(X))\leq p_B(X)<p_A(X)\leq a({\cal C}(X))$. Therefore, $X_{cl}\in\mathcal{L}$. We first check if $\mathcal{C}$ satisfies the definition of a clearing operator in Def.~\ref{def.clearing_absolute_general}.

If $X\in\mathcal{L}$, then $p_B(X)=b(X)$ and  $p_A(X)=a(X)$. Since $g_{{X}}(p_{B}({X}))+{X}^+_{p_{B}({X})}=-B_X(p_B(X))+X^+_{p_B(X)}=0$ and $g_{{X}}(p_{A}({X}))-{X}^+_{p_{A}({X})}=S_X(p_B(X))-X^-_{p_A(X)}=0$, we have $$g_{{X}}(p_{B}({X}))\leq-{X}^+_{p_{B}({X})}\quad\mbox{and}\quad g_{{X}}(p_{A}({X}))\geq {X}^-_{p_{A}({X})}.$$ As we are in the first case of Eqs.~(\ref{eq.discrete-clearingrule_buy}) and (\ref{eq.discrete-clearingrule_sell}) for respectively ${\cal C}(X)^+$ and ${\cal C}(X)^-$, therefore ${\cal C}(X)^+=\tau_{  p_B(X)}(X^+)$. But because $p_B(X)=b(X)$, we have ${\cal C}(X)^+=\tau_{  p_B(X)}(X^+)=X^+$. The same reasoning holds for ${\cal C}(X)^-=X^-$. Therefore,  $X_{cl}=X$.
Conversely, if $\mathcal{C}(X)=X$, then $X^+={\cal C}(X)^+$. This together with Eqs.~(\ref{eq.discrete-clearingrule_buy}) and (\ref{eq.discrete-clearingrule_sell}) implies that $\forall i\in(p_B(X),d], X^+_i=0$. Thus $\sup{\rm supp}(X^+)\leq p_B(X)$. Using the same argument we have $\inf{\rm supp}(X^-)\geq p_A(X)$. And so we get 
\[
a(X)\leq p_B(X)<p_A(X)\leq b(X),
\]
which implies that $X\in\mathcal{L}$. Therefore, $\mathcal{C}$ is a clearing operator. Now we check conditions (i)-(iv):
\begin{enumerate}
\item[(i)] Let $X-X_{cl}=Z$. Then from Eqs.~(\ref{eq.discrete-clearingrule_buy}) and (\ref{eq.discrete-clearingrule_sell}) we have
\begin{equation}
\begin{cases}
Z^+=
\begin{cases}
\tau^{  p_{B}(X)+1}(X^+) &\mbox{   if }g_{X}(p_{B}(X))\leq-X^+_{p_{B}(X)},\\
\tau^{  p_{B}(X)}(X^+)+\min\{0,g_{X}(p_{B}(X))\}e_{p_{B}(X)} &\mbox{   if }g_{X}(p_{B}(X))>-X^+_{p_{B}(X)},\\
\end{cases}\\
Z^-=
\begin{cases}
\tau_{  p_{A}(X)-1}(X^-)&\mbox{   if }g_{X}(p_{A}(X))\geq X^-_{p_{A}(X)},\\
\tau_{  p_{A}(X)}(X^-)-\max\{0,g_{X}(p_{A}(X))\}e_{p_{A}(X)} &\mbox{   if }g_{X}(p_{A}(X))<X^-_{p_{A}(X)}.\\
\end{cases}
\end{cases}
\label{eq.clearing_executed_orders}
\end{equation}
We check that all entries of $Z^+$ and $Z^-$ are non-negative. For $Z^+$, the only case where it may not be true is if $g_{X}(p_{B}(X))>-X^+_{p_{B}(X)}$. But then $Z^+_{p_B(X)}=X^+_{p_B(X)}+\min\{0,g_{X}(p_B(X))\}=\min\{X^+_{p_B(X)},X^+_{p_B(X)}+g_{X}(p_B(X))   \}\geq 0$ and all other entries are non-negative. The same argument can be used for $Z^-$. Therefore, $Z\in E$ and condition (i) is satisfied.
\item[(ii-iii)] Conditions (ii) and (iii) can be checked together. If $X\in\mathcal{L}$, then $Z=(0,0)$. Therefore, $|Z^+|=|Z^-|$ and $\sup{\rm supp}(Z^-)\leq \inf{\rm supp}(Z^+)$ so that conditions (ii) and (iii) are satisfied. Now we assume that $X\not\in\mathcal{L}$. Consequently, $X\in E_+$. Define
\begin{equation}
g_{X}^+(k)=g_{X}(k)+X^+_k=S_{X}(k)-B_{X}(k+1),
\label{eq.clearing_criterion_buy}
\end{equation}
and
\begin{equation}
g_{X}^-(k)=g_{X}(k)-X^-_k=S_{X}(k-1)-B_{X}(k).
\label{eq.clearing_criterion_sell}
\end{equation}
Comparing $g_X^+(p_B(X))$ and 0, we have the following two cases:

\paragraph{Case 1:} $g_X^+(p_B(x))>0$. We have
\[
g_X(p_A(X))-X^-_{p_A(X)}=S_X(p_A(X)-1)-B_X(p_A(X)-1)\geq S_X(p_A(X))-B_X(p_A(X)-1)=g_X^+(p_B(X))>0,
\]
which implies that 
\[
Z^+=\tau^{  p_{B}(X)}(X^+)+g_{X}(p_{B}(X))e_{p_{B}(X)}\qquad\mbox{and}\qquad Z^-=\tau_{  p_{A}(X)-1}(X^-).
\]
Moreover, notice that $p_A(X)=p_B(X)+1$ as $g_X(p_B(X+1) = g^+_X(p_B(X)+1)+X^-_{p_B(X)+1}$. Therefore, we have
\[
|Z^+|=S_X(p_B(X))+B_X({p_B(X)})-B_X(p_B(X))=S_X(p_B(X))=S_X(p_A(X)-1)=|Z^-|,
\]
and
\[
\inf\mbox{supp}(Z^+)\geq p_B(X)=p_A(X)-1\geq \sup\mbox{supp}(Z^-),
\]
which shows that conditions (ii)-(iii) are satisfied.

\paragraph{Case 2:} $g_X^+(p_B(x))\leq0$. If $g_X^-(p_A(X))<0$, using the same argument as in Case~1, we can prove that conditions (ii)-(iii) are satisfied. Now suppose $g_X^-(p_A(X))\geq 0$. Then we have
\[
Z^+=\tau^{  p_{B}(X)+1}(X^+),\qquad\mbox{and}\qquad Z^-=\tau_{  p_{A}(X)-1}(X^-).
\]
It is easy to see that condition (iii) is satisfied. For condition (ii), notice that $|Z^+|=B_X(p_B(X)+1)$ and $|Z^-|=S_X(p_A(X)-1)$. Using the same reasoning as in Case~1 when proving $p_A(X)=p_B(X)+1$, we get $p_A(X)>p_B(X)+1$ in this situation. Now if $p_A(X)-1=p_B(X)+1$, since $g_X(p_A(X)-1)=0$ we have $|Z^+|=|Z^-|$. If $p_A(X)-1>p_B(X)+1$, we claim that $X^{\pm}_i=0$ for any $i\in( p_B(X)+1, p_A(X)-1)$. Indeed, w.l.o.g. suppose there exists $i\in( p_B(X)+1, p_A(X)-1)$ such that $X^+_i>0$. As $g_i(X)=0$, we have
\[
g_X(p_A(X)-1)=S_X(p_A(X)-1)-B_X(p_A(X)-1)\geq S_X(i)-B_X(i+1)=g_X(i)+X^+_i>0,
\]
which contradicts the definition of $p_A$. Therefore, for any $i\in( p_B(X)+1, p_A(X)-1)$, $X^{\pm}_i=0$ holds. Consequently, 
\[
|Z^+|=B_X(p_B(X)+1)=B_X(p_A(X)-1)=S_X(p_A(X)-1)=|Z^-|,
\]
implying that condition (ii) is satisfied.
\item[(iv)] This is a natural result from the fact that $X_{cl}+Z=X$, which finishes the proof.
\end{enumerate}

\medskip
\noindent
\textbf{Uniqueness.} Let $\mathcal{C}_1$ and $\mathcal{C}_2$ satisfy conditions (i) to (iv) in Prop.~\ref{prop.ordermatching}. Thus there exists $X\in E$, such that $W^1:=\mathcal{C}^1(X)\neq \mathcal{C}^2(X):=W^2$. We claim that $X\not\in\mathcal{L}$. Indeed, suppose $X\in\mathcal{L}$, then from Def.~\ref{def.clearing_absolute_general}, $W^1=\mathcal{C}^1(X)=X=\mathcal{C}^2(X)=W^2$, which contradicts the assumption that $W^1\neq W^2$. Let $Z^1=X-W^1$ and $Z^2=X-W^2$. Thus $Z^1\neq Z^2$ and both $| Z^{i+}|$ and $| Z^{i-}|$ are non-zero for $i=1, 2$. Consequently, $Z^i\in E_+$ for $i=1, 2$. We further claim that $|Z^{1\pm}|\neq |Z^{2\pm}|$. Indeed, Suppose $|Z^{1+}|=|Z^{2+}|$, then naturally we also have $|Z^{1-}|= |Z^{2-}|$. Since $Z^{i+}+W^{i+}=X^+$ and $\inf\mbox{supp}(Z^{i+})\geq\sup\mbox{supp}(W^{i+})$ when $i=1,2$ letting $\inf\mbox{supp}(Z^{i+})=k_i$ for $i=1,2$, we have
\[
Z^{i+} = \tau^{k_i}(X^+)+Z^{i+}_{k_i}e_{k_i}.
\]
We claim that $k_1=k_2$. If not, w.l.o.g. suppose $k_1<k_2$, then we get
\[
|Z^{1+}|=\sum_{j=k_1+1}^d X^+_j+Z^{1+}_{k_1}\geq\sum_{j=k_1}^d X^+_j>\sum_{j=k_2}^d X^+_j\geq \sum_{j=k_2+1}^d X^+_j+X^+_{k_2}\geq |Z^{2+}|,
\]
which contradicts the claim that $|Z^{1+}|=|Z^{2+}|$ and thus $k_1=k_2$. Now since $|Z^{1+}|=|Z^{2+}|$, we have
\[
|Z^{1+}|=\sum_{j=k_1+1}^d X^+_j+Z^{1+}_{k_1}=|Z^{2+}|=\sum_{j=k_2+1}^d X^+_j+Z^{2+}_{k_2}=\sum_{j=k_1+1}^d X^+_j+Z^{2+}_{k_1},
\]
which implies that $Z^{1+}_{k_1}=Z^{2+}_{k_2}$ and $Z^{1+}=Z^{2+}$. Using the same argument we can also show that $Z^{1-}=Z^{2-}$, which contradicts that that $Z^1\neq Z^2$. Therefore, $|Z^{1\pm}|\neq |Z^{2\pm}|$. 

Now w.l.o.g. suppose $|Z^{1\pm}|> |Z^{2\pm}|$. Using the same reasoning as in the previous proof, it is easy to see that $\inf\mbox{supp}(Z^{1+})<\inf\mbox{supp}(Z^{2+})$ and $\sup\mbox{supp}(Z^{1-})>\sup\mbox{supp}(Z^{2-})$. Letting $j_i:=\sup\mbox{supp}(W^i+)$, we have
\[
W^{1+}+Z^{1+}=\sum_{l=1}^{j_1-1}X^+_le_l+W^{1+}_{j_1}e_{j_1}+\sum_{l=k_1+1}^d X^+_le_l+Z^{1+}_{k_1}e_{k_1}=W^2+\sum_{l=k_2+1}^d X^+_le_l+Z^{2+}_{k_2}e_{k_2},
\]
which implies that
\[
W^{2+}=\sum_{l=1}^{j_1-1}X^+_le_l+W^{1+}_{j_1}e_{j_1}+\sum_{l=k_1+1}^{k_2}X^+_le_l-Z^{2+}e_{k_2}
\] 
and thus $\sup\mbox{supp}(W^2+)\geq\inf\mbox{supp}(Z^1+)$. Similarly, we get $\inf\mbox{supp}(W^2-)\leq\sup\mbox{supp}(Z^1-)$. Consequently, we have
\[
\inf\mbox{supp}(Z^1+)\leq\sup\mbox{supp}(W^2+)<\inf\mbox{supp}(W^2-)\leq\sup\mbox{supp}(Z^1-),
\]
which contradicts the property of the clearing operator that $\inf\mbox{supp}(Z^1+)\geq \sup\mbox{supp}(Z^1-)$. This finishes our proof.

\section{Proof of Proposition \ref{prop.operation_clearing_absolute} (order matching for elementary order flow event)}
\label{proof.2.5}
We start with a useful lemma:
\begin{lemma}
For any $z\in \mathbb{N}\backslash\{  0\}$ and $k\in \{1,\ldots,d  \}$, we have:
\begin{equation}
g_X(k)\leq S_X(k),
\label{eq.operation_clearing_lemma_1}
\end{equation}
\begin{equation}
X_{S_X^{-1}(z)}^->0,
\label{eq.operation_clearing_lemma_2}
\end{equation}
\begin{equation}
\mbox{If   } k\in (S_X^{-1}(z),S_X^{-1}(z+1)),\mbox{ then   } X_k^-=0, \mbox{  i.e.  } S_X(S_X^{-1}(z))=S_X(k),
\label{eq.operation_clearing_lemma_3}
\end{equation}
\begin{equation}
\mbox{If   } S_X^{-1}(z)<S_X^{-1}(z+1), \mbox{  then     }S_X(S_X^{-1}(z))=z.
\label{eq.operation_clearing_lemma_4}
\end{equation}
\label{lemma.operation_clearing}
\end{lemma}

\noindent
\textbf{Proof.}
Eqs.~(\ref{eq.operation_clearing_lemma_1}) and (\ref{eq.operation_clearing_lemma_2}) are just direct results from the definition of $S_X$, $g_X$ and $S_X^{-1}$. For Eq.~(\ref{eq.operation_clearing_lemma_3}), suppose it is not the case. Then there exists $i\in (S_X^{-1}(z),S_X^{-1}(z+1))$ such that $X^-_i>0$, which implies that 
        \[
        S_X(i)=\sum_{j=1}^i X^-_j\geq \sum_{j=1}^{S_X^{-1}(z)}X^-_j+X^-_i=S_X(S_X^{-1}(z))+X^-_i\geq S_X(S_X^{-1}(z))+1\geq z+1.
        \]
Therefore, from the definition of $S_X^{-1}$, $S_X^{-1}(z+1)\leq i$, which contradicts the fact that $S_X^{-1}(z+1)>i$. Eq.~(\ref{eq.operation_clearing_lemma_4}) can be proved using the same argument and we leave it to the reader. \endproof

Now we prove Proposition~\ref{prop.operation_clearing_absolute}:

The expressions of $X'$ and $(p_B(X'),p_A(X'))$ are derived from the expression of the clearing operator $\mathcal{C}$. The proofs for the four cases in Proposition~{\ref{prop.operation_clearing_absolute}} are similar. Therefore, we only give the detailed proof for Case~1, leaving the other cases for the reader. For Case~1, since the new order is a limit buy order with size $z$ at price $k$, the intermediate state $X\in E$ after the arrival of this new order is $X=(X^++ze_k,X^-)$. Therefore, $X'=\mathcal{C}(X)$. From the condition that $X^{\pm}\neq 0$ and $z<\min\{ B_X(1),S_X(d) \}$, we note that $p_A(X)>p_B(X)$. There are four alternatives according to the position of the price $k$ with respect to $p_B(X)$, $p_A(X)$ and $S_X^{-1}(z)$ (if $S_X^{-1}(z)>p_A(X)$).
\begin{itemize}
    \item \textbf{Alternative 1:} $k\leq p_B(X)$. Since $X=(X^++ze_k,X^-)$ then $\sup{\rm supp}(X^+)= p_B(X)<p_A(X)=\inf{\rm supp}(X^-)$. So $X\in \mathcal{L}$. Therefore, from Prop.~\ref{prop.ordermatching}~(i) we have $X'=\mathcal{C}(X)=(X^++ze_k,X^-)$. Moreover, 
    \[
    (p_B(X'),p_A(X'))=(\sup{\rm supp}(X'^+),\inf{\rm supp}(X'^-))=(\sup{\rm supp}(X^+),\inf{\rm supp}(X^-))=(p_B(X),p_A(X)).
    \]
    \item \textbf{Alternative 2:} $p_B(X)<k<p_A(X)$. Using the same argument as in Alternative~1, $\sup{\rm supp}(X^+)= k<p_A(X)=\inf{\rm supp}(X^-)$ and so $X\in\mathcal{L}$ and $X'=X=(X^++ze_k,X^-)$. However, since $p_B(X)<k$, $\sup{\rm supp}(X'^+)=\sup{\rm supp}(X^++ze_k)=k$. Therefore, $(p_B(X'),p_A(X'))=(k,p_A(X'))$.
    \item \textbf{Alternative 3:} $p_A(X)\leq k<S_X^{-1}(z)$. From the definition of $S^{-1}_X$ in Eq.~(\ref{eq.inverse_of_B_X_and_S_X}), we have $p_A(X)=S^{-1}_{X}(1)$. Then since $S_X^{-1}$ is non-decreasing and $z>0$, we get $S^{-1}_{X}(z)\geq p_A(X)$. Also since $S_{X}$ is non-deceasing, $k<S^{-1}_{X}(z)$ implies that $S_X(k)<z$. Then the number of sell orders at prices lower than or equal to $k$ is smaller than the volume $z$ of the new buy order. In this alternative, our first task will be to determine $p_B(X)$ and $p_A(X)$ from Eqs.~(\ref{eq.discrete-clearingrule_buy}) and (\ref{eq.discrete-clearingrule_sell}) respectively. For this, we will need $g_{X}$ from Eq.~(\ref{eq.g_x_absolute_coordinate}) and so $S_{X}$ and $B_{X}$ from Eq.~(\ref{eq.B_X_and_S_X}). From the definition of $B_X$, we have
    \[
    B_{X}(i)=B_{X'}(i)+z\mbox{  if  }i\leq k\quad\mbox{and}\quad
    B_{X'}(i)=B_X(i) \mbox{  if  }i> k.
    \]
    Since $X'^-=X^-$, $S_{X'}(i)=S_{X}(i)$ for $1\leq i\leq d$. Therefore, we get
    \begin{equation}
    g_{X'}(i)=g_X(i)-z\mbox{  if  }i\leq k\quad\mbox{and}\quad
    g_{X'}(i)=g_X(i) \mbox{  if  }i> k.
    \label{eq.operation_clearing_proof_g_tilde_X}
    \end{equation}
    Consequently, 
    \begin{equation}
        g_X(i)-z \leq g_{X'}(i)\leq g_X(i) \quad      \forall i\in\{1,\ldots,d  \}.
        \label{eq.operation_clearing_g_tilde_X}
    \end{equation}
Note that Eqs.~(\ref{eq.operation_clearing_proof_g_tilde_X}) and (\ref{eq.operation_clearing_g_tilde_X}) are not specific to Alternative~3 and are valid for all alternatives. Therefore with Eq.~(\ref{eq.operation_clearing_proof_g_tilde_X}) and the fact that $k\geq p_A(X)>p_B(X)$, we have $B_X(k)=B_X(k+1)=0$. Therefore, we get $g_{X'}(k)=g_{X}(k)-z=S_X(k)-z<0$ and $g_{X'}(k+1)=g_{X}(k+1)=S_X(k+1)>0$, which implies that $p_B(X')=k$ and $p_A(X')=k+1$. For the same reason, $X'^{+}_k=0$ and so $X'^+_k=z$. Also noting that $B_X(k)=0$ and so $g_X(k)>0$ since $k\geq p_A(X)$, we get
    \[
    g_{X'}(p_B(X'))=g_{X}(k)=g_{X}(k)-z>-z=-X'^+_k
    \]
    and 
    \[
    g_{X'}(p_A(X'))=g_{X}(k+1)=g_{X}(k+1)=S_X(k+1)\geq X^-_{k+1}=X'^-_{k+1}.
    \]
    So, this corresponds to the second case of Eq.~(\ref{eq.discrete-clearingrule_buy}) and to the first case of Eq.~(\ref{eq.discrete-clearingrule_sell}). Therefore, we get 
    \begin{equation}
    X'^+=\tau_{  k-1}(X^++ze_k)-g_{X}(k)e_k=X^++(z-S_X(k))e_k,\mbox{  } X'^-=\tau^{  k+1}(X^-)=\tau^{  k+1}(X^-).
    \label{eq.operation_clearing_altenative_3_final_state}
    \end{equation}
    Since $k>p_B(X)$ and $(z-S_X(k))>0$, $p_B(X')=\sup{\rm supp}(X'^+)=k$. Recall that $p_A(X')=S^{-1}_{X'}(1)$. From the definition of $S_{X'}$ and the expression of $X'$, for any $i\in \{1,...,d  \}$, we have:
    \[
    S_{X'}(i)=\max \{ 0,S_{X}(i)-S_X(k)  \}.
    \]
    Consequently, we get 
    \[
    p_A(X')=S_{X'}^{-1}(1)=S^{-1}_{X'}(S_{X}(k)+1).
    \]
    Therefore in this alternative, $X'=(X^++(z-S_X(k))e_k,X^-)$ and $(p_B(X'),p_A(X'))=(k,S^{-1}_X(S_X(k)+1))$.
    \item \textbf{Alternative 4:} $p_A(X)\leq S^{-1}_X(z)\leq k$. Note that this implies that $S_X(k)\geq z$. On the one hand, for $X'^+$, we are going to use Eqs.~(\ref{eq.operation_clearing_proof_g_tilde_X}) and (\ref{eq.operation_clearing_g_tilde_X}) which are valid for all alternatives. Since $p_B(X)\leq p_A(X)-1\leq S_X^{-1}(z)-1<k$, we get from Eqs.~(\ref{eq.operation_clearing_proof_g_tilde_X}), (\ref{eq.operation_clearing_lemma_1}) and the definition of $S_X^{-1}$: $$g_{X'}(S_X^{-1}(z)-1)=g_X(S_X^{-1}(z)-1)-z\leq S_X(S_X^{-1}(z)-1)-z<0,$$ and 
    \[
    g_{X'}(S_X^{-1}(z))=g_X(S_X^{-1}(z))-z\geq 0.
    \]
    From this, we deduce that $p_B(X')=S_X^{-1}(z)-1$. Moreover, since 
    $S_X^{-1}(z)-1<S_X^{-1}(z)\leq k$, $X'^+_{S_X^{-1}(z)-1}=X^+_{S_X^{-1}(z)-1}$. Consequently, we have
    \begin{align*}
        g_{X'}(S_X^{-1}(z)-1)&=g_X(S_X^{-1}(z)-1)-z=(S_X(S_X^{-1}(z)-1)-z)-B_X(S_X^{-1}(z)-1)\\
        &<-B_X(S_X^{-1}(z)-1)\leq -X^+_{S_X^{-1}(z)-1}=-X'^+_{S_X^{-1}(z)-1}
    \end{align*}
    which can be written as $g_{X'}(p_B(X'))<-X'^+_{p_B(X')}$. Hence we are in the second case of Eq.~(\ref{eq.discrete-clearingrule_buy}). With the fact that $k>S_X^{-1}(z)-1$, we have $$X'^+=\tau_{  S_X^{-1}(z)-1}(X^+)=\tau_{  S_X^{-1}(z)-1}(X^++ze_k)=X^+,$$ and so $p_B(X')=p_B(X).$
    
    On the other hand, from the definition of $g_{X'}$, we have
    \[
    g_{{X'}}(S_{X}^{-1}(z+1))\geq S_X(S_X^{-1}(z+1))-z>0,\qquad g_{X'}(S_{X}^{-1}(z+1)-1)<z+1-z=1,
    \]
    which implies that $p_A(X')=S_{X}^{-1}(z+1)$. Next we derive the expression for $X'^-$. If $S_{X}^{-1}(z)=S_X^{-1}(z+1)$, then $g_{X'}(p_A(X'))-X'^-_{p_A(X')}=S_X(S_X^{-1}(z)-1)-z<0$. This applies to the second line of Eq.~(\ref{eq.clearing_criterion_sell}) and thus we get
    \[
    X'^-=\tau^{  S_X^{-1}+1}(X^-)+(S_X(S_X^{-1}(z))-z)e_{S_X^{-1}(z)}.
    \]
    Otherwise, if $S_{X}^{-1}(z)<S_X^{-1}(z+1)$, then we get
    \[
    g_{X'}(p_A(X'))-X'^-_{p_A(X')}\geq S_X(p_A(X'-)-1)-z=S_X(S_X^{-1}(z+1)-1)-z\geq S_X(S_X^{-1}(z))-z=0,
    \]
    which applies to the first line of Eq.~(\ref{eq.clearing_criterion_sell}) and thus 
    \[
    X'^-=\tau^{  S_X^{-1}(z+1)}(X^-).
    \]
    With Eq.~(\ref{eq.operation_clearing_lemma_4}), we have $S_X(S_X^{-1}(z))-z=0$. Consequently, $X'^-$ can also be written as follows:
    \[
    X'^-=\tau^{  S_X^{-1}+1}(X^-)+(S_X(S_X^{-1}(z))-z)e_{S_X^{-1}(z)}.
    \]
    Again from the definition of $S_{X'}$ and the expression of $X'$, for any $i\in\{1,...,d  \}$, we have $S_{X'}(i)=\max \{ 0,S_X(i)-z \}$. This implies that $p_A(X')=S_{X'}^{-1}(1)=S_X^{-1}(z+1)$. To sum up, if $k\geq S_X^{-1}(z)$, then we have
    \[
    X'=(X^+,\tau^{  S_X^{-1}(z)+1}({X}^-)+(S_X(S_X^{-1}(z))-z)e_{S_X^{-1}(z)}),\quad \mbox{and}\quad (p_B(X'),p_A(X'))=(p_B(X),S_X^{-1}(z+1)).
    \]
    
\end{itemize}
Using the same arguments, the transitions from $X$ to $X'$ and from $(P_B(X),p_A(X))$ to $(P_B(X'),p_A(X'))$ can be derived when the new order is a limit sell order, a cancellation of a buy order or a cancellation of a sell order, which finishes our proof. \endproof

\section{Proof of Proposition \ref{prop.pre_image_of_clearing_operator_absolute} (Determination of $\mathcal{C}^{-1}(\{X\})$)}
\label{proof.preimage}

For $X\in \mathcal{L}$, denote by $\Gamma(X)$ the right hand side of \eqref{eq.pre_image_set_expression}. We first show the following lemma:
\begin{lemma}
For any $X$, $X'\in\mathcal{L}$, $X\neq X' \Longrightarrow \Gamma(X)\cap \Gamma(X')=\emptyset$. Moreover, for  $Y\in E$, we have $Y\in\Gamma(\mathcal{C}(Y))$ and so $\cup_{X\in\mathcal{L}} \Gamma(X)=E$.
\label{lemma.pre_image_set_expression}
\end{lemma}

\noindent
\textbf{Proof.}
We first show that for any $Y\in E$, we have $Y\in\Gamma(\mathcal{C}(Y))$ and $\cup_{X\in\mathcal{L}} \Gamma(X)=E$. Indeed, for any $Y\in E$, $Z:=Y-\mathcal{C}(Y)$ satisfies conditions (i) to (iv) in Def.~\ref{def.order-matching-clearing-operator-absolute} (with $X$ replaced by $Y$). Noting that $\mathcal{C}(Y)\in\mathcal{L}$ and using Eq.~(\ref{eq.ask_price_and_bid_price_by_g_x}), we have $\inf{\rm supp}(\mathcal{C}(Y))^-=p_A(X)$ and $\sup{\rm supp}(\mathcal{C}(Y))^+=p_B(X)$. 
Therefore, conditions (i) to (iv) in Def.~\ref{def.order-matching-clearing-operator-absolute} are indeed the conditions for $Y$ to belong to $\Gamma(\mathcal{C}(Y))$ as defined in Eq.~(\ref{eq.pre_image_set_expression}) and so $Y\in\Gamma(\mathcal{C}(Y))$. Consequently, $Y\in \cup_{X\in\mathcal{L}} \Gamma(X)$ and so $\cup_{X\in\mathcal{L}} \Gamma(X)=E$.

 Now we show for any $X$, $X'\in\mathcal{L}$, $X\neq X' \Longrightarrow \Gamma(X)\cap \Gamma(X')=\emptyset$. Suppose it is not the case. Then there exist $X^1$ and $X^2\in\mathcal{L}$ with $X^1\neq X^2$ such that $\Gamma(X^1)\cap\Gamma(X^2)\neq\emptyset$. Let $Y\in\Gamma(X^1)\cap\Gamma(X^2)$. Since $\mathcal{C}(Y)\in\mathcal{L}$ and $X^1\neq X^2$, $\mathcal{C}(Y)$ must be different from at least one of $X^1$ or $X^2$. Without loss of generality, suppose that $\mathcal{C}(Y)\neq X^1$. 
 If $Y\in\mathcal{L}$ then $Y=\mathcal{C}(Y)\in\Gamma(X^1)$. Let $Z:=Y-X^1=\mathcal{C}(Y)-X^1\neq 0$. Since $Y\in\Gamma(X^1)$, then we have $\sup{\rm supp}(Z^-)\leq p_A(X^1)$ so
\begin{equation}
a(Y)=\inf{\rm supp}(X^{1-}+Z^-)\leq \inf{\rm supp}(Z^-)\leq \sup{\rm supp}(Z^-).
\label{eq.pre_image_lemma_case_1}
\end{equation}
Similarly, since $\inf{\rm supp}(Z^+)\leq p_B(X^1)$, we have 
\[
b(Y)=\sup{\rm supp}(X^{1+}+Z^+)\geq \sup{\rm supp}(Z^+)\geq \inf{\rm supp}(Z^+).
\]
This, together with Eq.~(\ref{eq.pre_image_lemma_case_1}) and the fact that $\inf{\rm supp}(Z^+)\geq \sup{\rm supp}(Z^-)$ from Eq.~(\ref{eq.pre_image_set_expression}) implies that 
\[
b(Y)\geq \inf{\rm supp}(Z^+) \geq \sup{\rm supp}(Z^-) \geq a(Y),
\]
which contradicts   $Y\in\mathcal{L}$.   Therefore, $Y\not\in\mathcal{L}$.
Define $\mathcal{B}:E\rightarrow \mathcal{L}$ by
\begin{equation}
\mathcal{B}(X)=
\begin{cases}
\mathcal{C}(X) & \mbox{  if   }X\neq Y,\\
X^1 & \mbox{   if   } X=Y.
\end{cases}
\end{equation}
Then $\mathcal{B}$ satisfies Def.~\ref{def.clearing_absolute_general}  and   conditions (i)-(iv) in Def.~\ref{def.order-matching-clearing-operator-absolute}:
\begin{itemize}
\item Suppose $\mathcal{B}(X)=X$. Then by the definition of $\mathcal{B}$, $X\in\mathcal{L}$. Conversely, for any $X\in\mathcal{L}$, since $Y\not\in\mathcal{L}$, $Y\neq X$. Therefore $\mathcal{B}(X)=\mathcal{C}(X)=X$, which shows that   Definition~\ref{def.clearing_absolute_general} is satisfied.
\item \textbf{Conditions (i)-(iv):} Let $W\in E$. First suppose that $W\neq Y$, then $\mathcal{B}(W)=\mathcal{C}(W)$. Therefore, $Z'=W-\mathcal{B}(W)$ satisfies the conditions (i)-(iv) in Def.~\ref{def.order-matching-clearing-operator-absolute} (with $X$ replaced by $W$). Now suppose that $W=Y$. Let $Z:=Y-X^1$. Since $Y\in\Gamma(X^1)$, we have $Z\in E$, $|Z^+|=|Z^-|$, $\sup{\rm supp}(Z^-)\leq \inf{\rm supp}(Z^+)$, $\sup{\rm supp}(Z^-)\leq a(X^1)$ and $\inf{\rm supp}(Z^+)\geq b(X^1)$. This, together with the fact that $\mathcal{B}(Y)=X^1$, implies that $Z$ satisfies the conditions (i)-(iv) in Def.~\ref{def.order-matching-clearing-operator-absolute} (with $X$ replaced by $Y$ and $Z(X)$ replaced by $Y-X^1$). 
\end{itemize}
Therefore $\mathcal{B}$ is a clearing operator different from $\mathcal{C}$ defined by Eqs.~(\ref{eq.discrete-clearingrule_buy}) and (\ref{eq.discrete-clearingrule_sell}), which contradicts the uniqueness of the clearing operator as stated in Prop.~\ref{prop.ordermatching} and finishes the proof. \endproof

\paragraph{Proof of Proposition \ref{prop.pre_image_of_clearing_operator_absolute}}
 
We first show that $\mathcal{C}^{-1}(\{X\})\subset \Gamma(X)$. For any $Y\in\mathcal{C}^{-1}(\{\mathcal{C}(Y)\})=\mathcal{C}^{-1}(\{X\})$, $\mathcal{C}(Y)=X$. Therefore, from Lemma~\ref{lemma.pre_image_set_expression} we have $Y\in\Gamma(X)$. Consequently, $\mathcal{C}^{-1}(\{X\})\subset \Gamma(X)$.

Conversely, suppose $Y\in \Gamma(X)$. Showing that $Y\in\mathcal{C}^{-1}(\{X\})$ is equivalent to showing that $\mathcal{C}(Y)=X$. Suppose it is not the case. Then $\mathcal{C}(Y)=X'\neq X$. Therefore, $Y\in \mathcal{C}^{-1}(\{X'\})\subset\Gamma(X')$. And from the first part of the proof, we have that $\mathcal{C}^{-1}(\{X'\})\subset\Gamma(X')$. So $Y\in\mathcal{C}^{-1}(\{X'\})$, which implies that $\Gamma(X)\cap\Gamma(X')\neq\emptyset$ and contradicts Lemma~\ref{lemma.pre_image_set_expression}. Therefore $Y\in\mathcal{C}^{-1}(\{X\})$. \endproof

\section{Proof of Lemma~\ref{lemma.shifting_operator_image} (Image of the centering operator $J$)}\label{proof.lemma.shifting_operator_image}
Since the centering operator involves deleting volume at the boundary, we need to first check that the image is still in $E^m$ i.e. the bid and ask sides are non-empty:
\begin{lemma}
For any $(Y,p)\in\mathcal{L}^0\times\mathbb{Z}$, $J(Y,p)\in \mathcal{L}^m$.
\label{lemma.stability_of_state_space_under_shifting_operator}
\end{lemma}

\noindent
\textbf{Proof.}
By Eq.~(\ref{eq.expression_shifting_operator}), we have
\[
\begin{cases}
p'=p+\Delta p \quad\mbox{with} \quad\Delta p=a(Y)+b(Y)+\hat{p},\\
X=\sigma_{[p,p']}(Y).
\end{cases}
\]
We now prove that $X^+\neq 0$, and the same argument can be used to prove that $X^-\neq 0$. 
\begin{enumerate}
\item[Case 1]: If $[p,p']=0$, then $X^+=Y^+$. Using the fact that $Y\in\mathcal{L}^0$, $Y^+\neq 0$. Therefore, $X^+\neq 0$. 
\item[Case 2]: If $[p,p']<0$, then together with the definition of operator $\sigma$, we have
\begin{equation}
\begin{cases}
X^+_j=0 &\mbox{  if  } j<-d'-([p,p']),\\
X^+_j=Y^{\pm}_{j+[p,p']} &\mbox{  if  } j\geq -d'-([p,p']).
\end{cases}
\label{eq.lemma_shifting_non_empty_case_2}
\end{equation}
From the second line of Eq.~(\ref{eq.lemma_shifting_non_empty_case_2}), we note that $X^+_j\neq 0$ only when $j\geq -d'-([p,p'])$, which corresponds to $Y^+_{j+[p,p']}$ when $j+[p,p']\leq d'+[p,p']$. Therefore, showing that $X^+\neq 0$ is equivalent to showing that 
\begin{equation}
\inf{\rm supp}(Y^+)\leq d'+[p,p'].
\label{eq.lemma_shifting_non_empty_case_2_proof}
\end{equation}
Since 
\begin{equation}
[p,p']=\ceil{\frac{p+a(Y)+b(Y)+\hat{p}}{2}}-\ceil{\frac{p}{2}}=\ceil{\frac{a(Y)+b(Y)}{2}},
\label{eq.p_and_p_prime}
\end{equation}
and letting $k_1:=b(Y)$ and $k_2:=a(Y)$, we get 
\[
\inf{\rm supp}(Y^+)-([p,p'])=\inf{\rm supp}(Y^+)-\ceil{\frac{k_1+k_2}{2}}.
\]
Since $\inf{\rm supp}(Y^+)\leq k_1<k_2$, we have
\[
\inf{\rm supp}(Y^+)-\ceil{\frac{k_1+k_2}{2}}<0<d',
\]
from which we get that $X^+\neq 0$.
\item[Case 3]: If $[p,p']>0$, then together with the definition of operator $\sigma$, we have
\begin{equation}
\begin{cases}
X^+_j=0 &\mbox{  if  } j>d'-([p,p']),\\
X^+_j=Y^{\pm}_{j+[p,p']} &\mbox{  if  } j\leq d'-([p,p']).
\end{cases}
\label{eq.lemma_shifting_non_empty_case_3}
\end{equation}
Using the same argument as in Case 2, we are left to show that 
\begin{equation}
b(Y)\geq -d'+[p,p'].
\end{equation}
Letting $k_1:=b(Y)$ and $k_2:=a(Y)$, then together with Eq.~(\ref{eq.p_and_p_prime}) we can re-write the above inequality as:
\begin{equation}
k_1\geq -d'+\ceil{\frac{k_1+k_2}{2}}.
\label{eq.lemma_shifting_non_empty_case_3_proof}
\end{equation}
If $k_1=-d'$ and $k_2=d'$, then $k_1-\ceil{\frac{k_1+k_2}{2}}=-d'$ and Eq.~(\ref{eq.lemma_shifting_non_empty_case_3_proof}) is satisfied. Else if $k_2-k_1\leq 2d'-1$, then $$2(k_1-\ceil{\frac{k_1+k_2}{2}})=2k_1-(k_1+k_2)-\hat{(k_1+k_2)}=(k_1-k_2)-\hat{(k_1+k_2)}.$$ Since $\hat{(k_1+k_2)}\leq 1$, we get $2(k_1-\ceil{\frac{k_1+k_2}{2}})\geq -(2d'-1)-1=-2d'$. Therefore, Eq.~(\ref{eq.lemma_shifting_non_empty_case_3_proof}) is satisfied and hence $X^+\neq 0$.
\end{enumerate}
We conclude that $X^+\neq 0$ and $X^-\neq 0$ in these three cases, which finishes the proof.\endproof

The following lemma shows two properties of the centering operator $J$ which will be used to prove Lemma~\ref{lemma.shifting_operator_image}:
\begin{lemma}\label{lemma.shifting_operator_property}
For any $(Y,p)\in E^m\times\mathbb{Z}$, letting $(X,p'):=J(Y,p)$, then
\begin{itemize}
\item[(i).] $\inf\mbox{supp}(X^-)=\inf\mbox{supp}(Y^-)-[p,p']$, and $\sup\mbox{supp}(X^+)=\sup\mbox{supp}(Y^+)-[p,p']$.
\item[(ii).] $\inf\mbox{supp}(X^-)+\sup\mbox{supp}(X^-)+2\ceil{\frac{p'}{2}}=\inf\mbox{supp}(Y^-)+\sup\mbox{supp}(Y^+)+2\ceil{\frac{p}{2}}$.
\end{itemize}
\end{lemma}

\noindent
\textbf{Proof.}
These two properties of $J$ can be derived directly from its definition and Lemma~\ref{lemma.stability_of_state_space_under_shifting_operator}. \endproof

\medskip
\noindent
{\em Proof of Lemma~~\ref{lemma.shifting_operator_image}.} For any $(\hat{W},p)\in\mathcal{L}^0\times\mathbb{Z}$, let $(Z,p'):=J(\hat{W},p)$. We now prove that $(Z,p')\in\mathcal{L}^m$. Notice from Lemma~\ref{lemma.shifting_operator_property} (i) that the relative order of  $a(Z)$ and $b(Z)$ is the same as that of $\inf{\rm supp}(\hat{W}^-)$ and $\sup{\rm supp}(\hat{W}^+)$, which shows that the first condition of  Eq.~(\ref{eq.state_space_centred_order_book}) is satisfied. Then we are left to prove the second condition of Eq.~(\ref{eq.state_space_centred_order_book}), i.e. $a(Z)+b(Z)+\hat{p'}=0$. From Lemma~\ref{lemma.shifting_operator_property} (ii), we have
\[
a(Z)+b(Z)=\inf{\rm supp}(\hat{W}^-)+\sup{\rm supp}(\hat{W}^+)-2([p,p']).
\]
Notice that for any $a\in\mathbb{Z}$, $2\ceil{\frac{a}{2}}=a+\hat{a}$. Therefore, by the definition in Eq.~(\ref{eq.expression_shifting_operator}) of $\Delta p$, we have
\begin{align*}
&a(Z)+b(Z)=\inf{\rm supp}(\hat{W}^-)+\sup{\rm supp}(\hat{W}^+)-[p'+\hat{p'}-p-\hat{p}]\\
&=(\inf{\rm supp}(\hat{W}^-)+\sup{\rm supp}(\hat{W}^+)+\hat{p})-\Delta p-\hat{p'}=-\hat{p'},
\end{align*}
so that
\[
a(Z)+b(Z)+\hat{p}'=0,
\]
which shows that the second condition in Eq.~(\ref{eq.state_space_centred_order_book}) is also satisfied and finishes the proof. \endproof

\section{Adjoint operator and forward Kolmogorov equation}
\label{sec.adjoint_of_the_generator_of_the_order_book_centred}

To describe the adjoint operator $\emph{L}^{m*}$ of $\emph{L}^m$, we first describe the pre-image of a subset $A\subset\mathcal{L}^m$ by the centred clearing operator $\mathcal{C}^m$. Define the following sets:

\begin{definition}
For any $(X,p)\in\mathcal{L}^m$, define 
\begin{equation}
\begin{aligned}
\Lambda(X,p):=&\Bigg\{(Y,p'),\mbox{  with $p'$ s.t.   }\sup{\rm supp}(X^-)-d'\leq [p,p']\leq d'+\inf{\rm supp}(X^+),\\
& \mbox{  and $Y$ such that   }\exists (a_j)_{j=d'-[p,p']+1}^{d'}\in [0,m]^{[p,p']},  (b_j)_{j=-d'}^{-d'-[p,p']-1}\in [0,m]^{-[p,p']},\\
&\mbox{  with   }
Y^+=\sigma_{[p,p']}(X)^++\sum_{j=-d'}^{-d'-[p,p']-1}b_j\hat{e}_j \mbox{  and  }
Y^-=\sigma_{[p,p']}(X)^-+\sum_{j=d'-[p,p']+1}^{d'}a_j\hat{e}_j
\Bigg\}.
\end{aligned}
\label{eq.pre-image_shifting_operator}
\end{equation}
Moreover, for any $X\in\mathcal{L}^0$, define $\Gamma^m(X)$ as follows:
\begin{equation}
\begin{aligned}
\Gamma^m(X)=&\{Y\in E^m, \mbox{ s.t. } Y-X=Z\in\mathbb{N}^{2d'+1}\times\mathbb{N}^{2d'+1} , \quad|Z^+|=|Z^-|,\\
&\sup{\rm supp}(Z^-)\leq \inf{\rm supp}(Z^+),\quad\sup{\rm supp}(Z^-)\leq a(X), \quad\inf{\rm supp}(Z^+)\geq b(X)   \}.
\end{aligned}
\label{eq.pre_image_order_matching_clearing_centred}
\end{equation}
\end{definition}

In Eq.~(\ref{eq.pre-image_shifting_operator}) we assume that $[0,m]^k=\emptyset$ if $k\leq 0$, meaning that if $[p,p']\leq 0$, the set of coefficients $a_j$ is empty, and similarly for $b_j$ if $[p,p']\geq 0$. $\Lambda(X,p)$ denotes the pre-image of the centring operator $J$. In other words, it is the set of $(Y,p')\in E^m\times\mathbb{Z}$ that $(X,p)$ can shift back. Therefore the distance of shifting must be controlled, which is described the first line in the definition, the value at positions within the current coordinate must be the same as in $X$, compromising the second line in the definition, and it is free to choose any value for positions outside the current coordinate, as shown in the third line. $\Gamma^m$ denotes the pre-image of the operator related with clearing $C'$. It is easy to see it has a very similar expression to the pre-image of $\mathcal{C}$ the market clearing operator in the fixed coordinate. The next lemma gives the precise expression for the pre-image of $\mathcal{C}^m$:

\begin{lemma}
For any $(X,p)\in\mathcal{L}^m$, we have
\begin{equation}
(\mathcal{C}^m)^{-1}(\{ (X,p) \})=\bigcup_{(Y,p')\in\Lambda(X,p)}\bigcup_{Z\in\Gamma^m(Y)} \{(Z,p')\}.
\end{equation}
\label{prop.expression_pre_image_clearing_centred}
\end{lemma}

\begin{remark}
Notice that since $J$ is not injective, the set $\Lambda(X,p)$ in Eq.~(\ref{eq.pre-image_shifting_operator}) may contain other elements besides $(X,p)$. Therefore, the pre-image of $\{(X,p)\}$ by $\mathcal{C}^m$ contains more elements than the pre-image of $\{X\}\subset\mathcal{L}$ by $\mathcal{C}$ in the fixed coordinate framework.
\end{remark}

To prove this lemma, we use the result of the following two lemmas. From Eq.~(\ref{eq.decomposition_clearing_operator_centred}) it is easy to see that for any $A\subset\mathcal{L}^m$, 
\[
(\mathcal{C}^{m})^{-1}(A)=\{(Z,p')\in E^m\times\mathbb{Z}:Z\in  (\mathcal{C}')^{-1}(\{Y\}),(Y,p')\in (J)^{-1}(A)\}. 
\]
We first show that $J^{-1}(\{(X,p)\})=\Lambda(X,p)$ for any $(X,p)\in\mathcal{L}^m$.

\begin{lemma}
$\Lambda(X,p)\subset\mathcal{L}^0\times\mathbb{Z}$. Moreover, for any $(Y,p')\in\Lambda(X,p)$, we have $b(Y)=\sup{\rm supp}(\sigma_{[p,p']}(X)^+)$ and $a(Y)=\inf{\rm supp}(\sigma_{[p,p']}(X)^-)$.
\label{lemma.pre-image_shifting_operator_property}
\end{lemma}

\noindent
\textbf{Proof.}
Let $(Y,p')\in \Lambda(X,p)$. From (i) in Lemma~\ref{lemma.shifting_operator_property}, we get $\sup{\rm supp}(\sigma_{[p,p']}(X)^+)=b(X)-[p,p']$. Since $b(X)\geq -d'$, we 
\[
\sup{\rm supp}\bigg(\sum_{j=-d'}^{-d'-[p,p']-1}b_j\hat{e}_j\bigg)\leq -d'-[p,p']-1<-d'-[p,p']\leq b(X)-[p,p']=\sup{\rm supp}(\sigma_{[p,p']}(X)^+),
\]
which implies that $b(Y)=\sup{\rm supp}(\sigma_{[p,p']}(X)^+)$. Using the same argument, we derive that $a(Y)=\inf{\rm supp}(\sigma_{[p,p']}(X)^-)$. Since $(X,p)\in \mathcal{L}^m$, $a(X)>b(X)$. Consequently, again by Lemma~\ref{lemma.shifting_operator_property} (i), we get $$a(Y)=a(X)-[p,p']>b(X)-[p,p']=b(Y),$$ which implies that $(Y,p')\in \mathcal{L}^0\times\mathbb{Z}$ and finishes the proof.\endproof

The next lemma shows that $\Lambda(X,p)$ determines the pre-image of $\{(X,p)  \}$ by $J$ for any $(X,p)\in\mathcal{L}^m$:

\begin{lemma}
For any $(X,p)\in\mathcal{L}^m$, we have
\begin{equation}
(J)^{-1}(\{(X,p)\})=\Lambda(X,p).
\end{equation} 
\label{lemma.pre-image_shifting_operator_expression}
\end{lemma}

\noindent
\textbf{Proof.}
We first prove that $\Lambda(X,p)\subset (J)^{-1}(\{(X,p)\})$. For any $(Y,p')\in\Lambda(X,p) $, let $(Z,q):=J(Y,p')$. We want to show that $(Z,q)=(X,p)$. We first prove that $q=p$. Suppose $[p,p']\geq 0$ (the same argument can be used to prove that $(Z,q)=(X,p)$ when $[p,p']<0$). From Eq.~(\ref{eq.pre-image_shifting_operator}) we obtain that 
\begin{equation}
Y^+=\sigma_{[p,p']}(X)^+\mbox{   and   }Y^-=\sigma_{[p,p']}(X)^-+\sum_{j=d'-[p,p']+1}^{d'}a_j\hat{e}_j.
\label{eq.pre_image_shifting_operator_first_direction_expression_Y}
\end{equation}
From Lemma~\ref{lemma.pre-image_shifting_operator_property} we have $a(Y)=a(X)-[p,p']$ and $b(Y)=b(X)-[p,p']$. This together with the definition of $J$ implies that 
\[
\Delta p'=q-p'=a(Y)+b(Y)+\hat{p'}.
\]
Consequently, we get
\[
\begin{aligned}
\Delta p'&=a(X)+b(X)+\hat{p'}-2[p,p']=a(X)+b(X)+\hat{p'}-(p'+\hat{p'}-p-\hat{p})\\
&=[a(X)+b(X)+\hat{p}]+(p-p')=p-p',
\end{aligned}
\]
where the last equality is due to the fact that $(X,p)\in\mathcal{L}^m$. Therefore, $$q=p'+\Delta p'=p'+(p-p')=p.$$ 

We are left to prove that $Z=X$. From the definition of $J$ and the fact that $[p,p']=\ceil{\frac{p'}{2}}-\ceil{\frac{p}{2}} =-[p',p]$, $Z=\sigma_{-[p,p']}(Y)$, which together with the fact that $[p,p']\geq 0$ implies that
\begin{equation}
\begin{cases}
Z^+_j=0 &\mbox{  if  } j<-d'+[p,p'],\\
Z^+_j=Y^{+}_{j-[p,p']} &\mbox{  if  } j\geq -d'+[p,p'].
\end{cases}
\label{eq.lemma_pre_image_shifting_1}
\end{equation}
We also have, because of Eqs.~(\ref{eq.expression_sigma}) and (\ref{eq.pre_image_shifting_operator_first_direction_expression_Y}):
\begin{equation}
\begin{cases}
Y^+_j=0 &\mbox{  if  } j>d'-[p,p'],\\
Y^+_j=X^{+}_{j+[p,p']} &\mbox{  if  } j\leq d'-[p,p'].
\end{cases}
\label{eq.lemma_pre_image_shifting_2}
\end{equation}
Introducing $j':=j-[p,p']\leq d'-[p,p']$ and using the second case of Eq.~(\ref{eq.lemma_pre_image_shifting_2}), we get for $j\geq -d'+[p,p']$: $Z^+_j=Y^{+}_{j-[p,p']}=Y^+_{j'}=X^+_{j'+[p,p']}=X^+_j$. From the definition of the set $\Lambda(X,p)$, $\inf{\rm supp}(X^+)\geq [p,p']-d'$. Note that from Eq.~(\ref{eq.lemma_pre_image_shifting_1}) we also have $\inf{\rm supp}(Z^+)\geq [p,p']-d'$. So $X^+_j=Z^+_j=0$ for $j<[p,p']-d'$. Consequently, we deduce that $X^+=Z^+$. The same argument can be used to prove $X^-=Z^-$. Therefore, $X=Z$.

We then prove that $(J)^{-1}(\{(X,p)\})\subset  \Lambda(X,p)$. For any $(Y,p')\in (J)^{-1}(\{(X,p)\})$, recalling that $[p,p']=-[p',p]$ and using Eq.~(\ref{eq.expression_shifting_operator}) to compute $J(Y,p')=(X,p)$, we have
\[
\begin{cases}
p=p'+\Delta p' \quad\mbox{with} \quad\Delta p'=a(Y)+b(Y)+\hat{p'},\\
X=\sigma_{[p',p]}(Y)=\sigma_{-[p,p']}(Y).
\end{cases}
\]
Suppose $[p,p']>0$ (the same argument can be applied to the case when $[p,p']\leq 0$). Since $X=\sigma_{-[p,p']}(Y)$, from Eq.~(\ref{eq.expression_sigma}) we get
\begin{equation}
\begin{cases}
X^{\pm}_j=0 &\mbox{  if  } j<-d'+[p,p'],\\
X^{\pm}_j=Y^{\pm}_{j-[p,p']} &\mbox{  if  } j\geq -d'+[p,p'].
\end{cases}
\label{eq.lemma_pre_image_shifting_3}
\end{equation}
We first prove that the condition for $p'$ in the definition of $\Lambda$ is satisfied. On the one hand, from the first case of Eq.~(\ref{eq.lemma_pre_image_shifting_3}) we get $\inf{\rm supp}(X^+)\geq -d'+[p,p']$, which implies that $0<[p,p']\leq d'+\inf{\rm supp}(X^+)$. On the other, since $b(X)\leq d'$, the condition that $b(X)-d'\leq [p,p']$ is trivially satisfied. Consequently, the condition involved for $p'$ in Eq.~(\ref{eq.pre-image_shifting_operator}) is satisfied. 

We now prove that $Y$ also satisfies the condition involved in the definition of the set $\Lambda(X,p)$. Let $j':=j-[p,p']\leq d'-[p,p']$. From the second line of Eq.~(\ref{eq.lemma_pre_image_shifting_3}) we have $Y^{\pm}_{j'}=X^{\pm}_{j'+[p,p']}$ when $j'\leq d'-[p,p']$. This implies that 
\[
Y^{\pm}=\sum_{j'=-d'}^{d'}Y^{\pm}_{j'}\hat{e}_{j'}=\sum_{j'=-d'}^{d'-[p,p']}X^{\pm}_{j'+[p,p']}\hat{e}_{j'+[p,p']}+\sum_{j'=d'-[p,p']+1}^{d'}Y^{\pm}_{j'}\hat{e}_{j'}.
\]
From Eq.~(\ref{eq.expression_sigma}) we note that $\sum_{j'=-d'}^{d'-[p,p']}X^{\pm}_{j'+[p,p']}\hat{e}_{j'+[p,p']}=\sigma_{[p,p']}(X)^{\pm}$ when $[p,p']>0$. Therefore, 
\[
Y^{\pm}=\sigma_{[p,p']}(X)^{\pm}+\sum_{j'=d'-[p,p']+1}^{d'}Y^{\pm}_{j'}\hat{e}_{j'}.
\]
Moreover, from Lemma~\ref{lemma.shifting_operator_property} we get $b(Y)=b(X)-[p,p']\leq d'-[p,p']$, which implies that $Y^-_{j'}=0$ when $j'>d'-[p,p']$. Consequently 
\[
Y^{+}=\sigma_{[p,p']}(X)^+=\sigma_{[p,p']}(X)^++\sum_{j'=-d'}^{-d'-[p,p']-1}Y^{+}_{j'}\hat{e}_{j'}.
\]
Therefore $(Y,p')$ satisfies Eq.~(\ref{eq.pre-image_shifting_operator}), which finishes the proof. \endproof

Now we deduce the set $(\mathcal{C}')^{-1}(\{X \})$ for any $X\in\mathcal{L}^0$. By a proof similar to that of Proposition~\ref{prop.pre_image_of_clearing_operator_absolute}, we can show that $\Gamma^m(X)$ in (\ref{eq.pre_image_order_matching_clearing_centred}) is the pre-image of $\{X\}$ by the order-matching operator $\mathcal{C}'$. More precisely, we state that:

\begin{lemma}
For any $X\in\mathcal{L}^0$, the set $(\mathcal{C}')^{-1}(\{ X \})$ is given by:
\begin{equation}
(\mathcal{C}')^{-1}(\{ X \})=\Gamma^m(X),
\end{equation}
where $\Gamma^m(X)$ is defined in Eq.~(\ref{eq.pre_image_order_matching_clearing_centred}).
\label{lemma.pre_image_order_matching_clearing_centred}
\end{lemma}

Now we prove Lemma~\ref{prop.expression_pre_image_clearing_centred}.

\noindent
\textbf{Proof.}
Lemma~\ref{prop.expression_pre_image_clearing_centred} is a natural result of Lemmas~\ref{lemma.pre-image_shifting_operator_property}-\ref{lemma.pre_image_order_matching_clearing_centred}. \endproof
 
Now, as mentioned earlier, the adjoint operator of $\emph{L}^m$ exists and can be described as follows. Let $\emph{L}^{m*}$ be the adjoint of $\emph{L}^m$, the generator of the centred order book. Using Eq.~(\ref{eq.decomposition_generator_centred_order_book}), $\emph{L}^{m*}$ can be written as the composition of the adjoint operators $\Xi^{m*}$, $\emph{L}_o^{m*}$ and $\tilde{\mathcal{C}}^{m*}$. Let $\mathcal{P}(E^m\times\mathbb{Z})$ and $\mathcal{P}(\mathcal{L}^m)$ be the set of probability measures on $E^m\times\mathbb{Z}$ and on $\mathcal{L}^m$. Then $\Xi^{m*}:\mathcal{P}(\mathcal{L}^m)\rightarrow\mathcal{P}(E^m\times\mathbb{Z})$, the adjoint operator of $\Xi^m$, is such that for any $\nu\in\mathcal{P}(\mathcal{L}^m)$ and any measurable subset $A$ of $E^m\times\mathbb{Z}$: 
\begin{equation}
\Xi^{m*}\nu(A):=\nu(A\cap\mathcal{L}^m);
\end{equation}
$\tilde{\mathcal{C}}^{m*}:\mathcal{P}(E^m\times\mathbb{Z})\rightarrow\mathcal{P}(\mathcal{L}^m)$, the adjoint operator for $\tilde{\mathcal{C}}^m$, is such that for any $\mu\in \mathcal{P}(E^m\times\mathbb{Z})$, and any measurable subset $A$ of $\mathcal{L}^m$:
\begin{equation}
\tilde{\mathcal{C}}^{m*}\mu(A):=\mu((\mathcal{C}^m)^{-1}(A)) =\sum_{(Z,p)\in A}\mu((\mathcal{C}^m)^{-1}(\{ (Z,p) \}));
\end{equation}
$\emph{L}_o^{m*}:\mathcal{P}(E^m\times\mathbb{Z})\rightarrow \mathcal{P}(E^m\times\mathbb{Z})$, the adjoint operator of $\emph{L}^m_o$, is such that for any measure $\mu\in\mathcal{P}(E^m\times\mathbb{Z})$ and any measurable subset $A$ of $E^m\times\mathbb{Z}$,
\begin{equation}
\label{eq.expression_adjoint_free_order_flow_centred}
\emph{L}_o^{m*}\mu(A)=\sum_{(X,p)\in A}\sum_{(Y,p')\in E^m\times\mathbb{Z}}[p_o^m((Y,p'),(X,p))\mu(\{(Y,p')\})-p_o^m((X,p),(Y,p'))\mu(\{(X,p))\})].
\end{equation}
The adjoint operator of $\emph{L}^m$ can thus be defined via the composition of the three operators $\Xi^*$, $\emph{L}_o^{m*}$ and $\tilde{\mathcal{C}}^{m*}$:
\begin{proposition}
We have
\begin{equation}
\emph{L}^{m*}=\tilde{\mathcal{C}}^{m*}\emph{L}_o^{m*}\Xi^{m*}.
\end{equation}
Moreover, $\emph{L}^{m*}$ is bounded on $\mathcal{P}(\mathcal{L}^m)$ and consequently is the generator of the group $e^{t\emph{L}^{m*}}$ for $t\geq 0$. For any fixed $T\in\mathbb{R}_+$, we have:
\begin{equation}
\lim_{\Delta t\rightarrow 0}\| (e^{-\Delta t\emph{L}^{m*}})^{\frac{T}{\Delta t}}- (\tilde{\mathcal{C}}^{m*}e^{-\Delta t\emph{L}_o^{m*}}\Xi^{m*} )^{\frac{T}{\Delta t}}\|_{\mathcal{P}(\mathcal{L}^m)}=0.
\end{equation}
\end{proposition}
\begin{remark}
The decomposition of $\emph{L}^{m*}$ through the adjoint of the restriction operator $\Xi^{m*}$, the adjoint of the order flow generator $\emph{L}_o^{m*}$ and the adjoint of the operator $\tilde{\mathcal{C}}^{m*}$ is similar to the decomposition of the operator $\emph{L}^*$ in the absolute coordinate framework. However, one should bear in mind that the expression of $\tilde{\mathcal{C}}^{m*}$ is different from that of the operator $\tilde{\mathcal{C}}^*$.
\end{remark}

\end{document}